\documentclass[pdflatex,sn-mathphys-num]{sn-jnl}

\usepackage{graphicx}%
\usepackage{multirow}%
\usepackage{amsmath,amssymb,amsfonts}%
\usepackage{amsthm}%
\usepackage{mathrsfs}%
\usepackage[title]{appendix}%
\usepackage{xcolor}%
\usepackage{textcomp}%
\usepackage{manyfoot}%
\usepackage{booktabs}%
\usepackage{algorithm}%
\usepackage{algorithmicx}%
\usepackage{algpseudocode}%
\usepackage{listings}%
\usepackage{mathtools, ucs, xparse, tikz, lmodern, anyfontsize, physics, varioref, cleveref, tensor}
\crefname{pluralequation}{eqs.}{eqs.}
\Crefname{pluralequation}{Eqs.}{Eqs.}
\crefformat{pluralequation}{eqs.~(#2#1#3)}
\Crefformat{pluralequation}{Eqs.~(#2#1#3)}
\crefname{table}{table}{tables}
\Crefname{table}{Table}{Tables}
\crefname{figure}{figure}{figures}
\Crefname{figure}{Figure}{Figures}
\renewcommand{\d}{\partial}

\newcommand{\Z}{\mathbb{Z}}

\newcommand\mperiod[1][\rlap]{#1{\, .}}	
\newcommand\mcomma[1][\rlap]{#1{\, ,}}
\newenvironment{eq}
    {\begin{equation}
    \begin{aligned}
    }
    { 
    \end{aligned}
    \end{equation}
    \ignorespacesafterend
    }

\theoremstyle{thmstyleone}%
\theoremstyle{thmstyletwo}%

\theoremstyle{thmstylethree}%

\begin{document}

\title[Aspects of strings without spacetime supersymmetry]{Aspects of strings

without spacetime supersymmetry}


\author*[1]{\fnm{Giorgio} \sur{Leone}}\email{giorgio.leone@sns.it}

\author*[2,3]{\fnm{Salvatore} \sur{Raucci}}\email{salvatore.raucci@uam.es}

\affil[1]{\orgname{Scuola Normale Superiore and INFN}, \orgaddress{\street{Piazza dei Cavalieri 7}, \city{Pisa}, \postcode{56126}, \country{Italy}}}

\affil[2]{\orgname{Instituto de F\'{i}sica Te\'{o}rica IFT-UAM/CSIC}, \orgaddress{\street{C/ Nicol\'{a}s Cabrera 13-15}, \city{Cantoblanco}, \postcode{28049}, \state{Madrid}, \country{Spain}}}

\affil[3]{\orgdiv{Departamento de F\'{i}sica Te\'{o}rica}, \orgname{Universidad Aut\'{o}noma de Madrid}, \orgaddress{\city{Cantoblanco}, \postcode{28049}, \state{Madrid}, \country{Spain}}}

\abstract{String theory relies on spacetime supersymmetry to guarantee the existence of stable vacua. In this review, we survey two features of non-supersymmetric strings that challenge both aspects: the appearance of tachyons and worldsheet tadpoles. We describe how tachyons arise, how to characterise their presence in closed strings and in their orientifold projections, and how off-shell approaches can be used to tackle them. We then turn to tachyon-free, non-supersymmetric strings. After introducing the simplest ten-dimensional models, we address the additional issues raised by tadpoles and the spacetime consequences of their cancellation. Finally, we discuss recent attempts to explore the non-supersymmetric string landscape.}




\maketitle

\newpage 

\tableofcontents

\section{Introduction}\label{sec:intro}

We understand string theory as a quantum theory of \emph{supergravity}. 
Despite considerable progress, the \emph{super} remains essential.

Spacetime supersymmetry is not required by the internal consistency of the theory. Nor is it observed in our nature. There are thus theoretical and phenomenological reasons why we must learn how to deal with the absence of supersymmetry in string theory. In fact, departing from the supersymmetric landscape reveals a richer mathematical structure of the theory, while providing insights into puzzles such as the cosmological evolution and the generation of hierarchies.

String theory was not born spacetime-supersymmetric, yet it grew supersymmetric in the last 50 years of research, and not for aesthetic reasons. Deep dynamical motivations underlie the imbalance between our understanding of supersymmetric and non-supersymmetric setups. 

Without the protection of supersymmetry, it is not clear whether stable vacua exist. This problem is overlooked in other contexts, such as when dealing with an effective field theory of gravity coupled to particle physics. However, it comes to light when gravity is treated in a fully quantum fashion. In string theory, it poses serious problems to the foundation of the whole framework because vacua are selected by the consistent propagation of strings and are not arbitrarily chosen.

Instabilities can manifest themselves in various ways: tachyonic modes, strong gravitational backreactions, and bubble nucleations.
In this review, we focus on the first two, tachyons in the string spectrum and string-generated scalar \emph{tadpole} potentials that jeopardise the stability of the vacuum. 

Tachyons, states with $m^2<0$, have appeared since the first formulation of the theory~\cite{Veneziano:1968yb} as the lightest modes in the bosonic string spectrum. Their role in quantum field theory is no mystery: they are interpreted as an indication that one is working in a vacuum that lies at the top of a potential hill. Indeed, they disappear when expanding around a minimum, as in the presence of spontaneous symmetry breaking. However, the would-be scalar potential in string theory is generally unknown, with a few exceptions. This motivated the search for tachyon-free strings.
To date, no general recipe has been developed to guarantee the absence of tachyons in non-supersymmetric setups.

This is by no means the only problem: a strong gravitational backreaction can arise in the form of scalar potentials generated by tadpole diagrams on the string worldsheet, even for tachyon-free models. Their physical interpretation is clear, but no complete and consistent scheme addresses them.

We again emphasise that even in tachyon-free theories, the \emph{existence} of a proper gravitational vacuum without supersymmetry is the problem: the hydrogen atom is stable because of quantum mechanics, and no supersymmetry is required. Similarly, non-BPS states in a supersymmetric string theory are fine, provided they exert negligible gravitational backreaction. The true concern is the global gravitational backreaction in the absence of supersymmetric protections.

Based on~\cite{Leone:2024xae,Raucci:2024fnp}, this review aims to address some of the key aspects behind these statements from the personal point of view and taste of the authors. 
The reader can consult~\cite{Schwarz:1982jn,Green:1987sp,Green:1987mn,DHoker:1988pdl,Polchinski:1996na,Dijkgraaf:1997ip,Polchinski:1998rq,Polchinski:1998rr,Sen:1998kr,Angelantonj:2002ct,Witten:2012bh,Mourad:2017rrl,Basile:2020xwi,Basile:2021vxh,Angelantonj:2024tns} for further details and~\cite{Abel:2018zyt,Itoyama:2020ifw,Angelantonj:2020pyr,Faraggi:2020hpy,Itoyama:2021fwc,Itoyama:2021itj,Avalos:2023mti,Nakajima:2023zsh,Avalos:2023ldc,Saxena:2024eil,Basaad:2024lno,Abel:2024vov,Larotonda:2024thv,Montero:2025ayi,Basile:2025mnj} for recent developments that will not be discussed here, although we consider them relevant to the overall picture.

The presentation follows two main routes. After introducing some background material in \cref{sec:background}, we discuss the features of tachyons in perturbative string theory in \cref{sec:closedtachyons}. We outline the original issue of tachyon condensation in \cref{ssec:tachyonsoffshell}, and review recent progress on characterising their presence in the spectrum in \cref{ssec:tachyonsworldsheet}.
\Cref{sec:stringtheory} introduces the ten-dimensional string models that we choose to focus on in this review. In particular, in \cref{ssec:closedstring10d} we present the closed-string models, while in \cref{ssec:orientifolds10d} we discuss the open strings obtained through the orientifold procedure.
Then, in \cref{sec:tadpoles}, we address the issue of gravitational backreaction in tachyon-free models, which arises from the divergences of \cref{ssec:sugimoto_divergence}. The generation of scalar potentials is discussed in \cref{ssec:FS} and their spacetime implications are presented in \cref{ssec:nonsusy_vacua}.

\section{String perturbation theory and vacuum amplitudes}
\label{sec:background}

String theory in the perturbative regime can be seen as a quantum field theory defined on {\em all} Riemann surfaces. Here, and in the rest of this review, we refer to a Riemann surface as a compact Euclidean manifold in two real dimensions equipped with a conformal structure. The compatibility with the latter demands diffeomorphisms and Weyl rescalings to be consistently gauged and hence to be anomaly-free symmetries preserved by quantum corrections.
As a result, choosing a particular gauge-fixing procedure, the theory can be described as a conformal field theory with a vanishing central charge, admitting a ghost sector and whose physical states are encoded in cohomology classes of the BRST operator.

This means that we can use the tools provided by conformal field theory to analyse the dynamics of the theory. Hence, we can properly define an amplitude in string theory, $\mathcal{A}_n$, as the correlator of $n$ vertex operators associated with physical states on all Riemann surfaces. In this sense, we can rearrange the expression of the correlator by showing the contribution of each Riemann surface, organised according to their Euler characteristic, $\chi$, controlled by the {\em string coupling} $g_s$: 
\begin{equation}\label{eq:stringyamplitudes}
    \mathcal{A}_n = \sum_{g,b,c=0}^{\infty} g_s^{-\chi} \mathcal{A}_{g,b,c,n} \, .
\end{equation}
In this expression, $n$ is the number of vertex operators\footnote{Each bulk vertex operator counts as $1$, while each vertex operator on the boundary as $\frac12$.}, $g$ denotes the genus, $b$ the number of boundaries and $c$ the number of crosscaps characterising the topology of the surface through $\chi=2-2g-b-c-n$\footnote{A manifold with a given genus, number of boundaries and crosscaps can always be described as the action of an involution on a compact closed complex curve with doubled Euler characteristic. In this spirit, the insertion of vertex operators can be seen as defining proper local coordinates around each insertion point \cite{Schweigert:2000ix,Schweigert:2001cu}, still defining a compact surface.}.

A convenient way to represent the amplitudes in \cref{eq:stringyamplitudes} is via the so-called {\em Polyakov path-integral} \cite{Polyakov:1981rd, Polyakov:1981re},
\begin{equation}\label{eq:pathintegral}
    \mathcal{A}_n \sim \sum_{g,b,c=0}^{\infty} g_s^{-\chi} \int \big [ D\phi \big ] \big [D\gamma \big ]_{g,b,c}  \  \phi(z_1) \dots \phi(z_n) \ e^{-S[\phi]} \, , 
\end{equation}
suitably regularised following the Faddeev-Popov procedure. In \cref{eq:pathintegral}, we have introduced the worldsheet metric $\gamma$, a set of fields $\phi$, and a proper action $S[\phi]$, with a vanishing Weyl anomaly. 
The latter, as mentioned above, is crucial for the theory to depend only on the conformal structure of the metric and not on the metric itself. In this sense, the dependence on $\gamma$ is redundant, since in a consistent theory only inequivalent conformal classes contribute to the amplitudes. These are obtained from the space of all the possible conformal structures modded out by the action of the automorphisms connected to the identity. This is the so-called {\em Teichm\"uller space}. We still have a non-trivial action of the automorphisms that are not connected to the identity, forming the so-called {\em modular group} (or {\em mapping class group}). The set of inequivalent conformal classes is thus obtained by taking the quotient of the Teichm\"uller space by the modular group. This is the definition of the {\em moduli space}, $\mathfrak{M}_{\chi}$, which is known to be finite-dimensional as a consequence of the {\em Riemann-Roch} theorem (see \cite{DHoker:1988pdl} for a proof of this statement). Indeed, according to the Riemann-Roch theorem, the difference between the dimension of the conformal Killing group and that of the moduli space should be a topological invariant given by $3\chi$. All in all, the amplitudes must be arranged into integrals over the moduli space of functions invariant under the action of the modular group. 

In the following, we will restrict our attention to the cases of free bosons and free fermions, describing the propagation of (super)strings\footnote{Actually, superstring perturbation theory is naturally defined on super-Riemann surfaces \cite{Witten:2012bh}. The description provided above requires the choice of a particular gauge-fixing for the $2d$ gravitino, which, however, does not hold globally for any genus, resulting in spurious singularities. For low genera, super-Riemann surfaces can be reduced to ordinary Riemann surfaces \cite{Witten:2012bg, Witten:2012ga}, so that in our discussion these subtleties do not play any role.} in flat Minkowski space, although in principle any theory satisfying the requirements explained above is a sensible string theory. Moreover, we will only discuss amplitudes with the insertion of the vertex operator associated with the vacuum of the theory (see \cite{DHoker:1988pdl} for a review on generic $n$-point scattering amplitudes). This plays the role of the identity in the fusion algebra. Since it is globally defined, no additional patch of coordinates is to be added. Hence, the amplitudes that we will consider in the following discussion correspond to $\mathcal{A}_0$.

The lowest order in this setting consists of the Riemann surface with $\chi=2$: the sphere. In this case, the Riemann-Roch theorem tells us that there are no moduli for the sphere since the associated conformal Killing group is $\text{PSL}(2,\mathbb{C})$, with real dimension $6$. Hence, the amplitude $\mathcal{A}_{0,0,0,0}$ can only be a constant built from characterising quantities of the theory. In particular, it has a natural interpretation as the cosmological constant of the spacetime theory under consideration. For instance, if the target space is Minkowski, this term is expected to vanish, while for AdS spaces it is not. These results in the cases of AdS spaces are well-motivated from an holographic perspective \cite{Henningson:1998gx, Eberhardt:2023lwd}, but no clear worldsheet derivation is yet available, aside from ad-hoc prescriptions, as in \cite{Liu:1987nz,Troost:2011ud}. Indeed, from a worldsheet perspective, the usual argument relies on the path-integral interpretation of the amplitudes where the Faddeev-Popov procedure leaves a residual redundancy, the ``volume" of the conformal Killing group (CKG). For the $0$-pt, $1$-pt, $2$-pt correlators on the sphere, the associated volume is infinite (the CKG of the sphere corresponds to $\text{PSL}(2,\mathbb{C})$). However, at this point one cannot simply conclude that the partition function vanishes, since infinite contributions may arise from locally defined conformal transformations, as is the case for the Liouville field theory \cite{Distler:1988jt, Maltz:2012zs, Anninos:2021ene, Mahajan:2021nsd}, or from delta function singularities imposing momentum conservation, as happens for the $2$pt correlator \cite{Erbin:2019uiz}. The sphere partition function is interpreted as the on-shell action at tree level of closed-string modes. In particular, this involves the gravitational piece for which it is unclear how to provide a proper definition without using a regularisation scheme that breaks Weyl invariance \cite{Gibbons:1976ue}.    

For $\chi=1$, we only have two possible surfaces: the disc $D_2$ and the projective plane $\mathbb{RP}^2$. The conformal Killing group of the projective plane is $\text{SO}(3)$, while that of the disc is $\text{PSL}(2,\mathbb{R})$, which is still non-compact. To corroborate the limited understanding of these integrals, the na\"if infinite-volume argument described for the sphere would lead us to conclude that all the $0$-pt, $1$-pt, and $2$-pt correlators vanish. However, the $0$-pt amplitudes are connected via soft theorems \cite{Shapiro:1975cz, Ademollo:1975pf} to the $1$-pt correlators, which, as will be seen in the following, are known to be non-vanishing from the factorisation of the annulus amplitude. Moreover, the one-point functions of bulk fields are interpreted as measurable features of suitable extended objects, D-branes and orientifold planes~\cite{Sagnotti:1987tw,Pradisi:1988xd,Bianchi:1990yu,Bianchi:1991eu, Polchinski:1995mt}. In \cite{Eberhardt:2021ynh}, a careful analysis showed how the disc partition function does not vanish, refining the argument provided in \cite{Liu:1987nz}, and reproduces the tension of D-branes of the critical bosonic string in flat Minkowski space. Nevertheless, the overall picture remains incomplete: the gauge-fixing procedure of \cite{Eberhardt:2021ynh} should be consistent for both the disc and the sphere partition function. In fact, the latter would naively imply a vanishing result even for theories on AdS spaces, contrary to the expected physical interpretation; therefore, the analysis on the sphere is yet to be settled. 

The disc partition function is interpreted as the tree-level on-shell action of additional degrees of freedom living on the D-brane. The projective-plane contribution encodes the effect of the presence of orientifold planes. These do not add any new degrees of freedom, at least in string perturbation theory, as they can generally have negative tension. They indeed appear as the fixed locus of the parity operator on the string worldsheet and therefore are only compatible with unoriented strings. 

For $\chi=0$, we have four possible surfaces: the torus $T^2$, the Klein bottle $\text{KB}$, the annulus $\mathcal{A}$, and the M\"obius strip $\mathcal{M}$. These latter three amplitudes have no handles but boundaries and crosscaps, so that they are absent if we restrict our attention to the case of oriented closed surfaces associated with the propagation of closed strings. The torus has one modulus and conformal Killing group $\text{U}(1)\times \text{U}(1)$, corresponding to the translations along its fundamental cycles. This is the first example with a non-trivial space of Teichm\"uller parameters, on which the mapping class group acts as a $\text{PSL}(2,\mathbb{Z})$ transformation,
\begin{equation}
    \tau \to \frac{a \tau+b}{c \tau +d} \, , \qquad \begin{pmatrix}
        a & b \\ c & d
    \end{pmatrix} \in \text{PSL}(2,\mathbb{Z}) \, ,
\end{equation}
where $\text{PSL}(2,\mathbb{Z})$ admits the presentation $\big \langle S,T, C \  | \ S^2=C, \, (ST)^3=C, \, C^2=\mathbf{1} \big \rangle $ with
\begin{equation}
    S \, : \, \tau \to -\frac{1}{\tau} \, , \qquad T \, : \, \tau \to \tau+1 \, .
\end{equation}
With this in mind, the torus amplitude must have the following form:
    \begin{equation}\label{eq:torusamplitude}
       \mathcal{A}_{1,0,0,0} \propto \int_{\mathcal{F}} d\mu \, \mathcal{T}(\tau,\bar{\tau}) \, ,
    \end{equation}
where $d \mu = d^2 \tau / \tau_2^2$ is the modular-invariant measure and $\mathcal{T}(\tau,\bar{\tau})$ is a modular-invariant function specifying the features of the vacuum. In \cref{eq:torusamplitude}, $\mathcal{F}$ is the moduli space (also known as fundamental domain), obtained by taking the closure of 
\begin{equation}
     \Big \{ \tau \in \mathbb{C} \, | \, | \text{Re}\, \tau| < \frac12 \, , \, \text{Im} \, \tau >0 \, , |\tau|>1 \, \Big \} \, ,
\end{equation}
and identifying the $\text{Re} \, \tau >0$ part of the boundary with the $\text{Re} \, \tau <0$ component.
It is worth mentioning that the UV divergences are absent since the point $\tau_2 \to 0$ is excluded from the fundamental domain. Hence, modular invariance guarantees the UV finiteness of string theory.
The modular function $\mathcal{T}(\tau,\bar{\tau})$ can always be expressed in terms of suitable combinations of Virasoro characters, defined as 
\begin{equation}\label{eq:characters}
    \chi_i(\tau)= \text{Tr}_{\mathcal{V}_i} \, q^{L_0 - \frac{c}{24}} \, ,
\end{equation}
where $q=e^{2\pi i \tau}$ and $\mathcal{V}_i$ is the Verma module associated to the Virasoro primary with conformal weight $h_i$. In general, string theory allows for irrational conformal field theories leading a continuous spectrum, so that the torus partition function takes the form
\begin{equation} \label{eq:modularfunction}
    \mathcal{T}(\tau, \bar{\tau})= \displaystyle{\int\mathllap{\sum}}_{i,\bar \iota} N_{i,\bar \iota} \, \chi_i(\tau) \bar \xi_{\bar{\iota}}(\bar\tau) \, .
\end{equation}
The matrix $N$ enforces a modular-invariant combination of the characters associated with the holomorphic (left-moving) sector $\chi(\tau)$ and anti-holomorphic (right-moving) sector $\bar \xi(\bar \tau)$. The coefficients may be described as the result of a projection imposed on a larger Hilbert space which allows to make the theory local. This is the well known {\em Gliozzi-Scherk-Olive (GSO) projection} \cite{Gliozzi:1976jf, Gliozzi:1976qd}. For the superstring, choosing a specific GSO projection means choosing a suitable way to sum over {\em all} the spin structures \cite{Seiberg:1986by} that we have to specify when introducing fermions on closed cycles. In general, the continuous spectrum arises from the zero modes associated with non-compact coordinates, resulting in non-trivial powers in $\tau_2$. The general form of the torus partition function is
\begin{equation}\label{eq:toruspartitionfunction}
    \mathcal{T}(\tau, \bar \tau)= \tau_2^{1-d/2} \sum_{i, \bar \iota} N_{i,\bar \iota} \, \chi_i(\tau) \bar \xi_{\bar{\iota}}(\bar\tau) \, ,
\end{equation}
where $d$ is the dimension of the non-compact space, and thus the presence of $\tau_2$ is a gentle reminder that we are dealing with an irrational conformal field theory with a continuous spectrum. 

The characters defined in \cref{eq:characters} admit a Fourier expansion of the form:
\begin{equation}\label{eq:characterq}
    \chi_i(\tau)= \sum_{n=0}^\infty d_i(n) \, q^{h_i+n-\frac{c}{24}} \, ,
\end{equation}
where $n$ indicates the level of the Verma module and $d_i(n)$ the associated degeneracy. These functions are, in general, meromorphic in the $q$-plane, as is the case for the character associated with the Virasoro primary of conformal weight $h_0=0$. In CFT, such a term is associated with the identity of the fusion algebra and it is always present in order to admit a modular-invariant partition function\footnote{However, in superstring theory, spin-statistics identifies the identity of the fusion algebra as the character with conformal weight $h=\frac12$. This apparent light modification is nonetheless crucial since it implies that gravity is always part of the string spectrum also in the superstring case.}. In flat Minkowski space, the exponent and coefficients in the $q$($\bar q$)-expansion of the characters are associated with the mass and the degeneracy of states in the (anti-)holomorphic sector. A consistent interpretation of the string spectrum requires the mass terms from the left- and right-moving sectors to be equal, $h_i+n-c/24=\bar{h}_{\bar \iota} + \bar{n}-\bar c/24$, enforcing the {\em level-matching} condition. Note that level-matching is a stronger requirement than the invariance of $\mathcal{T}(\tau,\bar \tau)$ under a $T$ modular transformation. In fact, $h_i - c/24 - \bar h_{\bar \iota} + \bar c/24 \in \mathbb{Z}$ is a {\em necessary} requirement for the level matching condition to have solutions. However, two states belonging to characters that satisfy $h_i - c/24 - \bar h_{\bar \iota} + \bar c/24 \in \mathbb{Z}$ while having $h_i+n-c/24 \neq \bar{h}_{\bar \iota} + \bar{n}-\bar c/24$ are not level-matched and do not contribute to the physical spectrum. 

Given the behaviour of the characters for $q \sim 0$ ($\tau_2 \to \infty$), the modular-invariant combination $\mathcal{T}(\tau,\bar{\tau})$ in \eqref{eq:modularfunction}, is expected to conform to one of the following scenarios:
\begin{itemize}
    \item the partition function converges as 
    \begin{equation}
        \mathcal{T}(\tau,\bar{\tau}) \sim \tau_2^{\alpha} \Big ( \kappa + O(e^{-4 \pi m^2 \tau_2}) \Big ) \, ,  
    \end{equation}
    where $\kappa$ is a constant, $\alpha\leq 0$, and $m^2 >0$. In this case, the meromorphic behaviour of the Virasoro characters in the $q$ plane is compensated by the choice of the GSO projection that guarantees a modular-invariant partition function. This is the case for the supersymmetric type II theories, for which $\kappa=0$. 
    
    \item the partition function diverges but the polar part is eliminated upon integrating over $\tau_1$:
    \begin{equation}
        \mathcal{T}(\tau,\bar{\tau}) \sim \tau_2^{\alpha} \Big ( \kappa \, e^{4 \pi i m^2 \tau} + O(1) \Big ) \, ,  
    \end{equation}
    where $\kappa$ is a constant, $\alpha \leq 0$ and $m^2 >0$. This situation describes theories with only {\em unphysical} tachyons: tachyons that do not satisfy the level-matching condition required to guarantee a consistent interpretation of the spectrum. Although the partition function $\mathcal{T}$ grows exponentially at the cusp, the value of the path-integral at the one-loop level, $\Gamma_1$, is finite when a proper order of integration is chosen \cite{Kutasov:1990sv}. An example of this kind is the $\text{Spin}(16) \times \text{Spin}(16) \rtimes \mathbb{Z}_2$ ten-dimensional heterotic theory.

    \item the partition function diverges with a non-trivial Fourier mode in $\tau_1$,
    \begin{equation}
        \mathcal{T}(\tau,\bar{\tau}) \sim \tau_2^{\alpha} \Big ( \kappa \, e^{4 \pi  m^2 \tau_2} + O(1) \Big ) \, ,  
    \end{equation}
    where $\kappa$ is a constant, $\alpha \leq 0$, and $m^2 >0$. In this case, {\em physical} tachyons are present in the spectrum, leading to a divergent contribution to $\mathcal{A}_{1,0,0,0}$. Examples of this kind are the type 0 superstring theories and all heterotic non-supersymmetric theories aside from the $\text{Spin}(16) \times \text{Spin}(16) \rtimes \mathbb{Z}_2$ one. 
\end{itemize} 

The other Riemann surfaces with $\chi=0$ are obtained by adding boundaries and crosscaps on the string worldsheet. Adding two crosscaps leads to the Klein bottle amplitude, which can be obtained from a double covering torus with Teichm\"uller parameter $\tau=2 i \tau_2$ through a suitable involution. The Hilbert space is obtained by acting on the oriented closed-string sector with the worldsheet parity operator $\Omega$, which exchanges the left-moving and right-moving states. This operator satisfies the condition $\Omega^2=\mathbf{1}$, and is used to project the spectrum onto states which are invariant under the parity transformation on the worldsheet. This is the so-called {\em orientifold projection} and leads to a theory of unoriented strings \cite{Sagnotti:1987tw,Pradisi:1988xd,Horava:1989vt,Bianchi:1990tb,Bianchi:1990yu,Bianchi:1991eu,Dudas:2000bn}. Therefore, the general shape of the Klein bottle amplitude has the following form:
\begin{equation}\label{eq:directklein}
    \mathcal{A}_{0,0,2,0} \propto \int_0^{\infty} \frac{d\tau_2}{\tau_2^2} \, \mathcal{K}(2i\tau_2) = \int_0^{\infty} \frac{d\tau_2}{\tau_2^2} \, \tau_2^{1-\frac{d}{2}} \sum_i K_i \ \chi_i(2i\tau_2)\, ,
\end{equation}
where $K_i=N_{ii}$ are defined from the torus amplitude. As we will see in the following discussion, the orientifold projection can be dressed with additional $\mathbb{Z}_2$ automorphisms \cite{Fioravanti:1993hf} of the fusion algebra that are compatible with the torus partition function.
It is worth stressing that the involution used to obtain the Klein bottle amplitude breaks modular invariance. Indeed, the modular group acts covariantly on the characters, and provides different descriptions of the same amplitude. In particular, the generator $S$ swaps proper length and time on the string worldsheet so that the amplitude can also describe a tree-level propagation of closed string states between two crosscap states:
\begin{equation} \label{eq:transverseklein}
     2^{\frac{d}{2}} \int_0^{\infty} d \ell  \,  \sum_i \tilde{K}_i \ \chi_i(i\ell)\, ,
\end{equation}
where $\ell=\frac{1}{2\tau_2}$ and $\tilde{K}_i= \big ( S \cdot K\big )_i$. In this spirit, the coefficients $\tilde{K}_i$ admit a natural interpretation as squares of $1$-pt functions of bulk fields in front of a crosscap, $\tilde{K}_i = \big ( \Gamma_i \big )^2$. Only fields appearing with their conjugates in the torus partition function can propagate in the transverse channel. As mentioned before, the orientifold action admits a non-trivial fixed locus which corresponds to extended objects known as {\em orientifold planes}. Therefore, the transverse channel of the Klein-bottle amplitude is interpreted as the propagation of closed-string states between a pair of orientifold planes \cite{Bianchi:1990yu, Bianchi:1991eu}. 

The annulus amplitude is a surface with no handles and two boundaries, and it is obtained via a non-trivial involution from a double covering torus with Teichm\"uller parameter $\tau=\frac{i \tau_2}{2}$. This means that we can generally write down partition functions that take the form
\begin{equation}\label{eq:directannulus}
    \mathcal{A}_{0,2,0,0} \propto \int_0^{\infty} \frac{d\tau_2}{\tau_2^2} \mathcal{A}(i\tau_2/2) = \int_0^{\infty} \frac{d\tau_2}{\tau_2^2} \, \tau_2^{1-\frac{d}{2}} \sum_i A^i \ \chi_i(i\tau_2/2) \, .
\end{equation}
This amplitude is interpreted as the one-loop propagation of open string whose endpoints lie at the boundary of the surface, which thus corresponds to D-branes. It is possible to consider a stack of $n$ parallel D-branes, which provide an additional degeneracy of the open-string ground state. This extra degeneracy encodes the possibility of open-string states to transform in a representation of a (non-)abelian symmetry indicated by the so-called {\em Chan-Paton} labels \cite{Paton:1969je}. When computing the traces of the one-loop partition function, there could be a non-trivial degeneracy $n$, corresponding to the dimension of the representation in which the open-string states transform,
\begin{equation}
    A^i=\sum_{a,b}A^i_{ab} n^a n^b \, ,
\end{equation}
with $n^a$ known as the {\em Chan-Paton} multiplicities. The transverse channel is obtained by performing an $S$ modular transformation:
\begin{equation}\label{eq:transverseannaulus}
     2^{-\frac{d}{2}} \int_0^{\infty} d \ell  \,  \sum_i \tilde{A}^i \ \chi_i(i\ell)\, .
\end{equation}
This is interpreted as resulting from closed strings propagating between D-brane boundary states, so that the only characters allowed in the transverse amplitude are bulk fields appearing with their conjugates in the torus partition function. Similarly to the case of the Klein bottle amplitude, we can interpret the coefficients $\tilde{A}^i$ as squares of the $1$-pt functions of bulk fields in front of a boundary. However, the presence of the Chan-Paton labels allows us to factorise the extra degeneracy coming from the Chan-Paton matrices, so that 
\begin{equation}
 \tilde{A}^i=\Big ( \sum_a B^i_a n^a \Big )^2 \, .
\end{equation}

The last amplitude with $\chi=0$ is the M\"obius strip amplitude. In this case, the double covering torus has the same imaginary part as the annulus, namely $\tau_2/2$, but it also has a non-trivial real part, $\tau_1=1/2$. This makes the characters complex. Since the partition function is real, it is convenient to express their contribution in terms of the real basis defined by 
\begin{equation}
    \hat{\chi}_i(1/2 +i \tau_2/2)= e^{-\pi i(h_i-c/24)} \chi_i (1/2 +i \tau_2/2) \, .
\end{equation}
The partition function is thus encoded in
\begin{equation}\label{eq:directmoebius}
    \mathcal{A}_{0,1,1,0} \propto \int_0^{\infty} \frac{d\tau_2}{\tau_2^2} \mathcal{M}(1/2+i\tau_2/2) =  \int_0^{\infty} \frac{d\tau_2}{\tau_2^2} \, \tau_2^{1-\frac{d}{2}} \sum_{i,a} \varepsilon_i M^i_a n^a  \ \hat \chi_i(1/2 + i\tau_2/2) \, .
\end{equation}
where $\varepsilon_i$ are reflection coefficient signs associated with each character $\hat{\chi}_i$.
The modular transformation 
\begin{equation}
    P=T^{\frac12}ST^2ST^{\frac12}
\end{equation}
inverts proper time and length on the string worldsheet, thus leading to
\begin{equation} \label{eq:transversemoebius}
     2  \int_0^{\infty} d \ell  \,  \sum_{i,a} \varepsilon_i \tilde{M}^i_a n^a  \ \hat{\chi}_i(1/2+i\ell)\, .
\end{equation}
The M\"obius strip amplitude has no handles, one boundary, and one crosscap. The interpretation in the transverse channel consists of closed strings propagating between a D-brane and an O-plane. This means that we can naturally interpret $\tilde{M}^i_a=B^i_a \Gamma_i$, with $\varepsilon_i$ being the relative sign between the two. The factor of $2$ reflects the two possible choices at our disposal: a closed string can start from a boundary and end up on a crosscap, and vice-versa. The sum of the annulus and the M\"obius strip partition functions allows one to read the spectrum of the (unoriented) open string, which thus requires the terms appearing in the direct and transverse channel of both amplitudes to enforce a consistent particle interpretation. We come back to this point with some concrete examples.

In the following, we shall not deal with amplitudes for $\chi <0$. Nonetheless, given the locality of the theory on the string worldsheet, one may wonder whether its consistency at lower orders suffices to ensure the consistency of the theory on all Riemann surfaces. In particular, as a consequence of locality, these amplitudes factorise when the Riemann surface degenerates (by means of the operator product expansion). In general, there could be many inequivalent ways for such factorisations to occur. Requiring the amplitudes to be independent of this cutting and sewing procedure yields the so-called {\em sewing constraints}. Two conditions suffice to guarantee the consistency of oriented closed strings on any oriented Riemann surface: crossing symmetry of the $4$-pt amplitude on the sphere~\cite{Sonoda:1988mf,Sonoda:1988fq, Moore:1988uz, Moore:1988qv} and modular invariance~\cite{Sonoda:1988mf,Sonoda:1988fq}. 

Modular invariance also plays a role in the cancellation of local spacetime anomalies \cite{Schellekens:1986xh, Schellekens:1986yi}, which in general require the implementation of the Green-Schwarz-Sagnotti mechanism \cite{Green:1984sg, Green:1984bx,Sagnotti:1992qw}. Recently, a deeper understanding of anomalies has paved the way for new types of global anomalies, emerging from the possibility that the topology of spacetime changes. These are the {\em Dai-Freed} anomalies \cite{Garcia-Etxebarria:2018ajm}. Up to date, there is no worldsheet property that ensures their cancellation, except for a few cases \cite{Tachikawa:2021mby, Tachikawa:2021mvw}, and one must proceed with a case-by-case analysis \cite{Debray:2021vob, Debray:2023yrs, Basile:2023knk, Basile:2023zng}.   

When boundaries and crosscaps are introduced, additional sewing constraints emerge from the $4$-pt and $5$-pt amplitudes, involving both bulk and boundary fields \cite{Pradisi:1996yd}, from the $4$-pt amplitudes of only boundary fields \cite{Pradisi:1996yd}, and from the $2$-pt correlator of bulk fields in the presence of a crosscap (the {\em crosscap constraint} \cite{Fioravanti:1993hf,Pradisi:1996yd}, see \cite{Stanev:2001na} for a detailed discussion). 
These constraints apply to a generic conformal field theory with boundaries and crosscaps. In string theory, consistency requires an additional condition. In orientifold theories, modular invariance is absent, so that the amplitudes are not protected against the dangerous UV limit $\tau_2 \to 0$. As we will see, not all divergences are born equal: only one type threatens the consistency of the theory. 

To better understand this point, we now discuss the divergences in string amplitudes in detail.

\subsection{Moduli space divergences}
\label{ssec:moduli_space}

In string amplitudes, divergences can arise when Riemann surfaces degenerate or, equivalently, at the boundary of their moduli space~\cite{Gava:1986ei}. From the Riemann surface perspective, these limits correspond to some topological features that shrink. In closed strings, this occurs when a cycle collapses. Collapsing cycles can be of two different types: non-separating cycles or separating cycles, depending on whether pinching leaves a connected Riemann surface. 

In the first case, the degeneration is conformally equivalent to a handle becoming infinitely long; therefore, in a quantum-field-theory interpretation, it can be viewed as a loop integral. This degeneration contributes with the large-$t$ behaviour of
\begin{eq}\label{eq:tachyon_divergence}
    \sim t^{-1-\frac{D}{2}}\text{Str}\left(e^{-t m^2}\right)\mperiod
\end{eq}
Divergences of this type arise from closed-string tachyons. Similar divergences are present in the case of open strings. These lead to exponential divergences in the $\tau_2 \to \infty$ limit (i.e.~$\ell\to 0$), and are responsible for the exponential divergence that we encountered in the previous section. In this degeneration limit, the genus of the Riemann surface decreases by 1 and two tachyon vertex operators are inserted. This channel also contributes to another physical effect that we shall not consider: mass renormalisation---of the tachyon in this specific case---and its influence on the vacuum energy.

The second case of separating-cycle degeneration is conformally equivalent to an infinitely long cylinder that connects two sub-diagrams. Open-closed duality maps this geometry into two sub-diagrams that exchange a particle, each of which contributes with the propagator
\begin{eq}\label{eq:tachyon_and_dilaton_divergence}
    \sim \int^\infty e^{-t m^2}\mperiod
\end{eq}
States with $m^2 > 0$ give vanishing contributions, while those with $m^2\leq0$---tachyons and massless modes---generate divergences.
Note that this is also relevant for open strings, in which the propagation of tachyons implies an exponential divergence of the amplitudes in the transverse channel as $\ell \to \infty$. For this type of degeneration, the genus of the original Riemann surface is split between the two sub-diagrams, and a single additional vertex operator is inserted in each of them.

For the superstring, divergences of the type of \cref{eq:tachyon_and_dilaton_divergence} involving massless scalars arise from two different sectors, which are {\em a priori} unrelated unless spacetime supersymmetry is present, and have a completely different interpretation.
Massless modes coming from the R-R closed-string sector identify the propagation of a top form without a kinetic term. These are a true challenge to the consistency of the theory; in fact, local spacetime anomalies develop when R-R tadpoles are not cancelled\footnote{No analogous results are known for the cancellation of the global anomalies mentioned before.}  \cite{Bianchi:2000de, Aldazabal:1999nu}. 
Instead, the divergences arising from the propagation of massless NS-NS fields do not translate into inconsistencies. They admit a natural interpretation in terms of vacuum instability. We will postpone this discussion to \cref{sec:tadpoles}. 

In the following, we will focus on the divergences coming from tachyons. Conceptually, we should not be worried by their presence since they are not a threat to the consistency of the theory. Their interpretation is indeed inherited by our intuition in quantum field theory: in QFT, tachyons appear whenever the vacuum of the theory sits on a maximum of a given scalar potential. It is then well-known how to fix the problem: one must find the global minimum of the potential to properly quantise the theory. In the same spirit, the appearance of tachyons is interpreted as the instability of the background on which string theory is quantised. To corroborate such a picture, one can compute the one-loop vacuum-to-vacuum amplitude, $\mathcal{A}_{1,0,0,0}$, by performing an analytic continuation to the complexified moduli space and apply the $i \varepsilon$ prescription \cite{Witten:2013pra, Eberhardt:2022zay}. One then finds a non-trivial negative imaginary part implying the decay of the vacuum amplitude \cite{Marcus:1988vs, Baccianti:2025gll}, which thus reflects the instability of the vacuum. 
Ideally, such an interpretation of the issue indicates the proper way to proceed: one must find the true vacuum of the theory in which tachyons (and eventually other sources of instabilities that may arise from higher loops) are absent. 

However, answering this question is generally difficult. Indeed, a proper treatment of the issue would require an off-shell formulation of string theory from which one would read the potential characterising the dynamics. This is not accessible from the standard worldsheet formulation, which is necessarily on-shell: this would involve choosing a background that cancels the Weyl anomaly before computing the spectrum and the scattering amplitudes. 

There are two known approaches in the literature that provide an off-shell formulation of string theory. The first follows the sigma model approach \cite{Fradkin:1984pq, Fradkin:1985ys, Tseytlin:1986ti, Metsaev:1987ju} (see \cite{Tseytlin:1988rr, Tseytlin:1988mw} for reviews and \cite{Ahmadain:2022tew, Ahmadain:2024hdp} for a modern point of view), which produces an off-shell effective action once all massive fields are integrated out. The other approach follows string field theory, in which the string dynamics is encoded in a functional of the {\em string field}. In the following, we will mainly follow the first approach and refer the interested reader to \cite{Sen:2024nfd} for a detailed review on the latter formalism.

\section{The tachyon problem}\label{sec:closedtachyons}

\subsection{Off-shell approaches}\label{ssec:tachyonsoffshell}

On-shell string theories correspond to conformal field theories. A natural way to obtain an off-shell extension consists of starting with a general QFT on {\em all} two-dimensional Riemannian manifolds. In this picture, the on-shell vacua are fixed points of the renormalisation group flow and are thus obtained by solving for the vanishing of the $\beta$-functions. It was suggested in \cite{Fradkin:1984pq,Fradkin:1985ys} that such a quantum field theory would be obtained by coupling currents $J_i$ with all string vertex operators $V_i$, resulting in the so-called {\em sigma model}. This would provide a generalisation of the path-integral defined in \cref{eq:pathintegral}, whose expression corresponds the generating functional for string correlators. This corresponds to the {\em quantum effective action} for string theory, schematically
\begin{equation}
    S \big [ \{J_i \} \big ] \sim \mathcal{A}_{0} \big [ \{ J_i \} \big ] \, .
\end{equation}
To obtain a consistent description, the equations of motion of such an action must be compatible with the conditions for conformal invariance from the vanishing of the $\beta$-functions. Moreover, around an on-shell point, hence around a solution of the equation of motion, such an off-shell formalism should reproduce the on-shell string amplitudes.
In principle, one can start with the most general worldsheet action that is invariant under $2d$ diffeomorphisms. To make our discussion a bit more concrete we can take, for instance, the case of the bosonic string,
\begin{equation}
\begin{aligned}\label{eq:sigmamodel}
    S[\{\phi_i,J_i\}]= \frac{1}{4 \pi \alpha'} \int d^2\sigma \sqrt{\gamma} \bigg \{& T(X) + \gamma^{ab} \partial_a X^\mu \partial_b X^\nu G_{\mu \nu}(X) 
    \\
    &+ \varepsilon^{a b} \partial_a X^\mu \partial_b X^\nu B_{\mu \nu}(X) + \alpha' R^{(2)} \Phi(X) + \ldots  \bigg\} \, ,
    \end{aligned}
\end{equation}
with $G_{\mu \nu}, \, B_{\mu \nu}, \, T, \,  \Phi$ being the symmetric, anti-symmetric, and scalar tensors associated, as we will see, with the corresponding string states, and with the dots denoting the contributions from higher-order operators. Neglecting the latter terms, the theory is renormalisable\footnote{To be precise, the worldsheet theory, even with only massless and tachyonic modes, is not renormalisable in general: around a given background for $X$ we can expand the various currents leading to the presence of infinitely many operators which mix under the RG flow. Therefore, the previous statement holds if we assume to be close to the on-shell fixed point.}  in $\alpha'$ and the operators appearing in \cref{eq:sigmamodel} correspond to the massless and tachyonic modes in the string spectrum. Higher-order operators corresponding to massive string states would be relevant and would lead to a non-renormalisable theory, in which all operators mix under the RG flow. To obtain a feasible action, one should properly decouple these operators \cite{Tseytlin:1988rr, Tseytlin:1988mw}.

In \cref{eq:sigmamodel}, we have allowed for the presence of two operators which are incompatible with Weyl invariance. These two terms reflect the presence of the tachyon and of a massless scalar field, the {\em dilaton} $\Phi(X)$. The dilaton is coupled to the worldsheet curvature, $R^{(2)}$; if constant, it reproduces the string coupling $g_s^{-\chi}$ in \cref{eq:pathintegral}. Therefore, the string coupling has a natural interpretation as vacuum expectation value of the dilaton field. This is a universal feature of string theory: the only physical parameter is provided by the string length, $\alpha'$, while all the other parameters are vacuum expectation values of string states. 

Being off-shell, the presence of these two terms is not a problem in our treatment since conformal invariance is a property of the theory at the on-shell points, following from the vanishing of the $\beta$-functions. However, new issues emerge in the absence of manifest Weyl invariance. Indeed, in the conformal gauge, the path-integral does not factorise the aforementioned conformal Killing group and the associated divergences at the tree level are now translated into UV divergences of the sigma model. This means that a proper definition of the quantum effective action requires the path-integral to be renormalised\footnote{For closed strings, the situation is a bit more involved. For open strings, these UV divergences can be absorbed by a redefinition of the tachyon term. For closed strings, one has to subtract massless poles as well \cite{Tseytlin:1987ww, Tseytlin:1988tv}. In what follows, we will ignore this subtle technicality and focus on the general philosophy.} \cite{Tseytlin:1987ww, Tseytlin:1988tv}, raising non-trivial challenges to the foundation of the sigma-model whenever higher-order couplings are introduced. This means that a choice of a renormalisation scheme is required, which translates into different terms in the quantum effective action \cite{Tseytlin:1993df}. This ambiguity reflects the possibility to perform regular field redefinitions that leave the on-shell amplitudes invariant and correspond to different off-shell extensions. At first sight, this may look unpleasant, but this freedom actually reflects the background independence of string theory. Indeed, different extensions lead to different equations of motion in the perturbative $\alpha'$ expansion, admitting apparently different solutions. However, such solutions are related to each other via field redefinitions, thus reflecting the possibility to take any of the off-shell extensions as a starting point\footnote{For instance, in \cite{Tseytlin:1993df}, there exists a scheme admitting a $2d$ perturbative solution in $\alpha'$ which can be mapped into an exact solution via a field redefinition in which all $O(\alpha'^2)$ terms vanish.}. 

When the tachyon has a non-trivial profile, $T(X) \neq 0$, similar considerations hold and lead to a tachyon potential. However, this has been argued to be ambiguous~\cite{Banks:1991sg, Tseytlin:1991bu,Tseytlin:1993df,Tseytlin:2000mt}. 
Moreover, for the bosonic string, the quantum effective action in any chosen off-shell extension is incompatible with the tachyon $\beta$-function \cite{Tseytlin:2000mt}. Indeed, the effective action does not admit the standard Minkowski ``vacuum'' with $T=0$ as a solution of the equations of motion, which is incompatible with the vanishing of the $\beta$-function. To fix this, following a previous treatment in the context of string field theory~\cite{Witten:1992qy, Witten:1992cr, Shatashvili:1993kk, Shatashvili:1993ps}, \cite{Tseytlin:2000mt} suggested to perform a {\em singular} field redefinition according to which
\begin{equation}
    S=\mathcal{A}_0 \to \hat{S} \equiv S + \beta_T \frac{\delta S }{\delta T} \, .
\end{equation}
This shift can restore the compatibility between the equations of motion and the vanishing of $\beta$-functions since  \cite{Tseytlin:2000mt}
\begin{equation}
    \delta_i \hat S=\kappa_{ij} \beta^j= \delta_i S + \delta_i \beta^j \delta_j S + \beta^j \delta_i \delta_j S \to  ( \kappa_{ij} -\delta_i \delta_j S ) \beta^j = ( \delta_i^j + \delta_i \beta^j ) \delta_j S \, .
\end{equation}
Hence, $\beta^j=0$ is still compatible with $\delta_j S \neq 0$ if $\delta_i^j + \delta_i \beta^j=0$. This is precisely what happens for the open bosonic string tachyon. For the closed bosonic string a similar approach does not lead to a quantum effective action that is compatible with the $\beta$ function of the tachyon. The search for a tachyon-free vacuum of the closed bosonic string is still underway.  

For open strings, the situation is more under control. For the open bosonic string and for the open superstring, the exact scalar potentials for the tachyons are $e^{-T}(1+T)$ and $e^{-T^2}$. For the bosonic string, we have two solutions: $T=0$ and $T=\infty$. The effective action at the unstable point $T=0$ is related to the tension $T_{25}$ of the D$25$ brane~\cite{Kutasov:2000qp},
\begin{equation}
    \hat{S}[T=0] \sim T_{{25}} V_{26}\mcomma
\end{equation}
where $V_{26}$ is the spacetime volume. The on-shell action vanishes as $T \to \infty$, where the boundary operators flow into the identity of the fusion algebra and the open string poles in the propagators disappear \cite{Harvey:2000na}. The $T \to \infty$ solution is then interpreted as the tachyon condensate realising a pure bulk theory with no D-branes, as conjectured in~\cite{Sen:1998ii, Sen:1998sm}. This is consistent with the level-truncation approach in string field theory \cite{Sen:1999nx, Schnabl:2005gv} (see \cite{Sen:2024nfd} for a detailed analysis). Similar results hold for the open superstring, which admits $T=0, \, \pm\infty$ as solutions. $T=0$ corresponds to a Minkowski background with a non-trivial open-string tachyon, and represents the simultaneous presence of branes and anti-branes.

Recently, linear dilaton backgrounds have been discussed for the tachyonic heterotic string \cite{Hellerman:2007zz,Kaidi:2020jla}, in which tachyon condensation leads to meta-stable vacua in lower dimensions. These have been argued to represent the near-horizons of non-supersymmetric branes in supersymmetric heterotic theories \cite{Kaidi:2023tqo, Kaidi:2024cbx}.  

The above discussion reinforces the intuition that tachyons are instabilities of string backgrounds. In most cases, we do not have full technical control over these backgrounds, especially for closed strings. Even for open strings, we lack a systematic approach to include higher-order effects.

\subsection{Characterisation from the worldsheet} \label{ssec:tachyonsworldsheet}

From the previous discussion, it is clear that, although we understand the meaning of tachyonic backgrounds, finding the true vacuum of the theory remains a technically challenging task. This state of affairs ultimately reflects our poor understanding of tachyonic backgrounds. Therefore, one would like to focus on theories without tachyons, on which the background can be trusted at least at tree level. Actually, the requirement of absence of tachyonic contributions in one-loop open and closed string vacuum-to-vacuum diagrams dictates a precise behaviour of the corresponding one-loop partition function in the large-mass limit. This emerges from the well-known modular properties of the torus, Klein bottle, annulus, and M\"obius strip amplitudes that we discussed in the previous section. In the following, we will discuss separately the cases of oriented closed strings and unoriented closed and open strings, emphasising what such a large-mass limit can imply for the presence of tachyons in closed and open strings.

\subsubsection{Closed-string tachyons and large-mass behaviour of the torus amplitude} \label{sssec:misalignedsusyclosed}

Let us begin by discussing oriented closed string theories. Here, the large-mass regime is connected by modular invariance to the infrared features of the same amplitude. In this spirit, one would like to characterise the presence or the absence of tachyons by looking at an effective quantity that captures the behaviour of the partition function in such a limit: the {\em sector-averaged sum} \cite{Dienes:1994np},
\begin{equation} \label{eq:sectoraveragedsum}
   \big  \langle d(n) \big \rangle = \sum_{i \bar \iota}  N_{i \bar \iota} d_i(n) \, \bar{d}_{\bar \iota}(n+h_i-c/24 -\bar{h}_{\bar \iota} + \bar c/24) \, .
\end{equation}
Using the definition in \eqref{eq:characterq}, the limit $m^2 \to \infty $ reads 
\begin{equation}
\begin{aligned}
    \mathcal{T}(\tau, \bar \tau)&= \tau_2^{1-d/2} \sum_{i \bar \iota}  N_{i \bar \iota} \, \chi_i \bar \xi_{\bar \iota} 
    \\
    &\sim \tau_2^{1-d/2} \sum_{i \bar \iota}  N_{i \bar \iota} d_i(n) \, \bar{d}_{\bar \iota}(n+h_i-c/24 -\bar{h}_{\bar \iota} + \bar c/24) \, , 
    \end{aligned}
\end{equation}
where the argument of $\bar d$ encodes the level-matching condition. To investigate the behaviour of the partition function in such a limit, we can take advantage of Rademacher's exact formula \cite{Rademacher:1937a, Rademacher:1937b, Rademacher:1938b} adapted to the case of vector-valued modular forms \cite{Dijkgraaf:2000fq, Manschot:2007ha} of weight $w=1-d/2$, for which the degeneracy of the character at a given level reads
\begin{equation}
\begin{aligned}\label{eq:Rademacherformula}
    d_i(n) = 2 \pi \sum_{\ell=1}^{\infty} \sum_{ \substack{ j \ | \\ h_j-c/24<0}} & Q_{ij}^{(\ell,n)} d_j(0) \bigg ( \frac{|h_j-c/24|}{n+h_i-c/24} \bigg )^{\frac{1-w}{2}} 
    \\
    &I_{1-w}\Big (\frac{4\pi}{\ell} \sqrt{|h_j-c/24|(n+h_i-c/24)} \Big ) \, ,
    \end{aligned}
\end{equation}
where $I_{\nu}(x)$ is the {\em modified Bessel function of the first kind} and
\begin{equation} \label{eq:genKloosterman}
    Q_{ij}^{(\ell,n)}=i^w  \, \sum_{\substack{p=0\\ (p,\ell )=1}}^{\ell -1} e^{\frac{2\pi i}{\ell} \left( (h_j-c/24) p' - p (n+h_i-c/24 )\right)}\, (M^{-1}_{\ell ,p})_{i j} \, .
\end{equation} 
The latter is the {\em generalised Kloosterman sum} that encodes the modular transformations $M$, relating the behaviour of the character around the essential singularities lying at the boundary of the unit circle in the $q$-plane, as follows from the {\em Circle Method} \cite{Hardy, Kani:1989im}. Using such an analytic continuation of the degeneracies\footnote{To be precise, the analysis of the subleading terms in the series in $\ell$ involves additional technical difficulties that should be treated via a more refined analytic continuation following \cite{Cribiori:2020sct}. We refer to \cite{Angelantonj:2023egh} for a detailed discussion.}, we can take the large-$n$ limit and study the behaviour of the sector averaged sum. The final result is encoded in \cite{Angelantonj:2023egh}:
\begin{equation}
\begin{aligned}
    \big  \langle d(n) \big \rangle \sim  \sum_{\substack{j, \bar j \ | \\ h_j-c/24=\bar h_{\bar j}-\bar c/24<0}} & 4 \pi^2 N_{j \bar j} \, d_j(0) \bar d_{\bar j}(0) \frac{\big ( |h_j-c/24|\big )^{1/2-w}}{2 n^{3/2-w}} 
    \\
    &\sum_{\ell=1}^{\infty} \ell \varphi(\ell) \, e^{\frac{8\pi}{\ell} \sqrt{|h_j-c/24|n}}  \, ,   
\end{aligned}
\end{equation}
where $\varphi(\ell)$ is the {\em Euler totient function}, which counts the number of relative primes with respect to $\ell$. We sketch the derivation of the formula above by focusing on the leading-order term. The subleading terms require a more involved technical treatment, but the logic remains the same. The generalised Kloosterman sum at $\ell=1$ is given by
\begin{equation}
    Q_{ij}^{(1,n)} = i^w S_{ij} \, ,
\end{equation}
which means that, at the leading order, the sector averaged sum is 
\begin{equation} \label{eq:sasleading}
    \big  \langle d(n) \big \rangle \sim  \sum_{i \bar \iota} N_{i \bar \iota} \sum_{\substack{j, \,  \bar j \ | \\ h_j<c/24 \\ \bar h_{\bar j}<\bar c/24 }} S_{ij} \bar{S}_{\bar \iota \bar j} e^{\frac{4 \pi }{\ell} \Big ( \sqrt{ |h_j -c/24| n} + \sqrt{ |\bar h_{\bar j} -\bar c/24| n }\Big )} \, ,
\end{equation}
where we have omitted powers in $|h_j-c/24|$, $|\bar h_{\bar j}-\bar c/24|$, and $n$, which are irrelevant for our discussion. Since the matrix enforcing the GSO projection, $N$, is by definition modular invariant, the sum over $i, \bar \iota$ reduces to the entries of the GSO matrix $N_{j,\bar j}$, where $j$ and $\bar j$ only span the tachyonic characters appearing in the fusion algebra. Such contributions in string theory cannot be lower than $-1$, which means that $N_{j \bar j}$ is non-vanishing only if the ground states of the corresponding characters are level-matched. This means that non-vanishing entries for $N_{j \bar j}$ are present if only if tachyons are level-matched, hence $h_j-c/24=\bar h_{\bar j}-\bar c/24$.  
For the subleading term, we need a more refined analytic continuation to deal with the contributions of the generalised Kloosterman sums that involve a non-trivial dependence on $\ell$ and $n$. As shown in \cite{Angelantonj:2023egh}, the net result is similar to the expression in \eqref{eq:sasleading}, where the modular transformations dictated by $S$ and $\bar S$ are replaced by the relevant transformations appearing in the generalised Kloosterman sum \eqref{eq:genKloosterman}. For instance, at $\ell=2$, the element of the modular group is $P=TST^2S$. Regardless of the particular modular transformation, once we obtain an expression of the form given by eq. \eqref{eq:sasleading}, we can apply the same line of reasoning as before and conclude that all subleading terms survive if and only if level-matched tachyons are present. 

  Before showing this result at work in concrete examples, let us make a few more remarks concerning the leading term in the sector averaged sum. Looking at the expression for the individual character, the leading order is given by the identity with conformal weight $h=0$,  which means
\begin{equation}
    d_i(n) \sim e^{4\pi \sqrt{n c/24}} + \ldots \, .
\end{equation}
The previous formula shows that there is no dependence on the specific character, so that it is a universal contribution for all the characters. This means that the sum over $i, \, \bar \iota$ is factorised,
\begin{equation} \label{eq:sasleading0}
    \big  \langle d(n) \big \rangle \sim e^{4\pi \sqrt{n c/24} + 4\pi \sqrt{n \bar c/24} } \sum_{i , \, \bar \iota} N_{i \bar \iota} + \ldots \, .
\end{equation}
Hence, this sum is not vanishing if and only if the tachyon corresponding to $h=0$ is level-matched. This happens only if the theory does not involve spacetime fermions, which is consistent with the property of a general CFT that requires the identity to be level-matched. In fact, when spacetime fermions are present, the role of the identity is played by characters with conformal weight $h=\frac12$ in the fusion algebra, so that the $h=0$ character can be absent. All in all, the leading term vanishes if and only if spacetime fermions are present.

The leading growth of the sector averaged sum is dictated by the effective central charge $c_{\text{eff}}$ appearing in the exponential of \eqref{eq:sasleading0}, which corresponds to the mass of the lightest state in the spectrum. For instance, if fermions are absent, this amounts to
\begin{equation}
    c_{\text{eff}}= 4\pi \Big ( \sqrt{ c/24} + 4\pi \sqrt{ \bar c/24} \Big ) \, ,
\end{equation}
in agreement with the Cardy formula \cite{Cardy:1981fd}, while if fermions are present it satisfies
\begin{equation}
     c_{\text{eff}} < 4\pi \Big ( \sqrt{ c/24} + 4\pi \sqrt{ \bar c/24} \Big ) \, ,
\end{equation}
and vanishes if there are no level-matched tachyons in the theory. 

The result above shows that if we are only counting the number of degrees of freedom, there is an overall cancellation among all the bosons and fermions in the spectrum. This is somewhat reminiscent of the properties of supersymmetric theories, in which the net number of degrees of freedom vanishes at each mass level; hence the name {\em misaligned supersymmetry} \cite{Dienes:1994np}, in which the cancellation takes place among degrees of freedom at different mass levels. 
It is worth stressing that the cancellation occurring in the large-mass regime between the bosonic and fermionic degrees of freedom should be interpreted as a property of string theories without tachyons, but does not imply the vanishing of the overall partition function. Indeed, tachyon-free theories have in general a non-vanishing contribution to the path-integral at one loop (see \cref{sec:tadpoles}). 

The result just described relies on having a finite number of characters once the bosonic zero modes associated with the non-compact directions are integrated out. Whenever the underlying CFT induces the presence of an infinite (discrete) number of characters---for instance, in compactifications---the argument should be refined, although the result is still expected to hold by the analytic continuation of the formula valid for rational points \cite{Angelantonj:2023egh}. 

Finally, the result does not hold if $d=2$. In such a case, the Rademacher sum does not converge to the desired value of $d_i(n)$ since it can be defined up to an arbitrary constant. Indeed, in two dimensions, the modular forms become meromorphic functions, which can always be defined up to a constant. This is the usual ambiguity that we face when defining the Klein $j-$function. 

\subsubsection{Tachyons and large-mass behaviour of the \texorpdfstring{$\chi=0$}{chi=0} Riemann surfaces with boundaries and cross-caps} \label{sssec:misalignedsusyopen}

The extension of the previous analysis to the case of orientifold theories is natural. However, in such a case, the amplitudes that we are interested in do not form a modular-invariant combination of characters, but the modular group connects different descriptions of the same amplitude into one another. In particular, the generator $S$ switches the proper length and time on the string worldsheet for the Klein bottle and annulus amplitudes, describing a propagation of closed-string states between boundaries and cross-caps (or D-branes and O-planes). The same holds for the M\"obius strip amplitude and the $P$ transformation. It is then natural to expect that the large-mass behaviour of such amplitudes in a given channel be tied to the presence of tachyons in the dual channel. 
This can be explicitly seen by adapting the analysis of \ref{sssec:misalignedsusyclosed} to our present case. To this end, we define the {\em sector-averaged sums} associated with each one-loop amplitude,
\begin{equation}
    \langle d(n) \rangle \big ( \mathcal{S} \big )= \sum_i \mathcal{S}_i \, d_i(n) \, ,
\end{equation}
where $\mathcal{S}=\mathcal{K}, \, \mathcal{A}$, and $\mathcal{M}$, and the corresponding coefficients $\mathcal{S}_i$ encode the expressions described in \cref{eq:directklein,eq:directannulus,eq:directmoebius}. At this point, we can directly insert the Rademacher formula \eqref{eq:Rademacherformula} and compute the asymptotic behaviour for $n\to \infty$ via the refined analytic continuation \cite{Cribiori:2020sct} employed in the previous section. In the closed-string case, all the subleading terms give a non-trivial contribution to the final result coming from the simultaneous presence of the holomorphic and anti-holomorphic sectors. For open strings, only one sector is present, implying that all the subleading terms vanish \cite{Leone:2023qfd}. Therefore,
\begin{equation}
\begin{aligned}
    \langle d(n) \rangle \big ( \mathcal{S} \big ) \sim & \sum_i \mathcal{S}_i \, \sum_{\substack{j \, | \\ h_j-c/24<0}}S^i_{j} \  d_j(0) \ e^{4 \pi \sqrt{n|h_j-c/24|}} 
    \\
    &= \sum_{\substack{j \, | \\ h_j-c/24<0}}\tilde{\mathcal{S}}_{j} \  d_j(0) \ e^{4 \pi \sqrt{n|h_j-c/24|}} \, , \qquad \mathcal{S}=\mathcal{K}, \, \mathcal{A} \, ,
    \end{aligned}
\end{equation}
where $\tilde{\mathcal{S}}_j$ denotes the coefficient of the corresponding amplitude in the transverse channel appearing in \cref{eq:transverseklein,eq:transverseannaulus,eq:transversemoebius}. A similar result holds when analysing the transverse channel of the Klein bottle and annulus amplitudes, for which
\begin{equation}
    \langle d(n) \rangle \big ( \tilde{\mathcal{S}} \big ) \sim \sum_{\substack{j \, | \\ h_j-c/24<0}}\mathcal{S}_{j} \  d_j(0) \ e^{4 \pi \sqrt{n|h_j-c/24|}} \, .
\end{equation}
For the M\"obius strip amplitude, the situation is a bit different. Indeed, the presence of a non-trivial real part of the Teichm\"uller parameter requires the introduction of the hatted characters, for which the generalised Kloosterman sum \eqref{eq:genKloosterman} is modified as
\begin{equation}
    Q_{ij}^{(\ell,n)} \to (-1)^n Q_{ij}^{(\ell,n)}\equiv \hat{Q}_{ij}^{(\ell,n)} \, .
\end{equation}
The extra phase has the simple consequence that the subleading term with $\ell=2$ is the only one that survives. The modular transformation associated with $\ell=2$ is the $P$ transformation, thus implying that also for the M\"obius strip amplitude the sector-averaged sum can be expressed in terms of the coefficients of the dual description,
\begin{equation}
\begin{aligned}
    &\langle d(n) \rangle \big ( \mathcal{M} \big ) \sim  \sum_{\substack{j \, | \\ h_j-c/24<0}}\tilde{\mathcal{M}}_{j} \  d_j(0) \ e^{2 \pi \sqrt{n|h_j-c/24|}} \, , 
    \\
    &\langle d(n) \rangle \big ( \tilde{\mathcal{M}}\big ) \sim  \sum_{\substack{j \, | \\ h_j-c/24<0}}\mathcal{M} _{j} \  d_j(0) \ e^{2 \pi \sqrt{n|h_j-c/24|}} \, .
    \end{aligned}
\end{equation}

We conclude that a non-trivial growth is connected to the presence of tachyonic contributions in the dual channel for all Riemann surfaces with vanishing Euler characteristic with boundaries and cross-caps. 
What can we infer about the presence of tachyons in orientifold theories? Contrary to the case of oriented closed strings, the conclusions that we can draw are not as powerful. Indeed, a vanishing growth for the direct-channel sector-averaged sums implies that the closed-string tachyons do not couple to D-branes or O-planes. This is of course necessary if closed-string tachyons are absent, but it is not sufficient to exclude closed-string tachyons from the spectrum. This actually happens for a specific orientifold of type 0B string theory~\cite{Bianchi:1990tb} that we introduce in the next section. The vanishing of the transverse-channel sector-averaged sums (along with the torus sector-averaged sum) suffices to guarantee the absence of all tachyons, but it is not necessary since the tachyons can be eliminated via a non-trivial orientifold projection \cite{Sagnotti:1996qj}, so that the growth of the corresponding transverse-channel amplitudes is non-trivial even when tachyons are absent. One can still hope to obtain an overall vanishing growth when adding the various amplitudes, but this is prevented by the powers in the exponentials that make it impossible to compare the contributions of the amplitudes corresponding to different Riemann surfaces\footnote{It is possible to argue that the transverse channel sector-averaged sums also reflect the signs in the direct channel amplitude, thus containing information about the nature of the projection involved. However, this does not add any new information which was not already contained in the amplitudes themselves, since the growth is still non-trivial.}.

\section{Ten-dimensional strings and tachyonic instabilities}\label{sec:stringtheory}

We can now describe some concrete examples of ten-dimensional strings. Before studying the large-mass behaviour of some examples, we use this opportunity to display the construction and classification of all known string theories. The discussion involves both theories of oriented closed strings and, eventually, their orientifold projections.  
The first step is to specify the ingredients to build the one-loop partition functions \eqref{eq:toruspartitionfunction}, from which we read the spectrum once modular invariance is enforced. We will first understand which characters and modular invariants are allowed in \eqref{eq:toruspartitionfunction}, and then we will apply the orientifold procedure developed in \cite{Sagnotti:1987tw, Pradisi:1988xd, Horava:1989vt, Bianchi:1990yu, Bianchi:1990tb, Bianchi:1991eu}, being careful to cancel R-R tadpoles. 

Let us start with the description of oriented closed strings. In flat Minkowski space, the underlying worldsheet theory realises $\mathcal{N}=(1,1)$ supersymmetry and it is described, in the superconformal gauge, by $10$ non-compact free bosons, $10$ free fermions, the $(b,c)_{(2,-1)}$ system, and $(\beta, \gamma)_{(\frac32,-\frac12)}$ system. This is known as the {\em Ramond-Neveu-Schwarz (RNS) superstring}. The ghost fields enter the path integral as two directions that effectively cancel two contributions coming from the free bosons and free fermions, both in the holomorphic and the anti-holomorphic sectors. In the following, we will also consider worldsheet theories with $\mathcal{N}=(1,0)$ supersymmetry, whose holomorphic sector is that of the RNS superstring, while the anti-holomorphic sector is purely bosonic and thus comprises a CFT with $c=26$ to compensate the contribution from the $(b,c)_{(2,-1)}$ system. In this scenario, the CFT consists of $10$ free non-compact bosons that identify the ten-dimensional Minkowski space, with a further internal CFT with $c=16$. In order for the holomorphic and anti-holomorphic sectors to be treated independently, the internal CFT must be purely chiral. Note that this is not necessarily the case for free (non-)compact bosons because, although the holomorphic and the anti-holomorphic oscillators are completely independent, their zero modes act on the same Hilbert space. Only for specific values of the internal moduli, compact free bosons form a purely chiral CFT; this corresponds to a Euclidean lattice. In the following, we will consider this case, although a priori one can consider any internal chiral CFT with $c=16$. 

We are looking for modular invariants, which can take the schematic form
\begin{equation}
     \Big ( \text{RNS}_8 \Big ) \otimes  \Big ( \overline{\text{RNS}}_8\Big ) \, ,
\end{equation}
where RNS$_8$ stands for $8$ non-compact free bosons and $8$ free fermions, or 
\begin{equation}
     \Big ( \text{RNS}_8 \Big ) \otimes  \Big ( \overline{8 \text{ nc free bosons } \oplus 16 \text{ c free bosons }} \Big ) \, .
\end{equation}
The non-compact free-bosons give rise to a CFT which realises an $\mathbb{R}^8$ Wess-Zumino-Witten (WZW) model, thus described by $8$ copies of a $\mathfrak{u(1)}_1$ Ka$\check{\text{c}}$-Moody algebra\footnote{To take into account the non-compact nature of group manifold we should consider thee universal cover of $\text{U}(1)^8$, thus allowing zero modes to give rise to a continuous spectrum.}, the $8$ (anti-)holomorphic free fermions realise a $\text{SO}(8)$ WZW model, thus described by an $\mathfrak{so}(8)_1$ Ka$\check{\text{c}}$-Moody algebra, and the compact boson should form a lattice of rank $16$ and signature $(0,16)$. The spinorial classes in the RNS superstring arising from the $8$ free fermions denote spacetime Majorana-Weyl spinors, so that in the partition function they appear with a minus sign. This means that the true characters that we must consider when describing the (anti-)holomorphic sector of the RNS superstring are
\begin{equation}
    \chi=( O_8, V_8, -S_8, -C_8 ) \, .
\end{equation}
The definition of the characters of the underlying $\mathfrak{so}(2n)_1$ can be read from \eqref{eq:characters}, 
\begin{equation} \label{eq:KMcharacters}
    \chi_i(\tau)= \frac{1}{\eta^4} \sum_{\lambda \in D_n + (i)} q^{\frac{1}{2}\lambda^2} \, ,
\end{equation}
where $(i)$ is a representative of the four conjugacy classes of $\text{SO}(2n)$\footnote{In our case, the conjugacy class of the adjoint $(o)=(0,\ldots,0)$, of the fundamental $(v)=(1,0,\ldots,0)$, and of the spinor classes $(s)=(\frac{1}{2},\ldots,\frac12)$ and $(c)=(-\frac{1}{2},\frac12, \ldots,\frac12)$.}. For $\mathfrak{so}(2n)_1$, the characters can be expressed in terms of the Jacobi theta functions,
\begin{equation}
\begin{aligned}\label{eq:so2ncharacters}
    &O_{2n}=  \frac{\theta \big [ \substack{0 \\ 0}\big ]^n + \theta \big [ \substack{0 \\ 1/2}\big ]^n}{2 \eta^n} \, , \qquad \qquad V_{2n}=  \frac{\theta \big [ \substack{0 \\ 0}\big ]^n - \theta \big [ \substack{0 \\ 1/2}\big ]^n}{2 \eta^n} \, ,
    \\
    &S_{2n}=  \frac{\theta \big [ \substack{1/2 \\ 0}\big ]^n + i^{-n} \theta \big [ \substack{1/2 \\ 1/2}\big ]^n}{2 \eta^n} \, , \qquad C_{2n}=  \frac{\theta \big [ \substack{1/2 \\ 0}\big ]^n - i^{-n} \theta \big [ \substack{1/2 \\ 1/2}\big ]^n}{2 \eta^n} \, ,
    \end{aligned}
\end{equation}
where 
\begin{equation}
    \theta \big [ \substack{a \\ b}\big ](z|\tau)=\sum_{n \in \mathbb{Z}} q^{\frac12(n+a)^2} e^{2\pi i (n+a)(z+b)} \, , \qquad a,b=0,1/2\, .
\end{equation}
To provide a general classification of all modular invariants, we follow the strategy described in \cite{Lerche:1986ae}. We define a map preserving the modular properties of the characters that are involved. The idea is to map the RNS superstring and the heterotic string to an even self-dual Euclidean lattice, which realises a modular-invariant partition function. This approach allows us to take advantage of a known classification discussed in the work of Niemeier \cite{Niemeier:1973}.  The map 
\begin{equation}
\begin{aligned}\label{eq:Gepnermap}
     \phi \, : \, &\text{RNS}_{d-2} \qquad  \longrightarrow \qquad \mathfrak{so}(d+6)_1 \oplus \mathfrak{e}_{8,1} 
     \\
     & (O_{d-2}, V_{d-2},-S_{d-2},-C_{d-2}) \to (V_{d+6}, O_{d+6},S_{d+6} \text{ or } C_{d+6},C_{d+6} \text{ or } S_{d+6}) \ \chi_{0} \, ,
\end{aligned}
\end{equation}
preserves the modular properties of the two sets of characters. In the definition above, we have crucially used the characters of the $\mathfrak{e}_{8,k}$ Ka$\check{\text{c}}$-Moody algebra. At level $1$, there is only one character,
\begin{equation}
    \chi_0(\tau)= \frac{1}{2\eta^8} \Big ( \theta \big [ \substack{0 \\ 0}\big ]^8 + \theta \big [ \substack{0 \\ 1/2}\big ]^8 +\theta \big [ \substack{1/2 \\ 0}\big ]^8 \Big ) \, .
\end{equation}
This plays the role of the identity in the fusion algebra, so that the action of the $S$ generator is trivial, while the action of $T$ gives the required phase $e^{-\frac{2\pi i}{3}}$.
Moreover, the modular properties of the characters $D_n$ are the same as those of $D_m$ whenever $n=m$ mod $8$. The expression makes sense also in cases where $m$ or $n$ are negative, implying that the signature of the lattice is opposite. In ten-dimensional Minkowski space, the characters are thus mapped to $\mathfrak{so}(16)_1$ and the signature is changed in the anti-holomorphic sector. If we are dealing with the superstring, a similar action can be defined on the RNS component in the anti-holomorphic sector, with the only difference that now we do not change the signature of the lattice. This means that for both the superstring and the heterotic string, the holomorphic sector is given by the character $\chi_0$, corresponding to the identity of $\mathfrak{e}_8$, while the anti-holomorphic sector is a rank-24 lattice admitting the embedding of $\mathfrak{so}(16)_1$ for the heterotic string and of $\mathfrak{so}(16)_1 \oplus \mathfrak{so}(16)_1 \oplus \mathfrak{e}_{8,1}$ for the superstring. The classification of rank-24 modular-invariant lattice had already been done in the mathematical literature, even before the realisation that string theory was a theory of quantum gravity, see~\cref{tab:Niemeier}.

\begin{table}[ht]
\centering
\begin{tabular}{| c | c|}
\hline
 Lattice &   glue vectors
 \\
\hline
$\textcolor{red}{D_{24}}$ & $(s)$
\\
$\textcolor{blue}{D_{16} \, E_8}$ & $(s,0)$
\\
$\textcolor{blue}{E_8^3}$ & $(0,0,0)$
\\
$\textcolor{red}{A_{24}}$ & $(5)$
\\
$\textcolor{red}{D_{12}^2}$ & $(s,v),(v,s)$
\\
$A_{17} \, E_7$ & $(3,1)$
\\
$\textcolor{red}{D_{10} \, E_7^2}$ & $(s,1,0),(c,0,1)$
\\
$\textcolor{red}{A_{15} \, D_9}$ & $(2,s)$
\\
$\textcolor{red}{D_{8}^3}$ & $([s,v,v])$
\\
$A_{12}^2$ & $(1,5)$
\\
$A_{11} \, D_7 \, E_6$ & $(1,s,1)$
\\
$E_{6}^4$ & $(1,[0,1,2])$
\\
$A_9^2 \, D_{6}$ & $(2,4,o),(5,0,s),(0,5,c)$
\\
$D_{6}^4$ & $(\{o,s,v,c\})$
\\
$A_{8}^3$ & $([1,1,4])$
\\
$A_7^2 \, D_{5}^2$ & $(1,1,s,v),(1,7,v,s)$
\\
$A_6^{4}$ & $(1,[2,1,6])$
\\
$A_5^4 \, D_{4}$ & $(2,[0,2,4],o),(3,3,0,0,s),(3,0,3,0,v),(3,0,0,3,c)$
\\
$D_{4}^6$ & $(s,s,s,s,s,s),(0,[0,v,c,c,v])$
\\
$A_{4}^6$ & $(1,[0,1,4,4,1])$
\\
$A_{3}^8$ & $(3,[2,0,0,1,0,1,1])$
\\
$A_2^{12}$ & $(2,[1,1,2,1,1,1,2,2,2,1,2])$
\\
$A_1^{24}$ & $(1,[0,0,0,0,0,1,0,1,0,0,1,1,0,0,1,1,0,1,0,1,1,1,1])$
\\
\hline
\end{tabular}
\caption{The Niemeier lattices with the associated glue vectors. The square and curly brackets denote a cyclic and even permutation of the representative of conjugacy classes. We have highlighted in red the entries admitting the embedding of $\mathfrak{so}(16)$, while in blue those also admitting the embedding of $\mathfrak{so}(16)\oplus \mathfrak{so}(16)\oplus \mathfrak{e}_8$.}
\label{tab:Niemeier}
\end{table}

  Adding boundaries and crosscaps breaks the product of the holomorphic and anti-holomorphic Virasoro algebras to the diagonal subalgebra. This is of course only possible if the two chiral algebras are the same. Hence, we can only obtain orientifold projections if we start with the bosonic string or the RNS superstring. We will focus on the latter case and we will also ask the supercurrent to be preserved so that the boundaries and crosscaps also preserve a diagonal subalgebra of the original $\mathcal{N}=(1,1)$ super-Virasoro algebra. This means that we can still build the partition functions in terms of linear combinations of the $\mathfrak{so}(8)_1$ characters that we have just described for the oriented closed-string case. The $\tau$ variable entering the definition of the characters in the direct channel describes the Teichm\"uller parameter of the double-covering torus associated with the Klein bottle, annulus, and M\"obius strip amplitudes: $\tau=2 i \tau_2 \, ,i \tau_2/2 $, and $1/2 +i \tau_2/2$ for the three cases.

\subsection{Closed string theories in \texorpdfstring{$10d$}{10d}}\label{ssec:closedstring10d}

  Let us take as an example the entry of the table \ref{tab:Niemeier}  corresponding to the $D_{16} \, E_8$ lattice with glue vector $(s,0)$. This latter piece corresponds to the generator of the abelian group of simple currents\footnote{For a more detailed discussion see~\cite{Lerche:1986ae, Blumenhagen:2009zz}.}. Once these are known, one obtains the whole partition function by completing the orbit with a repeated action on the character associated to the identity. In the present case, the fusion rules imply that the conformal family of the simple current is nilpotent or of order $2$, hence the sum of the identity and the action of the simple current on it gives the modular-invariant torus amplitude. The overall one-loop partition function with the spectator holomorphic $E_8$ character corresponds to\footnote{We are taking the bosons implicit since they form an independent modular-invariant partition function.}
\begin{equation}
    \mathcal{Z}_{D_{16}E_8}= \chi_0 \, \Big ( \bar{O}_{32}+\bar{S}_{32} \Big ) \bar{\chi}_0 \, .
\end{equation}
We have a natural embedding of $\text{SO}(16)$ into $\text{SO}(32)$ by decomposing the corresponding characters,
\begin{equation}
    O_{32}+S_{32}= O_{16} O_{16}+V_{16} V_{16} +S_{16} S_{16} + C_{16} C_{16} \, ,
\end{equation}
following from the definitions given in \eqref{eq:so2ncharacters}. This decomposition is appropriate for both superstrings and heterotic strings. For instance, using the same map \eqref{eq:Gepnermap} on both the holomorphic and anti-holomorphic sectors, we obtain
\begin{equation}
    \mathcal{Z}_{D_{16}E_8} \quad \substack{\phi \phi \\ \longrightarrow} \quad O_8 \bar O_8 + V_8 \bar V_8+ S_8 \bar S_8+ C_8 \bar C_8 \equiv \big ( \sqrt{\tau_2} \eta \bar \eta \big )^8 \ \mathcal{T}_{0B}(\tau, \bar \tau)  \, ,
\end{equation}
which identifies the partition function of the type 0B superstring \cite{Seiberg:1986by} in the light-cone gauge. We can then read the spectrum by looking at the the $q$-expansion of the partition function, which, up to the massless level, reads
\begin{equation}
\begin{aligned}
    (\sqrt{\tau_2})^8 \ \mathcal{T}_{0B}(\tau, \bar \tau) \sim (q \bar q)^{-\frac12} & \Big \{  1 + \ldots + 8_{v} \cdot 8_{v} (q \bar q)^{\frac12} + \ldots 
    \\
    &+ 8_{s} \cdot 8_{s} (q \bar q)^{\frac12} + \ldots + 8_{c} \cdot 8_{c} (q \bar q)^{\frac12} + \ldots  \Big \} \, .
    \end{aligned}
\end{equation}
Here, $1$ corresponds to the singlet representation, which occurs in the expansion of $O_8$ and thus identifies a level-matched tachyon with mass $\alpha' m^2=-\frac12$. The terms $8_{v,s,c}$ correspond to the dimensions of the fundamental and spinor representations of $\text{Spin}(8)$, coming from the characters $V_8, \, S_8$, and $C_8$. This means that the products appearing in the expansion must be decomposed in terms of irreducible representations of $\text{Spin}(8)$ according to their Clebsch-Gordan coefficients,
\begin{equation}
\begin{aligned}
    &\boldsymbol{8}_v \otimes \boldsymbol{8}_v= \boldsymbol{1} \oplus \boldsymbol{28} \oplus \boldsymbol{35} \, ,
    \\
    &\boldsymbol{8}_s \otimes \boldsymbol{8}_s= \boldsymbol{1} \oplus \boldsymbol{28} \oplus \boldsymbol{35}^+ \, ,
    \\
    &\boldsymbol{8}_c \otimes \boldsymbol{8}_c= \boldsymbol{1} \oplus \boldsymbol{28} \oplus \boldsymbol{35}^- \, .
    \end{aligned}
\end{equation}
The first line identifies the dilaton, a $2$-form field known as the Kalb--Ramond field, and the graviton. The second line identifies a $0$-form, a $2$-form, and a self-dual $4$-form\footnote{What is self-dual and antiself-dual is matter of conventions. We have chosen the $4$-form from $8_s \otimes 8_s$ to be the self-dual component.}. The third line identifies a $0$-form, a $2$-form, and an antiself-dual $4$-form. Higher powers allow us to read the massive spectrum, although it is not trivial to identify the various spacetime fields as the mass increases. Indeed, one needs to recombine the $\mathfrak{so}(8)$ representations in terms of the $\mathfrak{so}(9)$ ones on a level-by-level basis. Recently \cite{Markou:2023ffh, Basile:2024uxn}, a more efficient technology has been developed allowing to reorganise the massive spectrum in a covariant fashion.

  Similarly, we can use the map in \eqref{eq:Gepnermap} to act only on the holomorphic sector\footnote{The two choices for the map correspond to different possible chiralities for the spacetime fermions. For the superstring, we have two independent choices: spacetime fermions with the same chiralities and with opposite chiralities. The remaining two options are related to them by a parity transformation. For the heterotic theories, the two choices are equivalent.}.This means that
\begin{equation}
    \mathcal{Z}_{D_{16}E_8} \quad \substack{ \phi \\ \longrightarrow} \quad \Big (O_8 \bar V_{16} + V_8 \bar O_8- S_8 \bar S_{16}- C_8 \bar C_{16} \Big ) \bar{\chi}_0\equiv \big ( \sqrt{\tau_2} \eta \bar \eta \big )^8 \ \mathcal{T}_{16 \times 8}(\tau, \bar \tau) \, ,
\end{equation}
thus realising a heterotic theory with $\text{Spin}(16) \times E_8$ gauge group. After performing the $q$-expansion as before, one finds that the light spectrum comprises a tachyon of mass $\alpha' m^2=-\frac12$ in the $(\boldsymbol{16}, \boldsymbol{1})$ representation of $\text{Spin}(16) \times E_8$, a gauge vector in the adjoint representation, the dilaton, the graviton, and the Kalb--Ramond field in the trivial representation $(\boldsymbol{1},\boldsymbol{1})$, a left-handed Majorana-Weyl fermion in the $(\boldsymbol{128}_s,\boldsymbol{1})$ representation, and a right-handed Majorana-Weyl fermion in the $(\boldsymbol{128}_c,\boldsymbol{1})$ representation.

  In both of the above cases there is a level-matched tachyon, thus leading to a non-trivial exponential behaviour for the sector-averaged sum defined in \eqref{eq:sectoraveragedsum},
\begin{equation} \label{eq:sasclosed}
	\langle d(n) \rangle \sim \frac{d(0)}{(2n)^{11/2}} \Big (  e^{8 \pi \sqrt{n/2}} + e^{4 \pi \sqrt{n/2}} + \ldots \Big )  \ ,
\end{equation} 
where $d(0)=1$ for the type 0B superstring and $d(0)=16$ for the  $\text{Spin}(16) \times E_8$ heterotic string. The behaviour of \eqref{eq:sasclosed} is essentially universal for all the ten-dimensional theories, aside from an overall normalisation, since the level-matched tachyon has the same mass for all theories. $d(0)$ changes according to the dimension of the representation in which the tachyon transforms. 

  The examples that we have explored up to now have physical tachyons, but analogous results hold for all the other non-supersymmetric theories of \cref{tab:Niemeier}, with one important exception: the $\text{Spin}(16)\times \text{Spin}(16) \rtimes \mathbb{Z}_2$ heterotic string \cite{Alvarez-Gaume:1986ghj}. This theory will play an important role in \cref{sec:tadpoles}, so we now focus on this case. In the Niemeier classification, it corresponds to the lattice that realises a $D_8^3$ symmetry, and the glue vector is the cyclic permutations of $(s,v,v)$ conjugacy classes. This means that the associated modular-invariant partition function reads
\begin{equation}
\begin{aligned}
    \mathcal{Z}_{D_8^3}= &\chi_0 \Big ( \bar O_{16} \bar O_{16} \bar O_{16} + \bar S_{16} \bar V_{16} \bar V_{16} + \bar V_{16} \bar S_{16} \bar V_{16}+ \bar V_{16} \bar V_{16} \bar S_{16}
    \\
    & + \bar C_{16} \bar O_{16} \bar C_{16} + \bar O_{16} \bar C_{16} \bar C_{16} + \bar C_{16} \bar C_{16} \bar C_{16} + \bar V_{16} \bar V_{16} \bar V_{16} \Big) \bar{\chi}_0 \, .
\end{aligned}
\end{equation}
Using again the map \eqref{eq:Gepnermap}, we obtain the following partition function:
\begin{equation}\label{eq:so16so16_heterotic}
\begin{aligned}
    \mathcal{Z}_{D_8^3} \quad \substack{\phi \\ \longrightarrow} \quad &   V_{8} \big ( \bar O_{16} \bar O_{16} +  \bar C_{16} \bar C_{16} \big ) + O_{8} \big ( \bar S_{16} \bar V_{16}+ \bar V_{16} \bar S_{16} \big )
    \\
    & - C_{8} \big ( \bar O_{16} \bar C_{16} + \bar C_{16} \bar O_{16} \big ) - S_{8} \big ( \bar S_{16} \bar S_{16} + \bar V_{16} \bar V_{16} \big ) 
    \\
    &\equiv   \big ( \sqrt{\tau_2} \eta \bar \eta \big )^8 \ \mathcal{T}_{16 \times 16}(\tau, \bar \tau)  \, ,
\end{aligned}
\end{equation}
which realises a gauge group $\text{Spin}(16) \times \text{Spin}(16) \rtimes \mathbb{Z}_2$. The discrete factor encodes the possibility to switch the two factors in the gauge group. The light spectrum comprises the dilaton, the graviton, and the Kalb--Ramond field in the trivial representation of the gauge group, a gauge vector in the adjoint of $\text{Spin}(16) \times \text{Spin}(16)$, a left-handed Majorana-Weyl fermion in the $(\boldsymbol{16},\boldsymbol{16})$ representation, and a right-handed Majorana-Weyl fermion in the spinorial representation $(\boldsymbol{128}_c,\boldsymbol{1}) \oplus (\boldsymbol{1}, \boldsymbol{128}_c)$. The main novelty comes from the analysis of the $O_{8} \big ( \bar S_{16} \bar V_{16}+ \bar V_{16} \bar S_{16} \big )$ term, which can potentially describe a tachyon. The $q$-expansion,
\begin{equation}
   \begin{aligned}
\frac{1}{(\eta \bar \eta)^8} O_{8} \big ( \bar S_{16} \bar V_{16}+ \bar V_{16} \bar S_{16} \big ) \sim & q^{-\frac12} \bar q^{-1} \big ( 1+ (28 + 8_v) \ q + \ldots \big )    
\\
   &\big ( 1+ 8_v\ \bar q + (128_s \cdot 16  +  16 \cdot 128_s ) \ \bar q^{\frac{3}{2}} + \ldots \big ) \, ,       
   \end{aligned}
\end{equation}
shows that the first contribution occurs at level $O((q\bar q)^{\frac12})$. Therefore, there are no level-matched tachyons in this theory, which means that we can trust the theory at least up to one loop (see \cref{sec:tadpoles}). The absence of level-matched tachyons corresponds to the vanishing of the sector-averaged sum,
\begin{equation}
    \langle d(n) \rangle=0 \, .
\end{equation}

The same result holds for supersymmetric theories, in which case the computation is somewhat trivialised by the cancellation at each mass level. Nevertheless, it is instructive to discuss the explicit case of type IIB string theory, because it will be useful in the upcoming discussion. 

  Type IIB superstring theory can be obtained from the entry $E_8^3$  of the Niemeier classification with glue vector $(0,0,0)$. The partition function reads
\begin{equation}
    \mathcal{Z}_{E_8^3}= \chi_0 \bar \chi_0 \bar \chi_0 \bar \chi_0 \, .
\end{equation}
The decomposition of the identity $\chi_0$ in terms of $\text{SO}(16)$ characters is
\begin{equation}
    \chi_0= O_{16}+S_{16} \, ,
\end{equation}
so that the partition function becomes
\begin{equation}
    \mathcal{Z}_{E_8^3} \quad \substack{\phi \phi \\ \longrightarrow } \quad \big ( V_8 -S_8 \big ) \big ( \bar V_8 - \bar S_{8} \big ) \equiv   \big ( \sqrt{\tau_2} \eta \bar \eta \big )^8 \ \mathcal{T}_{\text{IIB}}(\tau, \bar \tau) \, .
\end{equation}
This expression vanishes, which follows from Jacobi's {\em aequatio identica satis abstrusa},
\begin{equation}
   \theta \big [ \substack{0 \\ 0}\big ]^4 - \theta \big [ \substack{0 \\ 1/2}\big ]^4 -\theta \big [ \substack{1/2 \\ 0}\big ]^4 \pm \theta \big [ \substack{1/2 \\ 1/2}\big ]^4=0 \, , 
\end{equation}
where the sign in front of $\theta \big [ \substack{1/2 \\ 1/2}\big ]^4$ enables us to apply the identity to both the $V_8-S_8$ and the $V_8-C_8$ combinations. Indeed, this reflects the triality of $\text{SO}(8)$, which is ultimately encoded in the relations
\begin{equation}
    V_8=S_8=C_8 \, .
\end{equation}
The partition function is formally vanishing, but we can still read the spectrum. At the massless level, it contains the dilaton, the graviton, the Kalb--Ramond field, a scalar, a $2$-form, and a self-dual $4$-form as the bosonic sector, and two left-handed Majorana-Weyl gravitini and two right-handed Majorana-Weyl fermions as the fermionic sector. Together, these form the gravity multiplet of $\mathcal{N}=(2,0)$ supergravity in ten dimensions.

  The analysis for the other closed-string theories follows a similar path, and we invite the interested reader to consult the relevant literature for the construction of the various models \cite{Kawai:1986vd, Dixon:1986iz, Lerche:1986ae} and for the implications of the presence of tachyons, along the lines of what we have discussed here \cite{Angelantonj:2023egh}. One final comment is in order: we have built the heterotic theory assuming the presence of an internal CFT with central charge $16$ corresponding to a Euclidean lattice with suitable conjugacy classes. In other words, we have assumed the internal CFT to form a chiral current algebra at level 1, and one may wonder if other scenarios are available relaxing such a condition. Indeed, allowing the level to be arbitrary, we can take advantage of the classification of \cite{Schellekens:1992db} to apply the map \eqref{eq:Gepnermap} and find new theories. There is one possibility, associated with the entry $B_{8,1} E_{8,2}$ with glue vector $(v,1)$ in \cite{Schellekens:1992db}, which corresponds to the tachyonic non-supersymmetric string theory with rank-reduced gauge group $E_8 \times \text{SO}(1)$\footnote{In the literature, such a theory is usually denoted as simply having $E_8$ as a gauge group. However, a simple computation of the central charges, $248 k/(k+30)+1/2 = 31/2+1/2=16$ for $k=2$, shows how the extra fermion is needed to reach criticality.} discussed in \cite{Kawai:1986vd}. With this model, the list of all possible closed string theories in ten dimensions is complete \cite{BoyleSmith:2023xkd}.

  The discussion so far has only involved the construction of string theories in ten dimensions, and it is natural to ask what happens for theories in lower (non-compact) dimensions. One may naively think that the same results and comments directly apply to these other scenarios. However, a new feature emerges which was absent in the critical ten-dimensional case. When some of the directions form a CFT with compact target space, additional marginal operators can be added to the theory, which may be used to deform the worldsheet sigma model. This is, for instance, the case of non-trivial Wilson lines which may jeopardise the classical stability of the theory, because they can connect tachyon-free to tachyonic theories \cite{Ginsparg:1986wr}. In this sense, in lower dimensions, it is not enough to realise a theory which is free of tachyons, but one should also make sure that such a theory cannot be deformed to any tachyonic theory. 
One possibility is to consider scenarios in which such marginal operators cannot be included in the theory by building a string theory with an internal CFT that has no geometrical interpretation in the target space. For instance, using asymmetric \cite{Narain:1986qm, Narain:1990mw, Aldazabal:2025zht} or quasi-crystalline \cite{Harvey:1987da, Baykara:2024tjr, Baykara:2024vss} orbifolds. Indeed, in this situation, the theory only exists at a specific point of the moduli space and cannot be deformed without losing a consistent particle interpretation of the spectrum \cite{Angelantonj:2024jtu}, thus undermining modular invariance at higher loops.
All these considerations apply to tachyons emerging in the tree-level spectrum. Tachyons may still be present when one-loop corrections are taken into account. See \cite{Basile:2018irz, Baykara:2022cwj} for a more detailed discussion.

\subsection{Ten-dimensional orientifolds}\label{ssec:orientifolds10d} 

We can now move to the description of known orientifolds in ten dimensions. We will focus on the theories built from orientifold projections of the type IIB and the type 0B superstring\footnote{There is also the possibility to build an orientifold projection out of the type 0A superstring \cite{Bianchi:1990tb}, but we will not discuss such a construction in this review. For the type IIA superstring in ten dimensions, no orientifold projection can be consistently implemented.}. Starting from the type IIB theory, there is only one possible projection that we can implement, corresponding to the standard worldsheet parity operator $\Omega$,
\begin{equation}
    \mathcal{K}(2i\tau_2)= \frac{1}{\big (\sqrt{\tau_2} \, \eta \big )^8} \big (V_8-S_8 \big ) \, ,
\end{equation}
which in the transverse channel becomes
\begin{equation}
    \tilde{\mathcal{K}}(i\ell)= \frac{2^{5}}{ \eta^8} \big (V_8-S_8 \big ) \, .
\end{equation}
In the transverse channel, we can read the propagators of the NS-NS and R-R closed-string fields. Indeed, by performing a $q$-expansion,
\begin{equation}\label{eq:kleintransverseI}
    \tilde{\mathcal{K}}(i\ell) \sim 2^5 \bigg \{ 8_v -8_s + O(e^{-2\pi \ell}) \bigg \} \, ,
\end{equation}
we can interpret the constant terms as the propagators of massless fields between space-filling O-planes, while the other terms are associated with massive modes.
This expression is integrated over $\ell$ in the range $(0,\infty)$. This means that, although higher-order terms vanish as $\ell \to \infty$, the massless contribution yields dangerous divergences. This precisely reflects the presence of on-shell propagators of massless fields, as advocated in the \cref{ssec:moduli_space}. At the level of Riemann surfaces, taking this limit factorises the cylinder into a propagator of massless fields along an infinitely long tube multiplying the square of the one-point function of such bulk fields.   
In \eqref{eq:kleintransverseI}, the term $8_v$ corresponds to the propagator of the dilaton and the graviton, which couples to the O-planes via a non-trivial tension, while $8_s$ corresponds to the propagator of the non-dynamical $10$-form, which couples to orientifold planes via a non-trivial charge. Therefore, the tension and charge of the orientifold are encoded in the tadpoles appearing in \eqref{eq:kleintransverseI}, and are determined up to a sign ambiguity since we can only read their value squared. A non-trivial value of the R-R tadpole is inconsistent, and indeed it implies perturbative anomalies for the spacetime theory \cite{Aldazabal:1999nu, Bianchi:2000de}. This means that we must add the other contributions occurring at $\chi=0$ from the annulus and M\"obius strip amplitudes to obtain a consistent construction. 
In this setup, the only terms that can appear in the transverse channel come from $V_8$ and $S_8$, in order to interpret the contribution from this channel as the propagator of physical fields that are present in the spectrum. This means that we can write only two independent terms,
\begin{equation}
\begin{aligned}
    \tilde{\mathcal{A}}(i\ell) &= \frac{2^{-5}}{ \eta^8} \bigg \{ \big (n_++n_- \big )^2 V_8-\big (n_+ - n_- \big )^2S_8 \bigg \}
    \\
    &\sim 2^{-5} \bigg \{ \big (n_++n_- \big )^2 8_v - \big (n_+ - n_- \big )^2 8_s + O(e^{-2\pi \ell}) \bigg \} \, ,
    \end{aligned}
\end{equation}
in which the signs of $n_{\pm}$ reflect the possibility to introduce D-branes with different charges with respect to the non-dynamical $10-$form, still having a positive tension as expected from dynamical objects carrying non-trivial degrees of freedom. The first terms in the second line are interpreted as the on-shell propagation of massless fields between two boundaries, where now the tadpoles correspond to the square of the one-point function of the bulk fields on the disc. This term is not sufficient to cancel the R-R tadpole of the Klein bottle, and we need to include the contribution coming from the propagation of closed strings between a boundary and a cross-cap, which identifies the M\"obius strip amplitude:
\begin{equation}
    \begin{aligned}
    \tilde{\mathcal{M}}(1/2+i\ell) &= \frac{2}{ \hat\eta^8} \bigg \{ \varepsilon_{\text{NS}}\big (n_++n_- \big ) \hat V_8-\varepsilon_{\text{R}}\big (n_+ - n_- \big ) \hat S_8 \bigg \}
    \\
    &\sim 2 \bigg \{ \varepsilon_{\text{NS}}\big (n_++n_- \big ) 8_v - \varepsilon_{\text{R}}\big (n_+ - n_- \big ) 8_s + O(e^{-2\pi \ell}) \bigg \} \, .
    \end{aligned}
\end{equation}
The factor of $2$ describes the two possible configurations of boundaries and cross-caps at our disposal, while the coefficient in front of the hatted characters gives the product of the charges of the orientifold planes and those of D-branes. Note the presence of the extra factors, $\varepsilon_{\text{NS,R}}=\pm1$, which reflect the sign ambiguity of the tadpoles of bulk fields on the projective plane and the disc that we have discussed in the transverse channel of the Klein bottle and the annulus. To cancel the R-R tadpole we must have
\begin{equation}
    2^{5}+\varepsilon_{\text{R}}(n_+-n_-)=0 \, .
\end{equation}
This equation has an infinite number of solutions. We will focus on two different configurations, $\varepsilon_{\text{R}}=\varepsilon_{\text{NS}}=-1$ and  $\varepsilon_{\text{R}}=\varepsilon_{\text{NS}}=1$\footnote{The other possible combinations of orientifold signs introduce anti-orientifold planes which lead to unstable configurations. We will omit them in the present discussion.}. In these configurations, the orientifold planes carry a negative tension in the former case and a positive tension in the latter case. We identify O$9_\mp$ planes as orientifold planes carrying negative(positive) tension and negative(positive) charge with respect to the non-dynamical $10$-form.  In the first case, the tadpole conditions become
\begin{equation}\label{eq:tadpolestypeIlike}
    2^{5}-(n_+-n_-)=0 \, , \qquad 2^{5}-(n_++n_-)=0 \, ,
\end{equation}
so that we can parametrise the solution as $n_-=n$ and $n_+=32+n$. The spectrum can be read as before by going to the direct channel
\begin{equation}
    \begin{aligned}\label{eq:annmoebiustypeIlike}
        & \mathcal{A}(i \tau_2/2)= \frac{1}{\big (\sqrt{\tau_2} \, \eta \big )^8} \bigg \{ \big (n_+^2+n_-^2 \big ) \big (V_8-S_8 \big ) + 2 n_+ n_- \big ( O_8 - C_8 \big ) \bigg \} \, ,
        \\
        & \mathcal{M}(1/2+i \tau_2/2)= -\frac{2}{\big (\sqrt{\tau_2} \, \hat \eta \big )^8} \bigg \{ \big (n_+ + n_- \big ) \hat V_8-\big (n_+ - n_- \big ) \hat S_8 \bigg \} \, .
    \end{aligned}
\end{equation}
In the closed sector, the projected spectrum can be read from $\big (\mathcal{T}+ \mathcal{K} \big )/2$ and comprises, at the massless level, the metric and the dilaton from the projected NS-NS sector, the R-R 2-form from the projected R-R spectrum, and a left-handed Majorana-Weyl gravitino with a right-handed Majorana-Weyl fermion, thus forming the gravity multiplet of ten-dimensional $\mathcal{N}=(1,0)$ supergravity. The open-string sector is obtained from \eqref{eq:annmoebiustypeIlike} and comprises a gauge boson transforming in the $(\boldsymbol{n_+}(\boldsymbol{n_+-1})/2,\boldsymbol{1})\oplus (\boldsymbol{1}, \boldsymbol{n_-}(\boldsymbol{n_-}-1)/2)$, thus identifying $\text{SO}(32+n)\times \text{SO}(n)$ as the gauge group, and a left-handed Majorana-Weyl fermion in the same representation. Moreover, we have a tachyon and a right-handed Majorana-Weyl fermion in the $(\boldsymbol{n_+},\boldsymbol{n_-})$ representation. The open-string spectrum contains a tachyon, denoting the instability of the brane configuration, which is reflected in a non-trivial growth of the sector-averaged sum in the transverse annulus amplitude,
\begin{equation}\label{eq:sasIlike}
    \langle d(n) \rangle \big (\tilde{\mathcal{A}} \big )= 2 n_+ n_- \frac{e^{4 \pi \sqrt{n/2}}}{(2n)^{11/4}} \, .
\end{equation}
Eliminating the tachyon means choosing $n=0$, so that we are left with the gauge group $\text{SO}(32)$ and a vector multiplet for $\mathcal{N}=(1,0)$ supergravity in ten dimensions. This is the type I superstring. This theory has spacetime supersymmetry, and one can immediately see that the NS tadpoles in \eqref{eq:tadpolestypeIlike} are automatically cancelled. 

  The other configuration that we are interested in is given by $\varepsilon_{\text{R}}=\varepsilon_{\text{NS}}=1$, which leads to the same annulus amplitude as in \eqref{eq:annmoebiustypeIlike}, subject to a different orientifold projection, $\Omega (-1)^{f_{\text{L}}}$. Here, $f_{\text{L}}$ is the worldsheet fermion number. This leads to the M\"obius strip amplitude
\begin{equation}
    \mathcal{M}(1/2+i \tau_2/2)= \frac{2}{\big (\sqrt{\tau_2} \, \hat \eta \big )^8} \bigg \{ \big (n_+ + n_- \big ) \hat V_8+\big (n_+ - n_- \big ) \hat S_8 \bigg \} \, .
\end{equation}
The massless open-string sector comprises a gauge boson in the adjoint of $\text{USp}(32+n)\times \text{USp}(n)$, with $n_+=n$ and  $n_-=32+n$, and a left-handed Majorana-Weyl fermion in the $(\boldsymbol{n_+}(\boldsymbol{n_+-1})/2,\boldsymbol{1})\oplus (\boldsymbol{1}, \boldsymbol{n_-}(\boldsymbol{n_-}-1)/2)$ representation. As before, we have a tachyon and a right-handed Majorana-Weyl fermion in the $(\boldsymbol{n_+},\boldsymbol{n_-})$ representation. Again, the only non-vanishing sector-averaged sum is the one coming from the transverse annulus amplitude, whose expression is still \eqref{eq:sasIlike}. 
Eliminating the tachyon thus means imposing $n=0$, which describes the  Sugimoto string \cite{Sugimoto:1999tx}.
In contrast to the type I superstring, the states do not form linear supermultiplets and the cancellation of the R-R tadpole does not imply the vanishing of the NS-NS one.
In the Sugimoto model, the breaking of supersymmetry occurs in the open-string sector. This is due to the simultaneous presence of O$9_+$ planes and $\overline{\text{D}9}$ branes, which are BPS objects but preserve different sets of supercharges. At a first glance, the model seems to be inconsistent, because the closed-string sector is still supersymmetric and the presence of a massless gravitino would appear to require unbroken supercurrents. The puzzle was solved in \cite{Dudas:2000nv}, where it was shown that supersymmetry is linearly realised in the closed-string sector but non-linearly realised in the open-string one. The left-handed fermion in the singlet of $\text{USp}(32)$, which comes from the reduction of the antisymmetric representation $\boldsymbol{496}=\boldsymbol{495} \oplus \boldsymbol{1}$, plays the role of the Volkov-Akulov field \cite{Volkov:1972jx}. 
This model still has an uncancelled NS-tadpole, which will be the subject of \cref{sec:tadpoles}.
In lower dimensions, additional solutions with non-linear realisations of supersymmetry are possible \cite{Antoniadis:1999xk, Aldazabal:1999jr, Angelantonj:2024iwi, Leone:2024hnr}. The cancellation of additional R-R tadpoles can sometimes imply rigid configurations \cite{Angelantonj:2024iwi}, which only accommodate discrete deformations \cite{Leone:2024hnr}.

  We now move to the possible orientifolds of the type 0B superstring. In this case, the orientifold projection is not unique. There are three alternatives (swapping right and left has no effect): $\Omega$, $\Omega(-1)^{F_R}$, and $\Omega (-1)^{f_R+1}$, where $F_R$ and $f_R$ are the right-moving spacetime and worldsheet fermion numbers. Let us begin with the standard orientifold projection $\Omega$, which identifies the Klein bottle amplitude
\begin{equation}
    \mathcal{K}_1(2i\tau_2)= \frac{1}{\big (\sqrt{\tau_2} \, \eta \big )^8} \big (O_8+V_8-S_8-C_8 \big ) \, .
\end{equation}
In the transverse channel, this becomes
\begin{equation}\label{eq:KBtype01}
    \tilde{\mathcal{K}}_1(i \ell)=  \frac{2^6}{ \eta^8} V_8 \, .
\end{equation}
From the transverse channel, we can read the orientifold planes that are present in such a theory. In particular, we can have non-trivial couplings arising from the dilaton, the metric, and two possible R-R $10$-forms from the $(v)$, $(o)$, $(s)$ and $(c)$ sectors flowing in the transverse channel. Therefore, the orientifold planes and D-branes \cite{Klebanov:1998yya, Dudas:2001wd} carry non-trivial couplings to all these contributions, identifying different extended objects. D$9_1$ branes carry a positive coupling to the closed-string tachyon and the two $10$-forms, while D$9_2$ branes carry a negative coupling to the closed-string tachyon and to the $(c)$-sector $10$-form, and a positive one to the $(s)$-sector $10$-form. We can introduce four types of orientifold planes: O$9^2_\pm$, which carry couplings of the same sign to all the aforementioned fields, and O$9^2_{\pm}$, for which the signs of the couplings to the tachyon and to the R-R $10$-forms in the $(c)$ sector are opposite with respect to the dilaton and the $10$-form in the $(s)$ sector. With this in mind, from the expression in \eqref{eq:KBtype01}, we can read that we have the following combination of O-planes:
\begin{equation}
    \text{O}9^1_\pm \oplus \text{O}9^2_{\pm} \oplus  \overline{\text{O}9^1_\pm} \oplus \overline{\text{O}9^2_\pm } \, .
\end{equation}
There are no R-R tadpoles, and it is not necessary to introduce an open sector. We will discuss this minimalist configuration and refer the interested reader to \cite{Angelantonj:2002ct} for a more detailed discussion of the more general case. 
In this minimalist scenario, we only have the closed-string sector whose lightest excitations comprise a tachyon, the metric, and the dilaton from the NS-NS sector, and two $2$-form fields from the R-R sector.  
The sector-averaged sum associated with the Klein bottle amplitude in the direct channel vanishes despite the presence of a closed-string tachyon, while the presence of the tachyonic character in the direct channel is reflected in the transverse-channel sector averaged sum
\begin{equation} \label{eq:sasKB0B1}
     \langle d(n) \rangle \big (\tilde{\mathcal{K}}_1 \big )=  \frac{e^{4 \pi \sqrt{n/2}}}{(2n)^{11/4}} \, .
\end{equation}
A similar story holds for the orientifold projection obtained by dressing the standard worldsheet parity operator $\Omega$ with $(-1)^{F_R}$. In this case, the Klein bottle amplitude encoding the orientifold projection reads
\begin{equation}
    \mathcal{K}_2(2i\tau_2)= \frac{1}{\big (\sqrt{\tau_2} \, \eta \big )^8} \big (O_8+V_8+S_8+C_8 \big ) \, .
\end{equation}
The transverse channel is obtained by applying the usual $S-$transformation,
\begin{equation}\label{eq:KBtype02}
    \tilde{\mathcal{K}}_2(i \ell)=  \frac{2^6}{ \eta^8} O_8 \, ,
\end{equation}
and thus only the coupling to the tachyon is non trivial, suggesting the following combinations of orientifold planes:
\begin{equation}
   \text{O}9^1_\pm \oplus \text{O}9^2_{\mp} \oplus  \overline{\text{O}9^1_\pm} \oplus \overline{\text{O}9^2_\mp }\, ,
\end{equation}
As in the previous case, since there are no R-R tadpoles, the model is consistent even without D-branes. We thus focus again on the minimalist setting and refer the interested reader to \cite{Angelantonj:2002ct} for further details. The NS-NS sector reproduces the spectrum obtained from the previous orientifold projection, thus comprising the tachyon, the metric, and the dilaton as the lightest excitations. From the R-R sector, the orientifold projection is flipped with respect to the previous case, and therefore it leads to two $0$-forms and a non-chiral $4$-form. In this model, there is still a non-trivial contribution from the tachyonic character to the Klein bottle amplitude in the direct channel, as can be seen from the large-mass behaviour of the dual description having the same expression as \eqref{eq:sasKB0B1}. However, the closed-string tachyon has now a non-trivial overall coupling to the orientifold planes, implying that the large-mass behaviour of \eqref{eq:KBtype02} reads
\begin{equation} \label{eq:sasKB0B2}
     \langle d(n) \rangle \big (\mathcal{K}_2 \big )=  2\frac{e^{4 \pi \sqrt{n/2}}}{(2n)^{11/4}} \, .
\end{equation}
The remaining orientifold projection is $\Omega (-1)^{f_R}$, which leads to the Klein bottle amplitude  
\begin{equation}
    \mathcal{K}_3(2i\tau_2)= \frac{1}{\big (\sqrt{\tau_2} \, \eta \big )^8} \big (-O_8+V_8+S_8-C_8 \big ) \, .
\end{equation}
In the transverse channel, we obtain
\begin{equation}\label{eq:KBtype03}
    \tilde{\mathcal{K}}_3(i \ell)=  \frac{2^6}{ \eta^8} \big ( -C_8 \big ) \, .
\end{equation}
In this model, the orientifold planes carry an overall vanishing coupling to the tachyon and the metric, but they have an overall non-trivial charge with respect to one of the two non-dynamical $10$-forms. This suggests the following combination of orientifold planes:
\begin{equation}
    \text{O}9^1_\pm \oplus \text{O}9^2_{\pm} \oplus  \overline{\text{O}9^1_\mp} \oplus \overline{\text{O}9^2_\mp } \, .
\end{equation}
We now need an open sector to cancel anomalies, introducing a suitable combination of D-branes:
\begin{equation}
   n_o \  \text{D}9_1 \oplus n_c \ \text{D}9_2 \oplus n_v \ \overline{\text{D}9_1} \oplus n_s \  \overline{\text{D}9_2} \, .
\end{equation}
From \cref{eq:KBtype03}, the only term flowing in the M\"obius strip amplitude corresponds to the character $-\hat{C}_8$, so that there is no projection for the gauge group, which has to be in general a product of unitary groups. There is only one possible way to write down the transverse channel of the annulus amplitude,
\begin{equation}\label{eq:trannulus0'B}
\begin{aligned}
     \tilde{\mathcal{A}}_3(i \ell)=  \frac{2^{-6}}{ \eta^8} \bigg \{ &\big ( n_o+n_v+n_s+n_c \big )^2 V_8 - \big ( n_o+n_v-n_s-n_c \big )^2 O_8 
     \\
     &- \big ( n_o-n_v-n_s+n_c \big )^2 C_8 + \big ( n_o-n_v+n_s-n_c \big )^2 S_8 \bigg \} \, ,
     \end{aligned}
\end{equation}
from which we obtain the expression in the direct channel via the $S$ transformation, 
\begin{equation}
\begin{aligned}
    \mathcal{A}_3(i\tau_2/2)= \frac{1}{\big (\sqrt{\tau_2} \, \eta \big )^8} \bigg \{&2\big ( n_o n_c + n_v n_s \big ) V_8+ 2\big ( n_o n_s + n_v n_c \big ) O_8
    \\
    &-\big ( n_o^2 +n_v^2+ n_s^2 + n_c^2 \big ) C_8-2\big ( n_o n_v + n_s n_c \big ) S_8 \bigg \} \, .
    \end{aligned}
\end{equation}
One can indeed see that the two expressions are compatible since a proper interpretation of the unitary groups requires $n_o=n$, $n_c=\bar n$, $n_v=m$ and $n_s=\bar m$, which cancels the unphysical contribution flowing in \eqref{eq:trannulus0'B} with the wrong spin-statistics. This means that the transverse channel of the M\"obius strip amplitude reads
\begin{equation}
     \tilde{\mathcal{M}}_3(1/2+i \ell)= - 2 ( n-m-\bar m+\bar n \big ) \big (- \hat{C}_8 \big ) \, ,
\end{equation}
leading to 
\begin{equation}
     \mathcal{M}_3(1/2+i \tau_2/2)= -  (  n-m-\bar m+\bar n \big )^2 \big (- \hat{C}_8 \big ) \, .
\end{equation}
We can read the uncancelled R-R tadpole, which corresponds to 
\begin{equation}
    \begin{aligned}
        & n-m-\bar m+\bar n =64 \, ,
    \end{aligned}
\end{equation}
giving a parametric solution $n=32+m$. As in the orientifolds obtained from the type IIB superstring, we can accommodate a configuration for which there are no open-string tachyons by choosing $m=0$. This choice identifies what is known as the type 0'B superstring \cite{Sagnotti:1995ga,Sagnotti:1996qj}. The associated massless spectrum comprises, in the open-string sector, a gauge boson in the adjoint representation of $\text{U}(32)$\footnote{Actually, the gauge boson associated with $\text{U}(1) \subset \text{U}(32)$ acquires a mass via the generalised Green-Schwarz-Sagnotti mechanism \cite{Sagnotti:1992qw}, leaving as a gauge group $\text{SU}(32)$.} and a left-handed Majorana-Weyl fermion in the $\boldsymbol{496}\oplus \overline{\boldsymbol{496}}$ representation. From the closed-string sector, the lightest spectrum comprises the metric and the dilaton from the NS-NS sector, and a $2$-form, a $0$-form, and an anti-selfdual $4$-form from the R-R sector. This model has the virtue of being tachyon-free despite a non-trivial growth of the sector-averaged sum associated with the transverse-channel Klein bottle amplitude, which is given by \eqref{eq:sasKB0B1} with a minus sign in front. This growth cannot cancel the growth associated with the torus amplitude \eqref{eq:sasclosed}, thus suggesting that features of the string spectrum reflecting the absence of tachyons in orientifold vacua are yet to be found.

\section{Dilaton tadpoles and vacuum stability}
\label{sec:tadpoles}

The previous discussion has focused on tachyons as the primary consequence of the absence of spacetime supersymmetry. The key takeaway is that, to understand the physics of non-supersymmetric strings, we must either proceed with tachyon condensation---or any type of process that starts from a tachyonic vacuum and lands to a tachyon-free one---or pinpoint the conditions that guarantee the absence of tachyons and thus focus on tachyon-free strings. The latter approach appears safe and under control on first inspection: no growing instabilities are visible from the string spectrum. However, the story is much more involved and, at the same time, intriguing.

We think that it is best to proceed through examples; we choose to focus on ten-dimensional strings because these are the cases about which more can be said. The choice is dictated by the ease of presentation and the desire to avoid complications. Other lower-dimensional constructions, such as~\cite{Fraiman:2023cpa,DeFreitas:2024ztt,Baykara:2024tjr,Angelantonj:2024jtu,Antoniadis:2025twv}, lead to analogous issues, but the advantage of working in ten dimensions is that we can focus on the effects of the absence of supersymmetry without having to worry about the details of compactifications and internal CFT sectors.
In other words, we aim for generic features by looking at the most basic models.

We have already introduced the three non-tachyonic non-supersymmetric strings in ten dimensions, but it is worth recalling them all together here for the benefit of the reader.

The first such string theory is the heterotic $\text{Spin}(16) \times \text{Spin}(16) \rtimes \mathbb{Z}_2$ string~\cite{Alvarez-Gaume:1986ghj,Dixon:1986iz} of \cref{eq:so16so16_heterotic}, the only ten-dimensional non-supersymmetric tachyon-free heterotic theory~\cite{Dixon:1986iz,Kawai:1986vd,BoyleSmith:2023xkd}. From now on, we shall call it the SO$(16)\times$SO$(16)$ string to conform to the literature on tadpoles.
It can be obtained from the E$_8\times$E$_8$ string by a $(-1)^F \delta$ orbifold in the bosonic formulation, where $F$ is the spacetime fermion number and $\delta$ is an order-two internal lattice shift, or by a $(-1)^{F+F_1+F_2}$ orbifold in the fermionic formulation, where $F_{1,2}$ are the spacetime fermion numbers on the two E$_8$ factors. 
The one-loop vacuum amplitude from \cref{eq:so16so16_heterotic} reads
\begin{eq}\label{eq:so1616_vacuum_ampl}
    {\cal T}_{\text{SO$(16)$$\times$SO$(16)$}} & = \int_{\cal F} \frac{d^2\tau}{\tau_2^2}\frac{1}{\tau_2^4 \eta^8 \bar{\eta}^8}\Bigg[O_8 \left(\bar{V}_{16}\bar{C}_{16}+\bar{C}_{16}\bar{V}_{16}\right)+V_8 \left(\bar{O}_{16}\bar{O}_{16}+\bar{S}_{16}\bar{S}_{16}\right)\\
    & - S_8 \left(\bar{O}_{16}\bar{S}_{16}+\bar{S}_{16}\bar{O}_{16}\right)-C_8 \left(\bar{V}_{16}\bar{V}_{16}+\bar{C}_{16}\bar{C}_{16}\right)\Bigg] \mperiod
\end{eq}
Note that there are more fermions than bosons at the massless level.

The second string model is the USp$(32)$ Sugimoto model~\cite{Sugimoto:1999tx}, an orientifold of type IIB string theory with projection $\Omega (-1)^{f_L}$ in the open sector, where $f_L$ is the left-moving worldsheet fermion number. In the conventions of~\cite{Bianchi:1990yu,Angelantonj:2002ct}, displaying explicitly the complex structure choice for the unoriented and open contributions, the one-loop vacuum amplitudes are 
\begin{eq}\label[pluralequation]{eq:vacuum_ampl_Sugimoto}
    {\cal T}_{\text{USp$(32)$}} & = \int_{{\cal F}}\frac{d^2\tau}{\tau_2^2}\frac{(V_8-S_8)(\bar{V}_8-\bar{S}_8)}{\tau_2^4 \eta^8\bar{\eta}^8}  \mcomma \\
    {\cal K}_{\text{USp$(32)$}} & = \int_0^\infty \frac{d\tau_2}{\tau_2^2}\frac{V_8-S_8}{\tau_2^4 \eta^8}(2i\tau_2)\mcomma \\
    {\cal A}_{\text{USp$(32)$}} & = N^2\int_0^\infty \frac{d\tau_2}{\tau_2^2}\frac{V_8-S_8}{\tau_2^4 \eta^8}\left(i\frac{\tau_2}{2}\right)\mcomma \\
    {\cal M}_{\text{USp$(32)$}} & = -N\int_0^\infty \frac{d\tau_2}{\tau_2^2}\frac{-\hat{V}_8-\hat{S}_8}{\tau_2^4 \hat{\eta}^8}\left(\frac{1}{2}+i\frac{\tau_2}{2}\right)\mcomma
\end{eq}
with the Chan-Paton charge $N=32$ that corresponds to $N=32$ $\overline{{\text{D9}}}$ branes. The presence of an orientifold $\text{O9}^+$ plane follows from the $\Omega (-1)^{f_L}$ projection. This model is the primary example of a more general approach to supersymmetry breaking in string theory, namely brane supersymmetry breaking~\cite{Antoniadis:1999xk,Angelantonj:1999jh,Aldazabal:1999jr,Angelantonj:1999ms}. Much of what we shall discuss about this model also applies to other examples from brane supersymmetry breaking. The number of massless bosons exceeds that of massless fermions, among which a massless gravitino signals the non-linear realisation of supersymmetry~\cite{Dudas:2000nv,Pradisi:2001yv,Kitazawa:2018zys}. 

The last of the ten-dimensional non-supersymmetric models is type 0'B string theory~\cite{Sagnotti:1995ga,Sagnotti:1996qj}, which is the $\Omega (-1)^{f_R}$ orientifold of type 0B string theory. The one-loop amplitudes in this case are
\begin{eq} \label[pluralequation]{eq:vacuum_ampl_0primeB}
    {\cal T}_{\text{0'B}} & = \int_{{\cal F}}\frac{d^2\tau}{\tau_2^2}\frac{O_8\bar{O}_8+V_8 \bar{V}_8+ S_8\bar{S}_8+ C_8\bar{C}_8}{\tau_2^4 \eta^8\bar{\eta}^8} \mcomma \\
    {\cal K}_{\text{0'B}} & = \int_0^\infty \frac{d\tau_2}{\tau_2^2}\frac{-O_8+V_8+S_8-C_8}{\tau_2^4 \eta^8}(2i\tau_2)\mcomma \\
    {\cal A}_{\text{0'B}} & = \int_0^\infty \frac{d\tau_2}{\tau_2^2}\frac{2 N\bar{N} \, V_8 -\left(N^2 + \bar{N}^2\right) C_8}{\tau_2^4 \eta^8}\left(i\frac{\tau_2}{2}\right)\mcomma \\
    {\cal M}_{\text{0'B}} & = \left(N+\bar{N}\right)\int_0^\infty \frac{d\tau_2}{\tau_2^2}\frac{\hat{C}_8}{\tau_2^4 \hat{\eta}^8}\left(\frac{1}{2}+i\frac{\tau_2}{2}\right)\mcomma
\end{eq}
with complex CP charges $N=\bar{N}=32$, taking values in the (anti)fundamental representation of unitary groups. Once again, there are more bosons at the massless level.

These three examples of tachyon-free non-supersymmetric strings seem harmless. They evade all the problems carried by the tachyons and have a seemingly consistent worldsheet description.
However, carefully inspecting the vacuum amplitudes unveils a lurking problem that awaits us.

\subsection{Tadpole divergences}\label{ssec:sugimoto_divergence}

Consider the Sugimoto model. The one-loop vacuum amplitude from \cref{eq:vacuum_ampl_Sugimoto} is
\begin{eq}\label{eq:Sugimoto_divergence}
    {\cal T}_{\text{USp$(32)$}}+{\cal K}_{\text{USp$(32)$}}+{\cal A}_{\text{USp$(32)$}}+{\cal M}_{\text{USp$(32)$}}=32\int_0^\infty \frac{d\tau_2}{\tau_2^6}\frac{\hat{V}_8}{\hat{\eta}^8}\left(\frac{1}{2}+i\frac{\tau_2}{2}\right)\mperiod
\end{eq}
This diverges. The problematic behaviour is in the region $\tau_2\to 0$, which initially appears to be a UV divergence because $\tau_2$ is the would-be Schwinger parameter of quantum field theory. However, this amplitude should be read from the point of view of bulk (i.e., closed-string) modes using open-closed duality. To this end, by performing a $P=T^{\frac{1}{2}} S T^2 S T^{\frac{1}{2}}$ transformation~\cite{Angelantonj:2002ct}, the amplitude becomes
\begin{eq}\label{eq:sugimoto_tadpole_divergence}
    64\int_0^\infty d\ell \frac{\hat{V}_8}{\hat{\eta}^8}\left(\frac{1}{2}+i\ell\right)\mcomma
\end{eq}
in which the divergence has a clear IR interpretation, arising from the $\ell\to\infty$ region. Expanding the character, one finds that the amplitude is proportional to 
\begin{eq}\label{eq:sugi_div}
    \int_0^\infty d\ell\mcomma
\end{eq}
which can be interpreted as an IR divergence from massless state propagators since
\begin{eq}
    \int_0^\infty d\ell = \int_0^\infty d\ell e^{-2\pi \ell m^2}\Big|_{m^2=0}=\frac{1}{2\pi m^2}\Big|_{m^2=0} \mperiod
\end{eq}
We encountered divergences of this type in \cref{ssec:moduli_space}.
Physically, the contribution in \cref{eq:sugi_div} comes from integrating over a region of the moduli space of Riemann surfaces that contains an infinitely long cylinder, representing string modes propagating for an infinite Schwinger time. This cylinder/propagator must end; therefore, the presence of a divergence is a signal of non-zero tadpole amplitudes at half-loop. This is precisely what happens in this case: the dilaton\footnote{The dilaton is the only Lorentz-scalar physical field in this theory. However, a non-vanishing tadpole for the non-physical trace part of the metric also contributes to the divergence. This is important for consistency, and the reader can find more details in~\cite{Polchinski:1988jq}.} has a non-vanishing tadpole amplitude on the disc and on the crosscap.

The divergence that we have just found is the hallmark of the absence of spacetime supersymmetry in string theory. To explain this point and resolve the divergence issue, we now return to the discussion of \cref{ssec:moduli_space}.

The degenerations from non-separating cycles, corresponding to \cref{eq:tachyon_divergence}, are irrelevant for amplitudes in tachyon-free string theories. Instead, the divergence in \cref{eq:sugimoto_tadpole_divergence} is of the second type discussed in \cref{ssec:moduli_space}, originating from separating cycles that shrink (see \cref{eq:tachyon_and_dilaton_divergence}). More generally, this is the type of divergence that all non-tachyonic non-supersymmetric strings have.

These divergences should make the reader uncomfortable. The importance of string theory lies in its power in computing scattering amplitudes, the physical observables in a quantum theory of gravity, with \emph{finite} results, not plagued by the infinities of quantum field theories. We have argued that strings also contain divergences. 

However, these divergences represent physical states---massless scalars that we shall denote generically as \emph{dilatons}---that propagate for an infinite amount of Schwinger time. They are IR divergences. Similarly to quantum field theory, IR divergences are not inconsistencies. They contain physical messages, including the consequences of supersymmetry breaking. Before explaining how strings take care of IR divergences, let us analyse the structure of the divergent diagrams and the role of supersymmetry.

In tachyon-free setups, we only have the separating-cycle divergences. However, they can vanish if at least one of the two sub-diagrams vanishes. The simplest examples are the vacuum amplitudes that we have been exploring at the one-loop level. An $\ell$-loop vacuum amplitude degenerates into two tadpole diagrams for the dilaton with genera $g$ and $\ell-g$, connected by the divergent zero-momentum dilaton propagator:
\begin{eq}\label{eq:divergences_vacuum_ampl}
    \int_0^1 \frac{dq}{q}\sum_{g=0}^\ell \expval{V_\phi}_g \expval{V_\phi}_{\ell-g}\mcomma
\end{eq}
where $\expval{V_\phi}_g$ is a shorthand notation for $\mathcal{A}_{g,0,0,1}$ with an insertion of the dilaton vertex operator.
Tree-level tadpoles are expected to vanish (see the discussion in \cref{sec:background}), so that one-loop vacuum amplitudes cannot diverge. This applies to all closed-string models. An analogous argument starting from $\expval{V_\phi V_\phi}$ shows that the one-loop dilaton tadpole is finite; therefore, divergences can arise starting from the two-loop vacuum amplitude,
\begin{eq}
    \mathcal{A}_{2,0,0,0}\to \expval{V_\phi}_1^2 \log \epsilon\mcomma 
\end{eq}
with $\epsilon\to 0$.

This can be seen explicitly in the 26-dimensional bosonic string following, for instance,~\cite{Belavin:1986cy,Moore:1986rh,Kato:1986wj}. At the two-loop level, the modular parameter of the torus is replaced by the period matrix $\tau_{ij}$, which lives in the Siegel upper half-plane, $\tau_{ij}=\tau_{ji}$ and $\text{Im}\tau>0$. The fundamental domain of the Siegel half-plane, namely the quotient by Sp$(4,\Z)$, is isomorphic to the moduli space of $g=2$ Riemann surfaces\footnote{Up to subtracting the diagonal period matrices, and ignoring a technical subtlety with the compactifications of the two spaces.}; therefore, the two-loop vacuum amplitude can be written in terms of the genus-two $\theta$ functions,
\begin{eq}
    \theta\smqty[\mathbf{a}\\\mathbf{b}](\mathbf{z}|\tau)=\sum_{\mathbf{n}\in \Z^2}e^{i\pi (\mathbf{n}+\mathbf{a})\cdot \tau \cdot (\mathbf{n}+\mathbf{a})+2\pi i (\mathbf{n}+\mathbf{a})(\mathbf{z}+\mathbf{b})}\mperiod
\end{eq}
The characteristics $(\mathbf{a},\mathbf{b})\in\left(\frac{1}{2}\Z/\Z\right)^4$ are called even or odd depending on whether the combination $4\mathbf{a}\cdot \mathbf{b}$ is even or odd~\cite{Alvarez-Gaume:1986rcs}. A basis of degree-2 modular forms can be built by generalising the Eisenstein series for $g=1$, obtaining 4 algebraically independent modular functions. The one-loop vacuum amplitude can be expressed in terms of the $E_4$ and $E_6$ Eisenstein series through the modular discriminant
\begin{eq}
    \Delta(\tau)=(2\pi)^{12}\eta(\tau)^{24} = (2\pi)^{12}\frac{1}{2^6 3^3}\left({E_6(\tau)}^3-{E_6(\tau)}^2\right)\mcomma
\end{eq}
where $\eta$ can be written in terms of $\theta$ functions. Similarly, the two-loop case uses the generalised Eisenstein series
\begin{eq}
    \Psi_{10}=\frac{1}{2^{12}}\prod_{(\mathbf{a},\mathbf{b})\text{ even}}\theta\smqty[\mathbf{a}\\\mathbf{b}](0|\tau)^2\mperiod
\end{eq}
The two-loop vacuum amplitude, up to overall factors, is
\begin{eq}\label{eq:bosonic_2_loop}
    \mathcal{A}_{2,0,0,0}\sim \int_{\mathcal{F}_2}\frac{d^2\tau_{ij}}{\left(\det\text{Im}\tau\right)^{13}} \left|\Psi_{10}\right|^{-2} \mperiod
\end{eq}
The region of moduli space integration that corresponds to the degeneration of the genus-two Riemann surface into two genus-one sub-surfaces has\footnote{$\alpha$ is a complex constant that depends on the choice of homology basis. Its modulus can be reabsorbed into a $t$ redefinition at $\mathcal{O}(t)$.} 
\begin{eq}\label{eq:bosonic_two_loop_degeneration}
    \tau\to\begin{pmatrix}
        \tau_1 & 0 \\ 0 & \tau_2
    \end{pmatrix}+t \begin{pmatrix}
        0 & \alpha \\ \alpha & 0
    \end{pmatrix} +\mathcal{O}(t^2)
\end{eq}
as $t\to0$, where $t$ measures the length of the pinching cycle,
\begin{eq}
    -\frac{2\pi^2}{\log |t|}\mcomma
\end{eq}
or equivalently the length of the connecting cylinder in the conformally equivalent picture, 
\begin{eq}
    s=-\frac{\log |t|}{2\pi}\mperiod
\end{eq}
Expanding $\Psi_{10}$ in powers of $t$, or equivalently in powers of $\tau_{12}$, 
\begin{eq}
    \Psi_{10}\to-\frac{\Delta(\tau_1)\Delta(\tau_2)}{(2\pi)^{22}}\tau_{12}^2\left[1-\frac{\tau_{12}^2}{6\pi^2}\frac{\theta'''\smqty[1/2\\1/2](0|\tau_1)}{\theta'\smqty[1/2\\1/2](0|\tau_1)}\frac{\theta'''\smqty[1/2\\1/2](0|\tau_2)}{\theta'\smqty[1/2\\1/2](0|\tau_2)}\right]^2\mperiod
\end{eq}
Expressing the zero-momentum propagator of level-$n$ states in terms of the $s$-variable,
\begin{eq}
    \int \frac{d^2\tau_{12}}{|\tau_{12}|^2}|\tau_{12}|^{2n}=(2\pi)^2|\alpha|^{2n}\int ds e^{-4\pi n s}\mcomma
\end{eq}
\cref{eq:bosonic_2_loop} becomes
\begin{eq}
    \mathcal{A}_{2,0,0,0}\sim \expval{V_T}_1 |\alpha|^{-2}\Pi_{T}(0)\expval{V_T}_1+\expval{V_\phi}_1 \Pi_{\phi}(0)\expval{V_\phi}_1 + \dots \mcomma
\end{eq}
where $V_{T,\phi}$ are the vertex operators of the tachyon and of the dilaton at zero momentum and $\Pi_{T,\phi}(0)$ are the divergent zero-momentum propagators. This explicitly shows how divergences appear in the bosonic two-loop vacuum amplitude.
Analogous divergences are present in all scattering amplitudes, following the pattern of \cref{eq:divergences_vacuum_ampl}, with the inclusion of additional vertex operators. 
Note that the $t$-expansion procedure is not modular invariant, and care must be taken when using these results beyond this simple evidence for infinities. We will elaborate on this issue when we discuss the tadpole-subtraction procedure.

As previously mentioned, all these divergences must disappear in the presence of unbroken spacetime supersymmetry. The reason is purely dynamical: the dangerous regions of moduli spaces are still there, but the sub-diagrams accompanying the divergent propagators vanish or conspire so that the final result vanishes once all string states are considered~\cite{Martinec:1986wa}. Different diagrams vanish depending on the amount of spacetime supersymmetry, and different infinities are lifted from the amplitudes~\cite{Friedan:1985ge}. A particularly simple case is the vanishing of $g$-loop vacuum amplitudes $\mathcal{A}_{g,0,0,0}$. Consider a genus-$g$ vacuum amplitude and cut a handle, gluing it again through the insertion of a sum over a complete set of states. $\mathcal{A}_{g,0,0,0}$ can then be expressed as a supertrace of a genus-$(g-1)$ diagram,
\begin{eq}\label{eq:vacuum_ampl_susy}
    \mathcal{A}_{g,0,0,0}=\sum_{\ket{B}}\bra{B} ((g-1)\text{-loop})\ket{B}-\sum_{\ket{F}}\bra{F} ((g-1)\text{-loop})\ket{F}\mcomma
\end{eq}
where $\ket{B}$ and $\ket{F}$ represent bosonic and fermionic states.
Unbroken supersymmetry implies that one can replace a (spacetime) fermionic operator on the worldsheet with a bosonic operator surrounded by the integral of the supersymmetry current---a global symmetry current on the worldsheet. The current integral is a topological operator that can sweep over the Riemann surface until it collapses to a point. The second term of \cref{eq:vacuum_ampl_susy} is equal to the first, and hence 
\begin{eq}
    \mathcal{A}_{g,0,0,0}=0\mperiod
\end{eq}
A similar argument holds for tadpoles: fermionic tadpoles are not allowed by Lorentz invariance and bosonic tadpoles can be mapped to fermionic ones by a contour integral of the supersymmetry current. 

In more complicated open-string diagrams, the GSO projection makes the sum of the genus-$g$ diagrams vanish. The one-loop vacuum amplitude of type I string theory is the simplest example of this cancellation mechanism, in which the open and unoriented amplitudes do not vanish by themselves---they actually diverge---but the complete one-loop contribution cancels because of the relative phases. Schematically, the degeneration limit becomes
\begin{eq}\nonumber
    (\text{disc } - \text{ crosscap} )   \text{---propagator---}(\text{disc } -\text{ crosscap} )\mcomma
\end{eq}
and the overall contribution vanishes because the disc and crosscap tadpoles are equal.
The reader can find more details in~\cite{Angelantonj:2002ct,Raucci:2024fnp}.

We have seen how IR divergences are present in string amplitudes whenever there is no dynamical protection. Some of these divergences have resolutions that are analogous to those of quantum field theory. In particular, tachyon divergences can be regularised, and their effect is to make scattering amplitudes complex. For instance, the decay rate of bosonic string theory in the Minkowski vacuum can be extracted by regularising the tachyon divergence in the one-loop vacuum amplitude~\cite{Marcus:1988vs,Baccianti:2025gll}. Similarly, one can extract mass renormalisation and decay rates of non-BPS states, even in supersymmetric theories.
However, it remains unclear how to deal with dilaton tadpole divergences in this manner. Currently, there is no clear procedure that regularises tadpole divergences in a quantum-field-theory fashion. Subtracting these divergences appears to be a purely quantum-gravitational effect, which is the main subject of the second part of this section: the physical consequences of tadpole divergences.

\subsection{The Fischler--Susskind mechanism}\label{ssec:FS}

Tadpoles in quantum field theories are removed by adding background fields. These contribute to the scattering amplitudes with additional tadpole terms, and cancellation occurs when the original tadpoles and those from the background fields are equal and opposite. In string theory, something analogous happens, albeit with a completely different mechanism.

In string perturbation theory, tadpoles can be subtracted by implementing a procedure known as the Fischler--Susskind mechanism, originally formulated in~\cite{Fischler:1986ci,Fischler:1986tb,Lovelace:1986kr} and then developed in~\cite{Callan:1986bc,Das:1986dy,Metsaev:1987ju,Callan:1988st,Callan:1988wz,Tseytlin:1988mw,Russo:1989kq,Tseytlin:1990mv}. The idea is to extend the $\alpha'$ renormalisation to the $g_s$ expansion by adding counterterms that cancel the divergences in the $g_s$ expansion. The subtraction procedure, which can be viewed as an improved RG for the string, employs the same worldsheet cutoff for both $\sigma$-model divergences and moduli-space divergences. This cuts off the length of the pinching necks at the degeneration points and then renormalises the fields in the $g_s$ expansion. For instance, at the lowest order, the bare dilaton reads
\begin{eq}
    \phi^{(0)}\sim \phi-g_s^2 T_1 \log\epsilon\mcomma 
\end{eq}
where $T_1$ is the one-loop tadpole and $\epsilon$ is the cutoff. Renormalising fields has the effect of cancelling tadpole contributions at the cost of mixing different orders of string perturbation theory. The two-loop divergence in the vacuum amplitude is removed by one-loop counterterms. These are divergent by themselves but cancel once the one-loop and two-loop terms are both included.

To give a flavour of the cancellations, consider a 4-point scattering amplitude in a theory with tadpoles. The divergence comes from the degeneration
\begin{eq}
    \expval{V_\varphi^4}_1\to \expval{V_\varphi^4 V_\phi}_0 \int_\epsilon \frac{dq}{q}\expval{V_\phi}_1\mperiod
\end{eq}
This IR divergence is cancelled by a counterterm that renormalises the dilaton, which enters as a tree-level contribution: $\expval{V_\varphi^4 V_\times}$.

Taking the same regulator in both $\alpha'$ and $g_s$ expansions has a remarkable consequence: it becomes impossible to disentangle them. The standard $\sigma$-model renormalisation works on a surface of fixed topology and is actually independent of it. In fact, $\sigma$-model $\beta$ functions are local quantities, sensitive to the short-distance physics on the worldsheet. They do not feel the different global structures of the Riemann surfaces. Instead, the Fischler--Susskind mechanism spectacularly captures the effects of the sum over Riemann surfaces. 

IR divergences are interpreted as violations of Weyl invariance at a fixed order in string perturbation theory. The improved RG procedure reminds us that this is not necessarily a problem because string theory is a theory of Riemann surfaces, and not a theory of one Riemann surface. The genus expansion relaxes the genus-by-genus Weyl invariance, only requiring the full $g_s$ series to be Weyl invariant. In other terms,
\begin{eq}
    \mathcal{A}_{g,0,0,0}\sim\int_{\mathcal{M}_g}d\tau_g \mathcal{D}[\varphi]e^{-S[\varphi,\tau_g]}
\end{eq}
is not Weyl-invariant and only the full partition function of \cref{eq:stringyamplitudes},
\begin{eq}
    \mathcal{A}_0=\sum_{g=0}^\infty g_{s}^{2g-2}\mathcal{A}_{g,0,0,0}=\sum_{g=0}^\infty g_{s}^{2g-2} \int_{\mathcal{M}_g}d\tau_g \mathcal{D}[\varphi]e^{-S[\varphi,\tau_g]}\mcomma
\end{eq}
is well-defined.

Mixing different orders of string perturbation theory through the \emph{all-genera} renormalisation procedure means that $\beta$ functions receive contributions from different Riemann surfaces. Schematically,
\begin{eq}
    \beta\to\sum_{g}\beta_g\mperiod
\end{eq}
This is not exotic. In fact, it is expected in string theory. Scattering amplitudes on genus-$g$ Riemann surfaces are certainly non-vanishing, and there must be corresponding effects on the $\sigma$-model $\beta$ functions encoding them. The Fischler--Susskind procedure is precisely the prescription that computes $g_s$ corrections to the string equations of motion.

To make it even less exotic, the coupling of open strings to closed strings can be recast in the language of the Fischler--Susskind subtraction. The DBI action, which couples the gauge fields on the D-branes to gravity, can be obtained as a combination of half-loop and one-loop contributions~\cite{Callan:1988st}: adding a boundary to a scattering amplitude implies integrating over its position and length. The latter, taking into account the dependence of the propagator on the boundary length, leads to a contribution proportional to
\begin{eq}
    \int \frac{d\ell}{\ell}\mperiod
\end{eq}
This is the familiar logarithmic divergence from a dilaton tadpole. Using the zero-momentum dilaton propagator
\begin{eq}
        \d X_\mu \bar{\d} X^\mu \mcomma
\end{eq}
one can reconstruct the expansion of the DBI action. 
We refer the reader to~\cite{Callan:1988st} for details on the computation, hoping to have provided at least some hints of why the tadpole subtraction procedure is physically reasonable and gives the expected results in known setups.

\subsubsection{Tadpole potentials}

The modified $\beta$ functions---and thus the modified equations of motion for non-supersymmetric strings---are a complex topic when considered in full generality. Loop corrections, whether in $\alpha'$ or $g_s$, involve delicate issues of field redefinition and require consistent choices of schemes, which remain beyond our current limited understanding of tadpole subtraction.

A strikingly simple example of the difficulties is the extraction of the two-loop dilaton tadpole, which we have already stumbled across in \cref{eq:bosonic_two_loop_degeneration}.
The naive $\expval{V_\phi}_2$ contains an infinite part coming from the one-loop tadpole and another finite part that is the proper two-loop tadpole $T_2$:
\begin{eq}
    \expval{V_\phi}_2\sim \expval{V_\phi V_\phi}_1 \int \frac{dq}{q} \expval{V_\phi}_1+ T_2\mperiod
\end{eq}
This separation is not modular invariant; therefore, it is not clear whether $T_2$ correctly accounts for the two-loop tadpole that needs to be cancelled.

Notably, the simplest contribution to the loop-corrected $\beta$ functions is free of ambiguities and intricacies. This is the one-loop dilaton tadpole that we have discussed several times, which is also the most important consequence of the absence of spacetime supersymmetry. 
Subtracting the tadpole with the Fischler--Susskind procedure generates a scalar \emph{tadpole potential} for the dilaton, which in the string frame is just a constant,
\begin{eq}
    V(\phi)=T_1\mperiod
\end{eq}
From the functional form of the dilaton vertex operator, one finds that $T_1$ is equal to the one-loop vacuum amplitude~\cite{Green:1987mn}, so that we have computational control over the scalar potential from the worldsheet formulation.
For open strings, the analogous contribution comes at half-loop, as we have seen in \cref{ssec:sugimoto_divergence}, and the scalar potential is 
\begin{eq}
    V(\phi)=T_{\frac{1}{2}}e^{-\phi}\mperiod
\end{eq}
In general, the expected form is 
\begin{eq}\label{eq:tadpole_potential_string_frame}
    V(\phi)=\sum_g T_g e^{2(g-1)\phi}\mcomma 
\end{eq}
although the subtleties mentioned above make the explicit computation of the $T_g$ coefficients unclear.

Note that the tadpole potential is actually similar to what a quantum field theorist would predict. In an effective field theory for the massless modes of the string, there would be quantum contributions to the cosmological constant. The scalar potential plays the same role, and in fact it scales with the cutoff scale of the EFT, the string mass, effectively acting as a bare contribution to the quantum cosmological constant. The overall picture is therefore consistent from both high-energy and low-energy perspectives, and at the same time it highlights the role of string theory as the UV completion of gravitational effective field theories.

What differs for strings is that the cosmological constant is a constant in the string frame. After the field redefinition that isolates the graviton, namely in the Einstein frame, it becomes a scalar potential for the dilaton. This difference is the hallmark of working with a UV-complete theory of gravity. The inclusion of tadpole potentials replaces the cosmological constant problem in string theory.

\subsubsection{Open problems}

Tadpole divergences are analogous to the presence of tachyons. They represent instabilities, albeit of a different sort, and signal that one is not expanding around the true vacuum of the theory. The Fischler--Susskind mechanism replaces tachyon condensation.

As in the case of tachyon condensation, we understand the general philosophy but we lack a clear, systematic, and complete formulation of tachyon subtraction. Concretely, what is missing is a consistent subtraction scheme that allows one to extract the loop corrections, possibly in a modular-invariant way, at all orders. We have seen the simplest manifestation of this problem in the two-loop tadpole term.

A separate and even more puzzling business is to understand the endpoint of the Fischler--Susskind shift in the worldsheet formulation. If the true vacuum lies inside the framework of string perturbation theory~\cite{Dudas:2004nd,Pius:2014gza}, then it must have miraculous cancellations that make all massless tadpoles vanish, or it must have no massless scalars at all in the spectrum, which is at odds with the dilaton controlling the string coupling. Alternatively, the true vacuum may be strongly coupled, outside the realm of the string worldsheet, drastically reducing the level of understanding that string theory can provide about our universe. Ultimately, string field theory methods could be effective in dealing with the background shift, possibly leading to results similar to the supersymmetry restoration of~\cite{Sen:2015uoa}. This raises the question of whether string theory actually contains non-supersymmetric vacua.

Whatever the case, this is a thorny issue that demands more work.
Additional subtleties arise when the tadpole involves open strings, some of which have been explored in~\cite{Dudas:2004nd,Kitazawa:2008hv}.

\subsection{Vacua of non-supersymmetric strings}\label{ssec:nonsusy_vacua}

Given a string theory without spacetime supersymmetry and no tachyons in the spectrum, the preceding sections instruct us to add string-loop corrections to the equations of motion of a specific type. Among these, the one that is most under control is actually the most crucial: the tadpole potential. 

In general, the Fischler--Susskind procedure predicts a scalar potential for all massless fields in the theory, whose structure becomes extremely intricate almost immediately. For this reason, explicit results are typically known only in the simplest models that display such potentials: the three non-supersymmetric and non-tachyonic string theories in ten dimensions.

Two possibilities arise. The tadpole potential can be either a small correction to some classical string background or a fundamental building block, therefore mixing classical and quantum effects in the spacetime gravitational fabric. The latter possibility is present because the tree-level vacuum energy of the string vanishes, and the one-loop tadpole potential is the first term in the $g_s=e^\phi$ expansion.

Cases where the tadpole potential is a subleading correction do not occur for the three ten-dimensional non-supersymmetric strings\footnote{They can be found in lower-dimensional models, such as compactifications of ten-dimensional strings on stringy geometries. See, for instance,~\cite{Baykara:2022cwj,Fraiman:2023cpa} for the S$^1$ compactification of the heterotic non-supersymmetric string where the sphere radius is of stringy size.} because in all cases the spectra are not rich enough to contain vacua with all the scalars stabilised. From now on, we shall focus on the second possibility, where the tree-level terms compete with the one-loop tadpole potential.

Vacua can be divided into two categories: vacuum solutions and flux compactifications. We shall outline what is known for each of these cases separately.

\subsubsection{Vacuum solutions: codimension-one}\label{ssec:codim-1}

A vacuum solution is the gravitational backreaction of an empty universe. For supersymmetric strings, this means ten-dimensional Minkowski or compactifications to lower-dimensional Minkowski times Calabi-Yau spaces.\footnote{For the level of complexity that we are addressing here, we should replace Calabi-Yau with any Ricci-flat manifold. For supersymmetric strings, we know that Calabi-Yau manifolds are vacuum solutions even after adding higher-derivative corrections, albeit not with the same metric.} For a theory of pure gravity with a cosmological constant, this would select the maximally symmetric (A)dS space. 

For non-supersymmetric strings, none of these options apply. The Fischler--Susskind procedure generates a scalar potential, which turns out to be a \emph{runaway potential} for the dilaton, and therefore no maximally symmetric space and no Ricci-flat space can be a vacuum solution. In fact, given the potential in \cref{eq:tadpole_potential_string_frame} in the string frame, the counterpart in the ten-dimensional Einstein frame is 
\begin{eq}
    V(\phi)=\sum_g T_g e^{\left(2g+\frac{1}{2}\right)\phi}\mperiod
\end{eq}
A one-loop tadpole, which is the case for the non-supersymmetric heterotic string, contributes as
\begin{eq}
    V(\phi)=T_1 e^{\frac{5}{2}\phi}\mcomma
\end{eq}
while a disc/crosscap tadpole, which is the case for the two non-supersymmetric orientifolds, contributes as 
\begin{eq}
    V(\phi)=T_{\frac{1}{2}}e^{\frac{3}{2}\phi}\mperiod
\end{eq}
These runaway potentials are the spacetime manifestation of the absence of supersymmetry in a quantum theory of gravity. 
Hence, vacuum solutions without supersymmetry necessarily involve non-trivial dilaton profiles in addition to the gravitational background.

One proposal to identify vacuum solutions that are the counterparts of ten-dimensional Minkowski space for supersymmetric strings is to insist on maintaining the maximum number of isometries that are compatible with the equations. In the presence of tadpole runaway potentials, this means \emph{codimension-one solutions}, in which one direction is selected and the background takes the form
\begin{eq}
    ds^2=e^{2A(y)}dx_9^2 + dy^2\mcomma \qquad \phi=\phi(y)\mcomma
\end{eq}
where $dx_9^2$ is a nine-dimensional maximally symmetric space.

In the context of non-supersymmetric strings, these types of vacua were first explored in~\cite{Dudas:2000ff} with a flat nine-dimensional Minkowski and are known as Dudas--Mourad vacua. Vacua of this type share some common features that have recently been in the spotlight: they contain two curvature singularities, separated by either a finite or an infinite proper distance. The former case is physically more interesting because it leaves a nine-dimensional theory that can be interpreted as a spontaneous compactification of the ten-dimensional string, in which the backreaction of supersymmetry breaking drives the compactification. The string coupling can either diverge or go to weak coupling at the singular points.

Codimension-one vacua of non-supersymmetric strings were recently reconsidered after the Dudas--Mourad solution was shown to be perturbatively stable~\cite{Basile:2018irz}. Generalizations~\cite{Basile:2022ypo} and vacua with analogous structures~\cite{Blumenhagen:2000dc,Dudas:2002dg,Pelliconi:2021eak,Mourad:2021qwf,Mourad:2021roa,Mourad:2022loy,Mourad:2023ppi,Mourad:2023loc,Mourad:2024dur} have been explored, but the most important aspect of the story is the universal behaviour of the boundaries~\cite{Mourad:2020cjq,Raucci:2022jgw,Mourad:2023wjg}. The strongly-curved boundaries are the problematic regions that can make the solutions unphysical. In a complete quantum-gravitational picture, they must be replaced by some stringy description or by a completely different smooth geometrical structure. 

In~\cite{Raucci:2022jgw,Mourad:2023wjg,Mourad:2024dur}, it was observed that only a few types of boundaries appear, which resonates with arguments from dynamical cobordism~\cite{McNamara:2019rup,Antonelli:2019nar,Buratti:2021yia,Buratti:2021fiv,Angius:2022aeq,Blumenhagen:2022mqw,Angius:2022mgh,Blumenhagen:2023abk,Angius:2023xtu,Huertas:2023syg,Angius:2023uqk,Angius:2024zjv,Ruiz:2020jjz,Huertas:2024mvy,Angius:2024pqk}. In particular, the analysis of~\cite{Mourad:2023wjg} reformulated the problem in terms of a self-adjoint Schr\"{o}dinger system, 
\begin{eq}
    \left(\mathcal{A}\mathcal{A}^\dagger + b\right)\psi = m^2 \psi\mcomma
\end{eq}
indicating that the Dudas--Mourad vacua admit boundary conditions\footnote{The choice of boundary condition is actually unique for the massless graviton.} that are compatible with stability. These findings are analogous to less recent tests for the consistency of boundaries, built on probing singularities with point particles or on holography~\cite{Wald:1980jn,Horowitz:1995gi,Gubser:2000nd}, therefore strengthening the status of the singular codimension-one vacua.
Another consistency check was performed in~\cite{Mourad:2024mpg}, embedding branes in the Dudas--Mourad vacua, and finding that self-adjoint boundary conditions exist for all expected branes in non-supersymmetric strings~\cite{Dudas:2000sn,Dudas:2001wd}.

Codimension-one vacua can alternatively be interpreted as time-dependent backgrounds,
\begin{eq}
    ds^2=-dt^2+e^{2A(t)}dx_9^2\mcomma \qquad \phi=\phi(t)\mperiod 
\end{eq}
Cosmologies of this type have been found with flat~\cite{Dudas:2000ff} and curved~\cite{Raucci:2022jgw} spatial slices. The exponential potentials from the Fischler--Susskind mechanism applied to half-loop tadpoles lead to a phenomenon known as \emph{climbing scalar}, in which the dynamics forces a scalar field---the dilaton $\phi$ or a combination of the dilaton and the internal volume if spacetime is compactified---to climb the steep exponential potential~\cite{Dudas:2010gi,Condeescu:2013gaa}. Some consequences of this mechanism for the CMB power spectrum have been investigated in~\cite{Dudas:2012vv,Sagnotti:2013ica,Kitazawa:2014dya,Kitazawa:2014mca,Kitazawa:2015uda}.

Much is known about the codimension-one vacua, but their fate is still uncertain due to the singular boundaries. Even with strong indications that they should have physical meaning, only the discovery of a stringy resolution of the dangerous boundary behaviour will put an end to the doubts. 
Much less is known about higher-codimension vacuum solutions, which should represent the non-supersymmetric counterpart of Calabi-Yau compactifications.

\subsubsection{Compactifications}\label{ssec:tadpole_cpt}

Finding geometric vacua of non-supersymmetric strings is extremely complicated because one needs to solve systems of coupled, second-order differential equations of motion. Spacetime supersymmetry, in addition to the other features it entails, makes it possible to bypass this step. In fact, imposing unbroken supersymmetry and the Bianchi identities automatically solves the equations of motion in non-pathological settings. Adding the supersymmetry-breaking tadpole potential deformation means that the level of control on the landscape of vacua is necessarily more primitive. 

Therefore, the general procedure is to solve the ten-dimensional equations, check the perturbative stability, and then possibly understand the non-perturbative decay channels. Each of these steps requires either a significant amount of computing time or undeveloped machinery to engineer solutions and monitor their stability. Despite much effort, no concrete systematic methods facilitate the three-step procedure outlined above, and only relatively simple vacua are known.

A first family of vacua that has been explored involves treating the tadpole potential as a small perturbation, as we have already mentioned. This is natural from the perspective of adding string loop corrections, but it also means that the dilaton must be stabilised in the unperturbed solution and that the range of parameters must be chosen to avoid a flat decompactification limit. Otherwise, the scalar potential would contribute as a runaway after decompactification, which is incompatible with its role as a small correction.
This class of vacua is not particularly rich in ten dimensions, because both the non-supersymmetric heterotic string and the Sugimoto model only have a three-form flux and gauge fields. The lowest-order equations without the tadpole perturbation would be those of type I supergravity, and no appropriate vacua are known that satisfy all our requirements. On the other hand, the type 0'B string has more available fluxes and solutions of this sort do exist. A characteristic example is given by a deformation of AdS$_5\times$S$^5$~\cite{Klebanov:1998yya,Angelantonj:1999qg,Dudas:2000sn,Angelantonj:2000kh,Basile:2021mkd,Basile:2025lek}, valid only in a small region of spacetime, or equivalently in a small region of parameter space. Physically, it corresponds to the geometry of $N\to\infty$ D3 branes. The 't Hooft limit
\begin{eq}
    N e^\phi=\text{constant}\mcomma \qquad N\to\infty\mcomma
\end{eq}
is chosen so that the contribution from the $N\to\infty$ branes dominates the tadpole term. Schematically, the metric $g$, flux $N$, and the tadpole potential $T$ enter the equations as
\begin{eq}
    \d^2 g \sim N^2 e^{2\phi} \gg T e^\phi\mperiod
\end{eq}
This solution has been used to envisage the existence of a non-supersymmetric version of the AdS/CFT correspondence, with the tadpole perturbation playing the role of running couplings~\cite{Angelantonj:2000kh}:
\begin{eq}
    \alpha \leftrightarrow e^\phi \mperiod
\end{eq}

An alternative approach is to balance the effect of the tadpole potentials with the flux and curvature terms. These vacua are of the type
\begin{eq}
    ds^2=ds_{\text{AdS}}^2+ ds^2_M\quad \text{with $M$ an Einstein manifold}\mcomma \qquad \phi=\phi_0\mperiod 
\end{eq}
They are sometimes known as the Freund--Rubin class because the form fluxes are proportional to the internal volume form. The one-(or half-)loop tadpole can be balanced with tree-level effects without inconsistencies because of the absence of a tree-level cosmological constant. Solutions of this type and their stability have been explored in~\cite{Gubser:2001zr,Mourad:2016xbk,Basile:2018irz,Raucci:2022bjw,Baykara:2022cwj,Fraiman:2023cpa}. They are typically unstable, either perturbatively~\cite{Basile:2018irz,Raucci:2022bjw,Fraiman:2023cpa} or non-perturbatively~\cite{Antonelli:2019nar}. The two archetypal examples are the heterotic AdS$_7\times$S$^3$ and orientifold AdS$_3\times$S$^7$ of ~\cite{Mourad:2016xbk}. The former can be made perturbatively stable by replacing the sphere with a projective plane, while the latter, valid for both non-supersymmetric ten-dimensional orientifolds, is perturbatively unstable~\cite{Basile:2018irz}. They have been argued to correspond to the near-horizons of the NS5 and D1 branes~\cite{Basile:2021mkd,Basile:2022zee}.
This class of solutions engineers a sort of spontaneous compactification (with flux) in which vacuum energy curves spacetime. It belongs to the same family of vacua as more recent attempts that use Casimir energies~\cite{DeLuca:2021pej,Luca:2022inb,ValeixoBento:2025yhz,DallAgata:2025jii,Aparici:2025kjj}. Work is underway to produce less simple examples, but the general structure is the same as in the two main vacua of~\cite{Mourad:2016xbk}.

Warped compactifications remain largely unexplored, apart from the singular solutions of~\cite{Mourad:2021roa,Mourad:2024dur}. Remarkably, a no-go theorem can be formulated~\cite{Basile:2020mpt} along the lines of the Maldacena-Nunez one~\cite{Maldacena:2000mw}, ruling out Minkowski and de Sitter vacua even with the inclusion of the (leading) tadpole potential. The procedure is simple enough that we can review it here. Consider the following warped metric:
\begin{eq}
    ds^2=e^{2A(y)}g_{\mu\nu}^{(d)}(x)dx^\mu dx^\nu + e^{2B(y)}g_{ij}^{(10-d)}(y)dy^i dy^j\mcomma
\end{eq}
with a $d$-dimensional maximally symmetric spacetime $g_{\mu\nu}^{(d)}$ with
\begin{eq}
    R_{\mu\nu}^{(d)}=\Lambda g_{\mu\nu}^{(d)}\mcomma
\end{eq}
and a $y$-dependent dilaton $\phi(y)$. The Einstein-frame action is
\begin{eq}
    S=\int_{10} R-\frac{1}{2}(\d\phi)^2-\frac{1}{2}\sum_k e^{-2\beta_k\phi}F_k^2-T e^{\gamma\phi}\mperiod
\end{eq}
The equations of motion take a simple form in the gauge
\begin{eq}
    B=-\frac{d-2}{10-d}A\mperiod
\end{eq}
Using the dilaton equation in the $\mu\nu$ Einstein equations, together with the identity\footnote{$\nabla^2$ is the Laplace operator on the internal manifold with the metric $g_{ij}^{(10-d)}(y)$.}
\begin{eq}
    \nabla^2 A=\frac{1}{\alpha}e^{-\alpha A}\nabla^2 e^{\alpha A}-\alpha (\d A)^2 \qquad \text{with}\quad \alpha=\frac{16}{10-d}\mcomma
\end{eq}
and then integrating over the internal space, leads to
\begin{eq}\label{eq:no_go}
    \int_{10-d} \Lambda=\frac{1}{8}\sum_{k}\left(\frac{\beta_k}{\gamma}-\frac{k-1}{2}\right)\int_{10-d}e^{2A-2\beta_k\phi}F_k^2\mcomma 
\end{eq}
under the assumption that the boundary terms vanish. This is true if appropriate boundary conditions hold or if the internal manifold has no boundary\footnote{Note that for the Dudas--Mourad vacua of \cref{ssec:codim-1}, boundaries are indeed present, allowing the no-go theorem to be evaded.}.
For the three ten-dimensional tachyon-free non-supersymmetric strings, $\beta_k$ and $\gamma$ are such that the integral on the right of \cref{eq:no_go} is always negative, thus ruling out de Sitter and Minkowski warped compactifications. 

This is not necessarily a problem because the presence of scalar potentials can still produce interesting structures in the final vacuum, but it is remarkable how the dilaton dressings of the form fluxes and the values of the $\gamma$ coefficients in the Einstein frame conspire to produce only AdS spacetimes. An interesting related question is whether a similar statement holds for scale separation, both in static vacua and in dynamical time-dependent scenarios. After all, the latter are the true payoff of adding tadpole potentials from supersymmetry breaking.

\subsubsection{Vacuum stability}

The last topic that we want to mention is perhaps the most complex and delicate. This is the issue of vacuum stability without the protection of spacetime supersymmetry.

Stability has been the leitmotif of our discussion, first heard in the appearance of tachyons, which motivated the search for tachyon-free strings. We learned that even without tachyons, stability is not guaranteed, and a strong gravitational backreaction emerges in the form of worldsheet tadpoles and spacetime scalar potentials. The question that we face now is what happens after the Fischler--Susskind flow. Can we reach a stable non-supersymmetric vacuum with dynamical gravity and no runaway directions?

Answering this question requires a deeper understanding of string theory and gravity in general, which is beyond our current capabilities. The standard approach, which is far from exhaustive, proceeds along the three steps of \cref{ssec:tadpole_cpt}, checking perturbative stability by linearising perturbations of the gravitational background, and then non-perturbative stability by probing known decay channels. The former is another spacetime manifestation of the tachyonic and runaway instabilities that we have already encountered. The latter appears to be a completely different story, although some intersection with tachyons may be possible~\cite{Fabinger:2000jd,McGreevy:2005ci,Adams:2005rb,Delgado:2023uqk}. In fact, non-perturbative decays in terms of bubbles of nothing~\cite{Witten:1981gj} have been argued to be generic without the dynamical protection of supersymmetry~\cite{McNamara:2019rup,GarciaEtxebarria:2020xsr}. This casts doubt on any non-supersymmetric vacuum.

For these reasons, some attempts have been made to apply positive energy-like theorems to non-supersymmetric string theories; for instance, using the formalism of fake supersymmetry~\cite{Witten:1981mf,Nester:1981bjx,Boucher:1984yx,Townsend:1984iu,Skenderis:1999mm,Freedman:2003ax,Townsend:2007aw,Trigiante:2012eb,Danielsson:2016rmq,Raucci:2023xgx} and similar procedures~\cite{Giri:2021eob,Menet:2023rnt,Menet:2023rml,Menet:2025nbf} that mimic the technical aspects of unbroken supersymmetry without carrying its physical significance.

Research in this area is definitely at an early stage, and there may be physical principles and mechanisms that we have yet to understand. Indeed, quantum mechanics has taught us that anything that can happen does happen. Therefore, the stability of a non-supersymmetric gravitational vacuum cannot be a mere coincidence.
Since we appear to be living in one, we should be either concerned about the state of our universe or thrilled that there is still much to understand about the physics of gravity.

\backmatter

\bmhead*{Acknowledgements}

We thank Augusto Sagnotti and Ivano Basile for comments on the manuscript. This work was performed in part at the Aspen Center for Physics, which is supported by a grant from the Simons Foundation (1161654, Troyer). GL is supported in part by Scuola Normale and by INFN (I.S. GSS-Pi).

\bibliography{sn-bibliography}


\begin{thebibliography}{266}
\ifx \bisbn   \undefined \def \bisbn  #1{ISBN #1}\fi
\ifx \binits  \undefined \def \binits#1{#1}\fi
\ifx \bauthor  \undefined \def \bauthor#1{#1}\fi
\ifx \batitle  \undefined \def \batitle#1{#1}\fi
\ifx \bjtitle  \undefined \def \bjtitle#1{#1}\fi
\ifx \bvolume  \undefined \def \bvolume#1{\textbf{#1}}\fi
\ifx \byear  \undefined \def \byear#1{#1}\fi
\ifx \bissue  \undefined \def \bissue#1{#1}\fi
\ifx \bfpage  \undefined \def \bfpage#1{#1}\fi
\ifx \blpage  \undefined \def \blpage #1{#1}\fi
\ifx \burl  \undefined \def \burl#1{\textsf{#1}}\fi
\ifx \doiurl  \undefined \def \doiurl#1{\url{https://doi.org/#1}}\fi
\ifx \betal  \undefined \def \betal{\textit{et al.}}\fi
\ifx \binstitute  \undefined \def \binstitute#1{#1}\fi
\ifx \binstitutionaled  \undefined \def \binstitutionaled#1{#1}\fi
\ifx \bctitle  \undefined \def \bctitle#1{#1}\fi
\ifx \beditor  \undefined \def \beditor#1{#1}\fi
\ifx \bpublisher  \undefined \def \bpublisher#1{#1}\fi
\ifx \bbtitle  \undefined \def \bbtitle#1{#1}\fi
\ifx \bedition  \undefined \def \bedition#1{#1}\fi
\ifx \bseriesno  \undefined \def \bseriesno#1{#1}\fi
\ifx \blocation  \undefined \def \blocation#1{#1}\fi
\ifx \bsertitle  \undefined \def \bsertitle#1{#1}\fi
\ifx \bsnm \undefined \def \bsnm#1{#1}\fi
\ifx \bsuffix \undefined \def \bsuffix#1{#1}\fi
\ifx \bparticle \undefined \def \bparticle#1{#1}\fi
\ifx \barticle \undefined \def \barticle#1{#1}\fi
\bibcommenthead
\ifx \bconfdate \undefined \def \bconfdate #1{#1}\fi
\ifx \botherref \undefined \def \botherref #1{#1}\fi
\ifx \url \undefined \def \url#1{\textsf{#1}}\fi
\ifx \bchapter \undefined \def \bchapter#1{#1}\fi
\ifx \bbook \undefined \def \bbook#1{#1}\fi
\ifx \bcomment \undefined \def \bcomment#1{#1}\fi
\ifx \oauthor \undefined \def \oauthor#1{#1}\fi
\ifx \citeauthoryear \undefined \def \citeauthoryear#1{#1}\fi
\ifx \endbibitem  \undefined \def \endbibitem {}\fi
\ifx \bconflocation  \undefined \def \bconflocation#1{#1}\fi
\ifx \arxivurl  \undefined \def \arxivurl#1{\textsf{#1}}\fi
\csname PreBibitemsHook\endcsname

\bibitem[\protect\citeauthoryear{Veneziano}{1968}]{Veneziano:1968yb}
\begin{barticle}
\bauthor{\bsnm{Veneziano}, \binits{G.}}:
\batitle{{Construction of a crossing - symmetric, Regge behaved amplitude for linearly rising trajectories}}.
\bjtitle{Nuovo Cim. A}
\bvolume{57},
\bfpage{190}--\blpage{197}
(\byear{1968})
\doiurl{10.1007/BF02824451}
\end{barticle}
\endbibitem

\bibitem[\protect\citeauthoryear{Leone}{2024}]{Leone:2024xae}
\begin{botherref}
\oauthor{\bsnm{Leone}, \binits{G.}}:
{Aspects of Stability, Rigidity and Unitarity in String Vacua}.
PhD thesis,
Universit\`a degli Studi di Torino, Italy, Turin U.
(2024)
\end{botherref}
\endbibitem

\bibitem[\protect\citeauthoryear{Raucci}{2024}]{Raucci:2024fnp}
\begin{botherref}
\oauthor{\bsnm{Raucci}, \binits{S.}}:
{Spacetime aspects of non-supersymmetric strings}.
PhD thesis,
Pisa, Scuola Normale Superiore
(September 2024)
\end{botherref}
\endbibitem

\bibitem[\protect\citeauthoryear{Schwarz}{1982}]{Schwarz:1982jn}
\begin{barticle}
\bauthor{\bsnm{Schwarz}, \binits{J.H.}}:
\batitle{{Superstring Theory}}.
\bjtitle{Phys. Rept.}
\bvolume{89},
\bfpage{223}--\blpage{322}
(\byear{1982})
\doiurl{10.1016/0370-1573(82)90087-4}
\end{barticle}
\endbibitem

\bibitem[\protect\citeauthoryear{Green et~al.}{1988a}]{Green:1987sp}
\begin{bbook}
\bauthor{\bsnm{Green}, \binits{M.B.}},
\bauthor{\bsnm{Schwarz}, \binits{J.H.}},
\bauthor{\bsnm{Witten}, \binits{E.}}:
\bbtitle{Superstirng Theory. Vol. 1: Introduction}.
\bsertitle{Cambridge Monographs on Mathematical Physics},
(\byear{1988})
\end{bbook}
\endbibitem

\bibitem[\protect\citeauthoryear{Green et~al.}{1988b}]{Green:1987mn}
\begin{bbook}
\bauthor{\bsnm{Green}, \binits{M.B.}},
\bauthor{\bsnm{Schwarz}, \binits{J.H.}},
\bauthor{\bsnm{Witten}, \binits{E.}}:
\bbtitle{Superstring Theory. Vol. 2: Loop Amplitudes, Anomalies and Phenomenology},
(\byear{1988})
\end{bbook}
\endbibitem

\bibitem[\protect\citeauthoryear{D'Hoker and Phong}{1988}]{DHoker:1988pdl}
\begin{barticle}
\bauthor{\bsnm{D'Hoker}, \binits{E.}},
\bauthor{\bsnm{Phong}, \binits{D.H.}}:
\batitle{{The Geometry of String Perturbation Theory}}.
\bjtitle{Rev. Mod. Phys.}
\bvolume{60},
\bfpage{917}
(\byear{1988})
\doiurl{10.1103/RevModPhys.60.917}
\end{barticle}
\endbibitem

\bibitem[\protect\citeauthoryear{Polchinski}{1996}]{Polchinski:1996na}
\begin{bchapter}
\bauthor{\bsnm{Polchinski}, \binits{J.}}:
\bctitle{{Tasi lectures on D-branes}}.
In: \bbtitle{Theoretical Advanced Study Institute in Elementary Particle Physics (TASI 96): Fields, Strings, and Duality},
pp. \bfpage{293}--\blpage{356}
(\byear{1996})
\end{bchapter}
\endbibitem

\bibitem[\protect\citeauthoryear{Dijkgraaf}{1997}]{Dijkgraaf:1997ip}
\begin{bchapter}
\bauthor{\bsnm{Dijkgraaf}, \binits{R.}}:
\bctitle{{Les Houches lectures on fields, strings and duality}}.
In: \bbtitle{NATO Advanced Study Institute: Les Houches Summer School on Theoretical Physics, Session 64: Quantum Symmetries},
pp. \bfpage{3}--\blpage{147}
(\byear{1997})
\end{bchapter}
\endbibitem

\bibitem[\protect\citeauthoryear{Polchinski}{1998a}]{Polchinski:1998rq}
\begin{bbook}
\bauthor{\bsnm{Polchinski}, \binits{J.}}:
\bbtitle{String Theory. Vol. 1: An Introduction to the Bosonic String}.
\bsertitle{Cambridge Monographs on Mathematical Physics},
(\byear{1998}).
\doiurl{10.1017/CBO9780511816079}
\end{bbook}
\endbibitem

\bibitem[\protect\citeauthoryear{Polchinski}{1998b}]{Polchinski:1998rr}
\begin{bbook}
\bauthor{\bsnm{Polchinski}, \binits{J.}}:
\bbtitle{String Theory. Vol. 2: Superstring Theory and Beyond}.
\bsertitle{Cambridge Monographs on Mathematical Physics},
(\byear{1998}).
\doiurl{10.1017/CBO9780511618123}
\end{bbook}
\endbibitem

\bibitem[\protect\citeauthoryear{Sen}{1998}]{Sen:1998kr}
\begin{bchapter}
\bauthor{\bsnm{Sen}, \binits{A.}}:
\bctitle{{An Introduction to nonperturbative string theory}}.
In: \bbtitle{A Newton Institute Euroconference on Duality and Supersymmetric Theories},
pp. \bfpage{297}--\blpage{413}
(\byear{1998})
\end{bchapter}
\endbibitem

\bibitem[\protect\citeauthoryear{Angelantonj and Sagnotti}{2002}]{Angelantonj:2002ct}
\begin{barticle}
\bauthor{\bsnm{Angelantonj}, \binits{C.}},
\bauthor{\bsnm{Sagnotti}, \binits{A.}}:
\batitle{{Open strings}}.
\bjtitle{Phys. Rept.}
\bvolume{371},
\bfpage{1}--\blpage{150}
(\byear{2002})
\doiurl{10.1016/S0370-1573(02)00273-9}
{\href{https://arxiv.org/abs/hep-th/0204089}{{arXiv:hep-th/0204089}}}.
\bcomment{[Erratum: Phys.Rept. 376, 407 (2003)]}
\end{barticle}
\endbibitem

\bibitem[\protect\citeauthoryear{Witten}{2012}]{Witten:2012bh}
\begin{botherref}
\oauthor{\bsnm{Witten}, \binits{E.}}:
{Superstring Perturbation Theory Revisited}
(2012)
{\href{https://arxiv.org/abs/1209.5461}{{arXiv:1209.5461}}}
{[hep-th]}
\end{botherref}
\endbibitem

\bibitem[\protect\citeauthoryear{Mourad and Sagnotti}{2017}]{Mourad:2017rrl}
\begin{botherref}
\oauthor{\bsnm{Mourad}, \binits{J.}},
\oauthor{\bsnm{Sagnotti}, \binits{A.}}:
{An Update on Brane Supersymmetry Breaking}
(2017)
{\href{https://arxiv.org/abs/1711.11494}{{arXiv:1711.11494}}}
{[hep-th]}
\end{botherref}
\endbibitem

\bibitem[\protect\citeauthoryear{Basile}{2020}]{Basile:2020xwi}
\begin{botherref}
\oauthor{\bsnm{Basile}, \binits{I.}}:
{On String Vacua without Supersymmetry: brane dynamics, bubbles and holography}.
PhD thesis,
Pisa, Scuola Normale Superiore
(2020)
\end{botherref}
\endbibitem

\bibitem[\protect\citeauthoryear{Basile}{2021}]{Basile:2021vxh}
\begin{barticle}
\bauthor{\bsnm{Basile}, \binits{I.}}:
\batitle{{Supersymmetry breaking and stability in string vacua: Brane dynamics, bubbles and the swampland}}.
\bjtitle{Riv. Nuovo Cim.}
\bvolume{44}(\bissue{10}),
\bfpage{499}--\blpage{596}
(\byear{2021})
\doiurl{10.1007/s40766-021-00024-9}
{\href{https://arxiv.org/abs/2107.02814}{{arXiv:2107.02814}}}
{[hep-th]}
\end{barticle}
\endbibitem

\bibitem[\protect\citeauthoryear{Angelantonj and Florakis}{2024}]{Angelantonj:2024tns}
\begin{bbook}
\bauthor{\bsnm{Angelantonj}, \binits{C.}},
\bauthor{\bsnm{Florakis}, \binits{I.}}:
\bbtitle{{A Lightning Introduction to String Theory}},
(\byear{2024}).
\doiurl{10.1007/978-981-19-3079-9_53-1}
\end{bbook}
\endbibitem

\bibitem[\protect\citeauthoryear{Abel et~al.}{2019}]{Abel:2018zyt}
\begin{barticle}
\bauthor{\bsnm{Abel}, \binits{S.}},
\bauthor{\bsnm{Dudas}, \binits{E.}},
\bauthor{\bsnm{Lewis}, \binits{D.}},
\bauthor{\bsnm{Partouche}, \binits{H.}}:
\batitle{{Stability and vacuum energy in open string models with broken supersymmetry}}.
\bjtitle{JHEP}
\bvolume{10},
\bfpage{226}
(\byear{2019})
\doiurl{10.1007/JHEP10(2019)226}
{\href{https://arxiv.org/abs/1812.09714}{{arXiv:1812.09714}}}
{[hep-th]}
\end{barticle}
\endbibitem

\bibitem[\protect\citeauthoryear{Itoyama and Nakajima}{2020}]{Itoyama:2020ifw}
\begin{barticle}
\bauthor{\bsnm{Itoyama}, \binits{H.}},
\bauthor{\bsnm{Nakajima}, \binits{S.}}:
\batitle{{Stability, enhanced gauge symmetry and suppressed cosmological constant in 9D heterotic interpolating models}}.
\bjtitle{Nucl. Phys. B}
\bvolume{958},
\bfpage{115111}
(\byear{2020})
\doiurl{10.1016/j.nuclphysb.2020.115111}
{\href{https://arxiv.org/abs/2003.11217}{{arXiv:2003.11217}}}
{[hep-th]}
\end{barticle}
\endbibitem

\bibitem[\protect\citeauthoryear{Angelantonj et~al.}{2020}]{Angelantonj:2020pyr}
\begin{barticle}
\bauthor{\bsnm{Angelantonj}, \binits{C.}},
\bauthor{\bsnm{Bonnefoy}, \binits{Q.}},
\bauthor{\bsnm{Condeescu}, \binits{C.}},
\bauthor{\bsnm{Dudas}, \binits{E.}}:
\batitle{{String Defects, Supersymmetry and the Swampland}}.
\bjtitle{JHEP}
\bvolume{11},
\bfpage{125}
(\byear{2020})
\doiurl{10.1007/JHEP11(2020)125}
{\href{https://arxiv.org/abs/2007.12722}{{arXiv:2007.12722}}}
{[hep-th]}
\end{barticle}
\endbibitem

\bibitem[\protect\citeauthoryear{Faraggi et~al.}{2021}]{Faraggi:2020hpy}
\begin{barticle}
\bauthor{\bsnm{Faraggi}, \binits{A.E.}},
\bauthor{\bsnm{Matyas}, \binits{V.G.}},
\bauthor{\bsnm{Percival}, \binits{B.}}:
\batitle{{Type $\mathbf{\bar{0}}$ heterotic string orbifolds}}.
\bjtitle{Phys. Lett. B}
\bvolume{814},
\bfpage{136080}
(\byear{2021})
\doiurl{10.1016/j.physletb.2021.136080}
{\href{https://arxiv.org/abs/2011.12630}{{arXiv:2011.12630}}}
{[hep-th]}
\end{barticle}
\endbibitem

\bibitem[\protect\citeauthoryear{Itoyama and Nakajima}{2021}]{Itoyama:2021fwc}
\begin{barticle}
\bauthor{\bsnm{Itoyama}, \binits{H.}},
\bauthor{\bsnm{Nakajima}, \binits{S.}}:
\batitle{{Marginal deformations of heterotic interpolating models and exponential suppression of the cosmological constant}}.
\bjtitle{Phys. Lett. B}
\bvolume{816},
\bfpage{136195}
(\byear{2021})
\doiurl{10.1016/j.physletb.2021.136195}
{\href{https://arxiv.org/abs/2101.10619}{{arXiv:2101.10619}}}
{[hep-th]}
\end{barticle}
\endbibitem

\bibitem[\protect\citeauthoryear{Itoyama et~al.}{2022}]{Itoyama:2021itj}
\begin{barticle}
\bauthor{\bsnm{Itoyama}, \binits{H.}},
\bauthor{\bsnm{Koga}, \binits{Y.}},
\bauthor{\bsnm{Nakajima}, \binits{S.}}:
\batitle{{Target space duality of non-supersymmetric string theory}}.
\bjtitle{Nucl. Phys. B}
\bvolume{975},
\bfpage{115667}
(\byear{2022})
\doiurl{10.1016/j.nuclphysb.2022.115667}
{\href{https://arxiv.org/abs/2110.09762}{{arXiv:2110.09762}}}
{[hep-th]}
\end{barticle}
\endbibitem

\bibitem[\protect\citeauthoryear{Avalos et~al.}{2023}]{Avalos:2023mti}
\begin{barticle}
\bauthor{\bsnm{Avalos}, \binits{A.R.D.}},
\bauthor{\bsnm{Faraggi}, \binits{A.E.}},
\bauthor{\bsnm{Matyas}, \binits{V.G.}},
\bauthor{\bsnm{Percival}, \binits{B.}}:
\batitle{{Fayet{\textendash}Iliopoulos D-term in non-supersymmetric heterotic string orbifolds}}.
\bjtitle{Eur. Phys. J. C}
\bvolume{83}(\bissue{10}),
\bfpage{926}
(\byear{2023})
\doiurl{10.1140/epjc/s10052-023-12059-9}
{\href{https://arxiv.org/abs/2302.10075}{{arXiv:2302.10075}}}
{[hep-th]}
\end{barticle}
\endbibitem

\bibitem[\protect\citeauthoryear{Nakajima}{2023}]{Nakajima:2023zsh}
\begin{botherref}
\oauthor{\bsnm{Nakajima}, \binits{S.}}:
{New non-supersymmetric heterotic string theory with reduced rank and exponential suppression of the cosmological constant}
(2023)
{\href{https://arxiv.org/abs/2303.04489}{{arXiv:2303.04489}}}
{[hep-th]}
\end{botherref}
\endbibitem

\bibitem[\protect\citeauthoryear{Avalos et~al.}{2023}]{Avalos:2023ldc}
\begin{barticle}
\bauthor{\bsnm{Avalos}, \binits{A.R.D.}},
\bauthor{\bsnm{Faraggi}, \binits{A.E.}},
\bauthor{\bsnm{Matyas}, \binits{V.G.}},
\bauthor{\bsnm{Percival}, \binits{B.}}:
\batitle{{D-term uplifts in nonsupersymmetric heterotic string models}}.
\bjtitle{Phys. Rev. D}
\bvolume{108}(\bissue{8}),
\bfpage{086007}
(\byear{2023})
\doiurl{10.1103/PhysRevD.108.086007}
{\href{https://arxiv.org/abs/2306.16878}{{arXiv:2306.16878}}}
{[hep-th]}
\end{barticle}
\endbibitem

\bibitem[\protect\citeauthoryear{Saxena}{2024}]{Saxena:2024eil}
\begin{barticle}
\bauthor{\bsnm{Saxena}, \binits{V.}}:
\batitle{{A T-duality of non-supersymmetric heterotic strings and an implication for Topological Modular Forms}}.
\bjtitle{JHEP}
\bvolume{09},
\bfpage{056}
(\byear{2024})
\doiurl{10.1007/JHEP09(2024)056}
{\href{https://arxiv.org/abs/2405.19409}{{arXiv:2405.19409}}}
{[hep-th]}
\end{barticle}
\endbibitem

\bibitem[\protect\citeauthoryear{Basaad et~al.}{2025}]{Basaad:2024lno}
\begin{barticle}
\bauthor{\bsnm{Basaad}, \binits{E.}},
\bauthor{\bsnm{Detraux}, \binits{L.A.}},
\bauthor{\bsnm{Avalos}, \binits{A.R.D.}},
\bauthor{\bsnm{Faraggi}, \binits{A.E.}},
\bauthor{\bsnm{Percival}, \binits{B.}}:
\batitle{{Vacuum energy in non-supersymmetric quasi-realistic heterotic-string vacua with fixed moduli}}.
\bjtitle{Eur. Phys. J. C}
\bvolume{85}(\bissue{2}),
\bfpage{209}
(\byear{2025})
\doiurl{10.1140/epjc/s10052-024-13733-2}
{\href{https://arxiv.org/abs/2408.03928}{{arXiv:2408.03928}}}
{[hep-th]}
\end{barticle}
\endbibitem

\bibitem[\protect\citeauthoryear{Abel et~al.}{2025}]{Abel:2024vov}
\begin{barticle}
\bauthor{\bsnm{Abel}, \binits{S.}},
\bauthor{\bsnm{Basile}, \binits{I.}},
\bauthor{\bsnm{Matyas}, \binits{V.G.}}:
\batitle{{Banks-Zaks stabilisation of non-SUSY strings}}.
\bjtitle{JHEP}
\bvolume{04},
\bfpage{107}
(\byear{2025})
\doiurl{10.1007/JHEP04(2025)107}
{\href{https://arxiv.org/abs/2412.01914}{{arXiv:2412.01914}}}
{[hep-th]}
\end{barticle}
\endbibitem

\bibitem[\protect\citeauthoryear{Larotonda and Lin}{2025}]{Larotonda:2024thv}
\begin{barticle}
\bauthor{\bsnm{Larotonda}, \binits{V.}},
\bauthor{\bsnm{Lin}, \binits{L.}}:
\batitle{{Anomaly inflow and gauge group topology in the 10d Sugimoto string theory}}.
\bjtitle{JHEP}
\bvolume{06},
\bfpage{136}
(\byear{2025})
\doiurl{10.1007/JHEP06(2025)136}
{\href{https://arxiv.org/abs/2412.17894}{{arXiv:2412.17894}}}
{[hep-th]}
\end{barticle}
\endbibitem

\bibitem[\protect\citeauthoryear{Montero and Zapata}{2025}]{Montero:2025ayi}
\begin{barticle}
\bauthor{\bsnm{Montero}, \binits{M.}},
\bauthor{\bsnm{Zapata}, \binits{L.}}:
\batitle{{M-theory boundaries beyond supersymmetry}}.
\bjtitle{JHEP}
\bvolume{07},
\bfpage{090}
(\byear{2025})
\doiurl{10.1007/JHEP07(2025)090}
{\href{https://arxiv.org/abs/2504.06985}{{arXiv:2504.06985}}}
{[hep-th]}
\end{barticle}
\endbibitem

\bibitem[\protect\citeauthoryear{Basile and Larotonda}{2025}]{Basile:2025mnj}
\begin{barticle}
\bauthor{\bsnm{Basile}, \binits{I.}},
\bauthor{\bsnm{Larotonda}, \binits{V.}}:
\batitle{{Non-supersymmetric branes and discrete topological terms}}.
\bjtitle{JHEP}
\bvolume{10},
\bfpage{030}
(\byear{2025})
\doiurl{10.1007/JHEP10(2025)030}
{\href{https://arxiv.org/abs/2507.11610}{{arXiv:2507.11610}}}
{[hep-th]}
\end{barticle}
\endbibitem

\bibitem[\protect\citeauthoryear{Schweigert et~al.}{2000}]{Schweigert:2000ix}
\begin{bchapter}
\bauthor{\bsnm{Schweigert}, \binits{C.}},
\bauthor{\bsnm{Fuchs}, \binits{J.}},
\bauthor{\bsnm{Walcher}, \binits{J.}}:
\bctitle{{Conformal field theory, boundary conditions and applications to string theory}}.
In: \bbtitle{Eotvos Summer School in Physics: Nonperturbative QFT Methods and Their Applications},
pp. \bfpage{37}--\blpage{93}
(\byear{2000}).
\doiurl{10.1142/9789812799968_0002}
\end{bchapter}
\endbibitem

\bibitem[\protect\citeauthoryear{Schweigert and Fuchs}{2003}]{Schweigert:2001cu}
\begin{barticle}
\bauthor{\bsnm{Schweigert}, \binits{C.}},
\bauthor{\bsnm{Fuchs}, \binits{J.}}:
\batitle{{The World sheet revisited}}.
\bjtitle{Fields Inst. Commun.}
\bvolume{39},
\bfpage{241}--\blpage{249}
(\byear{2003})
{\href{https://arxiv.org/abs/hep-th/0105266}{{arXiv:hep-th/0105266}}}
\end{barticle}
\endbibitem

\bibitem[\protect\citeauthoryear{Polyakov}{1981a}]{Polyakov:1981rd}
\begin{barticle}
\bauthor{\bsnm{Polyakov}, \binits{A.M.}}:
\batitle{{Quantum Geometry of Bosonic Strings}}.
\bjtitle{Phys. Lett. B}
\bvolume{103},
\bfpage{207}--\blpage{210}
(\byear{1981})
\doiurl{10.1016/0370-2693(81)90743-7}
\end{barticle}
\endbibitem

\bibitem[\protect\citeauthoryear{Polyakov}{1981b}]{Polyakov:1981re}
\begin{barticle}
\bauthor{\bsnm{Polyakov}, \binits{A.M.}}:
\batitle{{Quantum Geometry of Fermionic Strings}}.
\bjtitle{Phys. Lett. B}
\bvolume{103},
\bfpage{211}--\blpage{213}
(\byear{1981})
\doiurl{10.1016/0370-2693(81)90744-9}
\end{barticle}
\endbibitem

\bibitem[\protect\citeauthoryear{Witten}{2019a}]{Witten:2012bg}
\begin{barticle}
\bauthor{\bsnm{Witten}, \binits{E.}}:
\batitle{{Notes On Supermanifolds and Integration}}.
\bjtitle{Pure Appl. Math. Quart.}
\bvolume{15}(\bissue{1}),
\bfpage{3}--\blpage{56}
(\byear{2019})
\doiurl{10.4310/PAMQ.2019.v15.n1.a1}
{\href{https://arxiv.org/abs/1209.2199}{{arXiv:1209.2199}}}
{[hep-th]}
\end{barticle}
\endbibitem

\bibitem[\protect\citeauthoryear{Witten}{2019b}]{Witten:2012ga}
\begin{barticle}
\bauthor{\bsnm{Witten}, \binits{E.}}:
\batitle{{Notes On Super Riemann Surfaces And Their Moduli}}.
\bjtitle{Pure Appl. Math. Quart.}
\bvolume{15}(\bissue{1}),
\bfpage{57}--\blpage{211}
(\byear{2019})
\doiurl{10.4310/PAMQ.2019.v15.n1.a2}
{\href{https://arxiv.org/abs/1209.2459}{{arXiv:1209.2459}}}
{[hep-th]}
\end{barticle}
\endbibitem

\bibitem[\protect\citeauthoryear{Henningson and Skenderis}{1998}]{Henningson:1998gx}
\begin{barticle}
\bauthor{\bsnm{Henningson}, \binits{M.}},
\bauthor{\bsnm{Skenderis}, \binits{K.}}:
\batitle{{The Holographic Weyl anomaly}}.
\bjtitle{JHEP}
\bvolume{07},
\bfpage{023}
(\byear{1998})
\doiurl{10.1088/1126-6708/1998/07/023}
{\href{https://arxiv.org/abs/hep-th/9806087}{{arXiv:hep-th/9806087}}}
\end{barticle}
\endbibitem

\bibitem[\protect\citeauthoryear{Eberhardt and Pal}{2024}]{Eberhardt:2023lwd}
\begin{barticle}
\bauthor{\bsnm{Eberhardt}, \binits{L.}},
\bauthor{\bsnm{Pal}, \binits{S.}}:
\batitle{{Holographic Weyl anomaly in string theory}}.
\bjtitle{SciPost Phys.}
\bvolume{16}(\bissue{1}),
\bfpage{027}
(\byear{2024})
\doiurl{10.21468/SciPostPhys.16.1.027}
{\href{https://arxiv.org/abs/2307.03000}{{arXiv:2307.03000}}}
{[hep-th]}
\end{barticle}
\endbibitem

\bibitem[\protect\citeauthoryear{Liu and Polchinski}{1988}]{Liu:1987nz}
\begin{barticle}
\bauthor{\bsnm{Liu}, \binits{J.}},
\bauthor{\bsnm{Polchinski}, \binits{J.}}:
\batitle{{Renormalization of the Mobius Volume}}.
\bjtitle{Phys. Lett. B}
\bvolume{203},
\bfpage{39}--\blpage{43}
(\byear{1988})
\doiurl{10.1016/0370-2693(88)91566-3}
\end{barticle}
\endbibitem

\bibitem[\protect\citeauthoryear{Troost}{2011}]{Troost:2011ud}
\begin{barticle}
\bauthor{\bsnm{Troost}, \binits{J.}}:
\batitle{{The $AdS_3$ central charge in string theory}}.
\bjtitle{Phys. Lett. B}
\bvolume{705},
\bfpage{260}--\blpage{263}
(\byear{2011})
\doiurl{10.1016/j.physletb.2011.10.007}
{\href{https://arxiv.org/abs/1109.1923}{{arXiv:1109.1923}}}
{[hep-th]}
\end{barticle}
\endbibitem

\bibitem[\protect\citeauthoryear{Distler and Kawai}{1989}]{Distler:1988jt}
\begin{barticle}
\bauthor{\bsnm{Distler}, \binits{J.}},
\bauthor{\bsnm{Kawai}, \binits{H.}}:
\batitle{{Conformal Field Theory and 2D Quantum Gravity}}.
\bjtitle{Nucl. Phys. B}
\bvolume{321},
\bfpage{509}--\blpage{527}
(\byear{1989})
\doiurl{10.1016/0550-3213(89)90354-4}
\end{barticle}
\endbibitem

\bibitem[\protect\citeauthoryear{Maltz}{2013}]{Maltz:2012zs}
\begin{barticle}
\bauthor{\bsnm{Maltz}, \binits{J.}}:
\batitle{{Gauge Invariant Computable Quantities In Timelike Liouville Theory}}.
\bjtitle{JHEP}
\bvolume{01},
\bfpage{151}
(\byear{2013})
\doiurl{10.1007/JHEP01(2013)151}
{\href{https://arxiv.org/abs/1210.2398}{{arXiv:1210.2398}}}
{[hep-th]}
\end{barticle}
\endbibitem

\bibitem[\protect\citeauthoryear{Anninos et~al.}{2021}]{Anninos:2021ene}
\begin{barticle}
\bauthor{\bsnm{Anninos}, \binits{D.}},
\bauthor{\bsnm{Bautista}, \binits{T.}},
\bauthor{\bsnm{M\"uhlmann}, \binits{B.}}:
\batitle{{The two-sphere partition function in two-dimensional quantum gravity}}.
\bjtitle{JHEP}
\bvolume{09},
\bfpage{116}
(\byear{2021})
\doiurl{10.1007/JHEP09(2021)116}
{\href{https://arxiv.org/abs/2106.01665}{{arXiv:2106.01665}}}
{[hep-th]}
\end{barticle}
\endbibitem

\bibitem[\protect\citeauthoryear{Mahajan et~al.}{2022}]{Mahajan:2021nsd}
\begin{barticle}
\bauthor{\bsnm{Mahajan}, \binits{R.}},
\bauthor{\bsnm{Stanford}, \binits{D.}},
\bauthor{\bsnm{Yan}, \binits{C.}}:
\batitle{{Sphere and disk partition functions in Liouville and in matrix integrals}}.
\bjtitle{JHEP}
\bvolume{07},
\bfpage{132}
(\byear{2022})
\doiurl{10.1007/JHEP07(2022)132}
{\href{https://arxiv.org/abs/2107.01172}{{arXiv:2107.01172}}}
{[hep-th]}
\end{barticle}
\endbibitem

\bibitem[\protect\citeauthoryear{Erbin et~al.}{2019}]{Erbin:2019uiz}
\begin{barticle}
\bauthor{\bsnm{Erbin}, \binits{H.}},
\bauthor{\bsnm{Maldacena}, \binits{J.}},
\bauthor{\bsnm{Skliros}, \binits{D.}}:
\batitle{{Two-Point String Amplitudes}}.
\bjtitle{JHEP}
\bvolume{07},
\bfpage{139}
(\byear{2019})
\doiurl{10.1007/JHEP07(2019)139}
{\href{https://arxiv.org/abs/1906.06051}{{arXiv:1906.06051}}}
{[hep-th]}
\end{barticle}
\endbibitem

\bibitem[\protect\citeauthoryear{Gibbons and Hawking}{1977}]{Gibbons:1976ue}
\begin{barticle}
\bauthor{\bsnm{Gibbons}, \binits{G.W.}},
\bauthor{\bsnm{Hawking}, \binits{S.W.}}:
\batitle{{Action Integrals and Partition Functions in Quantum Gravity}}.
\bjtitle{Phys. Rev. D}
\bvolume{15},
\bfpage{2752}--\blpage{2756}
(\byear{1977})
\doiurl{10.1103/PhysRevD.15.2752}
\end{barticle}
\endbibitem

\bibitem[\protect\citeauthoryear{Shapiro}{1975}]{Shapiro:1975cz}
\begin{barticle}
\bauthor{\bsnm{Shapiro}, \binits{J.A.}}:
\batitle{{On the Renormalization of Dual Models}}.
\bjtitle{Phys. Rev. D}
\bvolume{11},
\bfpage{2937}
(\byear{1975})
\doiurl{10.1103/PhysRevD.11.2937}
\end{barticle}
\endbibitem

\bibitem[\protect\citeauthoryear{Ademollo et~al.}{1975}]{Ademollo:1975pf}
\begin{barticle}
\bauthor{\bsnm{Ademollo}, \binits{M.}},
\bauthor{\bsnm{D'Adda}, \binits{A.}},
\bauthor{\bsnm{D'Auria}, \binits{R.}},
\bauthor{\bsnm{Gliozzi}, \binits{F.}},
\bauthor{\bsnm{Napolitano}, \binits{E.}},
\bauthor{\bsnm{Sciuto}, \binits{S.}},
\bauthor{\bsnm{Di~Vecchia}, \binits{P.}}:
\batitle{{Soft Dilations and Scale Renormalization in Dual Theories}}.
\bjtitle{Nucl. Phys. B}
\bvolume{94},
\bfpage{221}--\blpage{259}
(\byear{1975})
\doiurl{10.1016/0550-3213(75)90491-5}
\end{barticle}
\endbibitem

\bibitem[\protect\citeauthoryear{Sagnotti}{1987}]{Sagnotti:1987tw}
\begin{bchapter}
\bauthor{\bsnm{Sagnotti}, \binits{A.}}:
\bctitle{{Open Strings and their Symmetry Groups}}.
In: \bbtitle{NATO Advanced Summer Institute on Nonperturbative Quantum Field Theory (Cargese Summer Institute)}
(\byear{1987})
\end{bchapter}
\endbibitem

\bibitem[\protect\citeauthoryear{Pradisi and Sagnotti}{1989}]{Pradisi:1988xd}
\begin{barticle}
\bauthor{\bsnm{Pradisi}, \binits{G.}},
\bauthor{\bsnm{Sagnotti}, \binits{A.}}:
\batitle{{Open String Orbifolds}}.
\bjtitle{Phys. Lett. B}
\bvolume{216},
\bfpage{59}--\blpage{67}
(\byear{1989})
\doiurl{10.1016/0370-2693(89)91369-5}
\end{barticle}
\endbibitem

\bibitem[\protect\citeauthoryear{Bianchi and Sagnotti}{1990}]{Bianchi:1990yu}
\begin{barticle}
\bauthor{\bsnm{Bianchi}, \binits{M.}},
\bauthor{\bsnm{Sagnotti}, \binits{A.}}:
\batitle{{On the systematics of open string theories}}.
\bjtitle{Phys. Lett. B}
\bvolume{247},
\bfpage{517}--\blpage{524}
(\byear{1990})
\doiurl{10.1016/0370-2693(90)91894-H}
\end{barticle}
\endbibitem

\bibitem[\protect\citeauthoryear{Bianchi et~al.}{1992}]{Bianchi:1991eu}
\begin{barticle}
\bauthor{\bsnm{Bianchi}, \binits{M.}},
\bauthor{\bsnm{Pradisi}, \binits{G.}},
\bauthor{\bsnm{Sagnotti}, \binits{A.}}:
\batitle{{Toroidal compactification and symmetry breaking in open string theories}}.
\bjtitle{Nucl. Phys. B}
\bvolume{376},
\bfpage{365}--\blpage{386}
(\byear{1992})
\doiurl{10.1016/0550-3213(92)90129-Y}
\end{barticle}
\endbibitem

\bibitem[\protect\citeauthoryear{Polchinski}{1995}]{Polchinski:1995mt}
\begin{barticle}
\bauthor{\bsnm{Polchinski}, \binits{J.}}:
\batitle{{Dirichlet Branes and Ramond-Ramond charges}}.
\bjtitle{Phys. Rev. Lett.}
\bvolume{75},
\bfpage{4724}--\blpage{4727}
(\byear{1995})
\doiurl{10.1103/PhysRevLett.75.4724}
{\href{https://arxiv.org/abs/hep-th/9510017}{{arXiv:hep-th/9510017}}}
\end{barticle}
\endbibitem

\bibitem[\protect\citeauthoryear{Eberhardt and Pal}{2021}]{Eberhardt:2021ynh}
\begin{barticle}
\bauthor{\bsnm{Eberhardt}, \binits{L.}},
\bauthor{\bsnm{Pal}, \binits{S.}}:
\batitle{{The disk partition function in string theory}}.
\bjtitle{JHEP}
\bvolume{08},
\bfpage{026}
(\byear{2021})
\doiurl{10.1007/JHEP08(2021)026}
{\href{https://arxiv.org/abs/2105.08726}{{arXiv:2105.08726}}}
{[hep-th]}
\end{barticle}
\endbibitem

\bibitem[\protect\citeauthoryear{Gliozzi et~al.}{1976}]{Gliozzi:1976jf}
\begin{barticle}
\bauthor{\bsnm{Gliozzi}, \binits{F.}},
\bauthor{\bsnm{Scherk}, \binits{J.}},
\bauthor{\bsnm{Olive}, \binits{D.I.}}:
\batitle{{Supergravity and the Spinor Dual Model}}.
\bjtitle{Phys. Lett. B}
\bvolume{65},
\bfpage{282}--\blpage{286}
(\byear{1976})
\doiurl{10.1016/0370-2693(76)90183-0}
\end{barticle}
\endbibitem

\bibitem[\protect\citeauthoryear{Gliozzi et~al.}{1977}]{Gliozzi:1976qd}
\begin{barticle}
\bauthor{\bsnm{Gliozzi}, \binits{F.}},
\bauthor{\bsnm{Scherk}, \binits{J.}},
\bauthor{\bsnm{Olive}, \binits{D.I.}}:
\batitle{{Supersymmetry, Supergravity Theories and the Dual Spinor Model}}.
\bjtitle{Nucl. Phys. B}
\bvolume{122},
\bfpage{253}--\blpage{290}
(\byear{1977})
\doiurl{10.1016/0550-3213(77)90206-1}
\end{barticle}
\endbibitem

\bibitem[\protect\citeauthoryear{Seiberg and Witten}{1986}]{Seiberg:1986by}
\begin{barticle}
\bauthor{\bsnm{Seiberg}, \binits{N.}},
\bauthor{\bsnm{Witten}, \binits{E.}}:
\batitle{{Spin Structures in String Theory}}.
\bjtitle{Nucl. Phys. B}
\bvolume{276},
\bfpage{272}
(\byear{1986})
\doiurl{10.1016/0550-3213(86)90297-X}
\end{barticle}
\endbibitem

\bibitem[\protect\citeauthoryear{Kutasov and Seiberg}{1991}]{Kutasov:1990sv}
\begin{barticle}
\bauthor{\bsnm{Kutasov}, \binits{D.}},
\bauthor{\bsnm{Seiberg}, \binits{N.}}:
\batitle{{Number of degrees of freedom, density of states and tachyons in string theory and CFT}}.
\bjtitle{Nucl. Phys. B}
\bvolume{358},
\bfpage{600}--\blpage{618}
(\byear{1991})
\doiurl{10.1016/0550-3213(91)90426-X}
\end{barticle}
\endbibitem

\bibitem[\protect\citeauthoryear{Horava}{1989}]{Horava:1989vt}
\begin{barticle}
\bauthor{\bsnm{Horava}, \binits{P.}}:
\batitle{{Strings on World Sheet Orbifolds}}.
\bjtitle{Nucl. Phys. B}
\bvolume{327},
\bfpage{461}--\blpage{484}
(\byear{1989})
\doiurl{10.1016/0550-3213(89)90279-4}
\end{barticle}
\endbibitem

\bibitem[\protect\citeauthoryear{Bianchi and Sagnotti}{1991}]{Bianchi:1990tb}
\begin{barticle}
\bauthor{\bsnm{Bianchi}, \binits{M.}},
\bauthor{\bsnm{Sagnotti}, \binits{A.}}:
\batitle{{Twist symmetry and open string Wilson lines}}.
\bjtitle{Nucl. Phys. B}
\bvolume{361},
\bfpage{519}--\blpage{538}
(\byear{1991})
\doiurl{10.1016/0550-3213(91)90271-X}
\end{barticle}
\endbibitem

\bibitem[\protect\citeauthoryear{Dudas}{2000}]{Dudas:2000bn}
\begin{barticle}
\bauthor{\bsnm{Dudas}, \binits{E.}}:
\batitle{{Theory and phenomenology of type I strings and M theory}}.
\bjtitle{Class. Quant. Grav.}
\bvolume{17},
\bfpage{41}--\blpage{116}
(\byear{2000})
\doiurl{10.1088/0264-9381/17/22/201}
{\href{https://arxiv.org/abs/hep-ph/0006190}{{arXiv:hep-ph/0006190}}}
\end{barticle}
\endbibitem

\bibitem[\protect\citeauthoryear{Fioravanti et~al.}{1994}]{Fioravanti:1993hf}
\begin{barticle}
\bauthor{\bsnm{Fioravanti}, \binits{D.}},
\bauthor{\bsnm{Pradisi}, \binits{G.}},
\bauthor{\bsnm{Sagnotti}, \binits{A.}}:
\batitle{{Sewing constraints and nonorientable open strings}}.
\bjtitle{Phys. Lett. B}
\bvolume{321},
\bfpage{349}--\blpage{354}
(\byear{1994})
\doiurl{10.1016/0370-2693(94)90255-0}
{\href{https://arxiv.org/abs/hep-th/9311183}{{arXiv:hep-th/9311183}}}
\end{barticle}
\endbibitem

\bibitem[\protect\citeauthoryear{Paton and Chan}{1969}]{Paton:1969je}
\begin{barticle}
\bauthor{\bsnm{Paton}, \binits{J.E.}},
\bauthor{\bsnm{Chan}, \binits{H.-M.}}:
\batitle{{Generalized veneziano model with isospin}}.
\bjtitle{Nucl. Phys. B}
\bvolume{10},
\bfpage{516}--\blpage{520}
(\byear{1969})
\doiurl{10.1016/0550-3213(69)90038-8}
\end{barticle}
\endbibitem

\bibitem[\protect\citeauthoryear{Sonoda}{1988a}]{Sonoda:1988mf}
\begin{barticle}
\bauthor{\bsnm{Sonoda}, \binits{H.}}:
\batitle{{Sewing Conformal Field Theories}}.
\bjtitle{Nucl. Phys. B}
\bvolume{311},
\bfpage{401}--\blpage{416}
(\byear{1988})
\doiurl{10.1016/0550-3213(88)90066-1}
\end{barticle}
\endbibitem

\bibitem[\protect\citeauthoryear{Sonoda}{1988b}]{Sonoda:1988fq}
\begin{barticle}
\bauthor{\bsnm{Sonoda}, \binits{H.}}:
\batitle{{Sewing Conformal Field Theories, 2.}}
\bjtitle{Nucl. Phys. B}
\bvolume{311},
\bfpage{417}--\blpage{432}
(\byear{1988})
\doiurl{10.1016/0550-3213(88)90067-3}
\end{barticle}
\endbibitem

\bibitem[\protect\citeauthoryear{Moore and Seiberg}{1988}]{Moore:1988uz}
\begin{barticle}
\bauthor{\bsnm{Moore}, \binits{G.W.}},
\bauthor{\bsnm{Seiberg}, \binits{N.}}:
\batitle{{Polynomial Equations for Rational Conformal Field Theories}}.
\bjtitle{Phys. Lett. B}
\bvolume{212},
\bfpage{451}--\blpage{460}
(\byear{1988})
\doiurl{10.1016/0370-2693(88)91796-0}
\end{barticle}
\endbibitem

\bibitem[\protect\citeauthoryear{Moore and Seiberg}{1989}]{Moore:1988qv}
\begin{barticle}
\bauthor{\bsnm{Moore}, \binits{G.W.}},
\bauthor{\bsnm{Seiberg}, \binits{N.}}:
\batitle{{Classical and Quantum Conformal Field Theory}}.
\bjtitle{Commun. Math. Phys.}
\bvolume{123},
\bfpage{177}
(\byear{1989})
\doiurl{10.1007/BF01238857}
\end{barticle}
\endbibitem

\bibitem[\protect\citeauthoryear{Schellekens and Warner}{1987}]{Schellekens:1986xh}
\begin{barticle}
\bauthor{\bsnm{Schellekens}, \binits{A.N.}},
\bauthor{\bsnm{Warner}, \binits{N.P.}}:
\batitle{{Anomalies, Characters and Strings}}.
\bjtitle{Nucl. Phys. B}
\bvolume{287},
\bfpage{317}
(\byear{1987})
\doiurl{10.1016/0550-3213(87)90108-8}
\end{barticle}
\endbibitem

\bibitem[\protect\citeauthoryear{Schellekens and Warner}{1986}]{Schellekens:1986yi}
\begin{barticle}
\bauthor{\bsnm{Schellekens}, \binits{A.N.}},
\bauthor{\bsnm{Warner}, \binits{N.P.}}:
\batitle{{Anomalies and Modular Invariance in String Theory}}.
\bjtitle{Phys. Lett. B}
\bvolume{177},
\bfpage{317}--\blpage{323}
(\byear{1986})
\doiurl{10.1016/0370-2693(86)90760-4}
\end{barticle}
\endbibitem

\bibitem[\protect\citeauthoryear{Green and Schwarz}{1984}]{Green:1984sg}
\begin{barticle}
\bauthor{\bsnm{Green}, \binits{M.B.}},
\bauthor{\bsnm{Schwarz}, \binits{J.H.}}:
\batitle{{Anomaly Cancellation in Supersymmetric D=10 Gauge Theory and Superstring Theory}}.
\bjtitle{Phys. Lett. B}
\bvolume{149},
\bfpage{117}--\blpage{122}
(\byear{1984})
\doiurl{10.1016/0370-2693(84)91565-X}
\end{barticle}
\endbibitem

\bibitem[\protect\citeauthoryear{Green et~al.}{1985}]{Green:1984bx}
\begin{barticle}
\bauthor{\bsnm{Green}, \binits{M.B.}},
\bauthor{\bsnm{Schwarz}, \binits{J.H.}},
\bauthor{\bsnm{West}, \binits{P.C.}}:
\batitle{{Anomaly Free Chiral Theories in Six-Dimensions}}.
\bjtitle{Nucl. Phys. B}
\bvolume{254},
\bfpage{327}--\blpage{348}
(\byear{1985})
\doiurl{10.1016/0550-3213(85)90222-6}
\end{barticle}
\endbibitem

\bibitem[\protect\citeauthoryear{Sagnotti}{1992}]{Sagnotti:1992qw}
\begin{barticle}
\bauthor{\bsnm{Sagnotti}, \binits{A.}}:
\batitle{{A Note on the Green-Schwarz mechanism in open string theories}}.
\bjtitle{Phys. Lett. B}
\bvolume{294},
\bfpage{196}--\blpage{203}
(\byear{1992})
\doiurl{10.1016/0370-2693(92)90682-T}
{\href{https://arxiv.org/abs/hep-th/9210127}{{arXiv:hep-th/9210127}}}
\end{barticle}
\endbibitem

\bibitem[\protect\citeauthoryear{Garc{\'\i}a-Etxebarria and Montero}{2019}]{Garcia-Etxebarria:2018ajm}
\begin{barticle}
\bauthor{\bsnm{Garc{\'\i}a-Etxebarria}, \binits{I.}},
\bauthor{\bsnm{Montero}, \binits{M.}}:
\batitle{{Dai-Freed anomalies in particle physics}}.
\bjtitle{JHEP}
\bvolume{08},
\bfpage{003}
(\byear{2019})
\doiurl{10.1007/JHEP08(2019)003}
{\href{https://arxiv.org/abs/1808.00009}{{arXiv:1808.00009}}}
{[hep-th]}
\end{barticle}
\endbibitem

\bibitem[\protect\citeauthoryear{Tachikawa and Yamashita}{2023}]{Tachikawa:2021mby}
\begin{barticle}
\bauthor{\bsnm{Tachikawa}, \binits{Y.}},
\bauthor{\bsnm{Yamashita}, \binits{M.}}:
\batitle{{Topological Modular Forms and the Absence of All Heterotic Global Anomalies}}.
\bjtitle{Commun. Math. Phys.}
\bvolume{402}(\bissue{2}),
\bfpage{1585}--\blpage{1620}
(\byear{2023})
\doiurl{10.1007/s00220-023-04761-2}
{\href{https://arxiv.org/abs/2108.13542}{{arXiv:2108.13542}}}
{[hep-th]}.
\bcomment{[Erratum: Commun.Math.Phys. 402, 2131 (2023)]}
\end{barticle}
\endbibitem

\bibitem[\protect\citeauthoryear{Tachikawa}{2022}]{Tachikawa:2021mvw}
\begin{barticle}
\bauthor{\bsnm{Tachikawa}, \binits{Y.}}:
\batitle{{Topological modular forms and the absence of a heterotic global anomaly}}.
\bjtitle{PTEP}
\bvolume{2022}(\bissue{4}),
\bfpage{04}--\blpage{107}
(\byear{2022})
\doiurl{10.1093/ptep/ptab060}
{\href{https://arxiv.org/abs/2103.12211}{{arXiv:2103.12211}}}
{[hep-th]}
\end{barticle}
\endbibitem

\bibitem[\protect\citeauthoryear{Debray et~al.}{2022}]{Debray:2021vob}
\begin{barticle}
\bauthor{\bsnm{Debray}, \binits{A.}},
\bauthor{\bsnm{Dierigl}, \binits{M.}},
\bauthor{\bsnm{Heckman}, \binits{J.J.}},
\bauthor{\bsnm{Montero}, \binits{M.}}:
\batitle{{The anomaly that was not meant IIB}}.
\bjtitle{Fortsch. Phys.}
\bvolume{70}(\bissue{1}),
\bfpage{2100168}
(\byear{2022})
\doiurl{10.1002/prop.202100168}
{\href{https://arxiv.org/abs/2107.14227}{{arXiv:2107.14227}}}
{[hep-th]}
\end{barticle}
\endbibitem

\bibitem[\protect\citeauthoryear{Debray et~al.}{2023}]{Debray:2023yrs}
\begin{botherref}
\oauthor{\bsnm{Debray}, \binits{A.}},
\oauthor{\bsnm{Dierigl}, \binits{M.}},
\oauthor{\bsnm{Heckman}, \binits{J.J.}},
\oauthor{\bsnm{Montero}, \binits{M.}}:
{The Chronicles of IIBordia: Dualities, Bordisms, and the Swampland}
(2023)
{\href{https://arxiv.org/abs/2302.00007}{{arXiv:2302.00007}}}
{[hep-th]}
\end{botherref}
\endbibitem

\bibitem[\protect\citeauthoryear{Basile et~al.}{2024}]{Basile:2023knk}
\begin{barticle}
\bauthor{\bsnm{Basile}, \binits{I.}},
\bauthor{\bsnm{Debray}, \binits{A.}},
\bauthor{\bsnm{Delgado}, \binits{M.}},
\bauthor{\bsnm{Montero}, \binits{M.}}:
\batitle{{Global anomalies {\&} bordism of non-supersymmetric strings}}.
\bjtitle{JHEP}
\bvolume{02},
\bfpage{092}
(\byear{2024})
\doiurl{10.1007/JHEP02(2024)092}
{\href{https://arxiv.org/abs/2310.06895}{{arXiv:2310.06895}}}
{[hep-th]}
\end{barticle}
\endbibitem

\bibitem[\protect\citeauthoryear{Basile and Leone}{2024}]{Basile:2023zng}
\begin{barticle}
\bauthor{\bsnm{Basile}, \binits{I.}},
\bauthor{\bsnm{Leone}, \binits{G.}}:
\batitle{{Anomaly constraints for heterotic strings and supergravity in six dimensions}}.
\bjtitle{JHEP}
\bvolume{04},
\bfpage{067}
(\byear{2024})
\doiurl{10.1007/JHEP04(2024)067}
{\href{https://arxiv.org/abs/2310.20480}{{arXiv:2310.20480}}}
{[hep-th]}
\end{barticle}
\endbibitem

\bibitem[\protect\citeauthoryear{Pradisi et~al.}{1996}]{Pradisi:1996yd}
\begin{barticle}
\bauthor{\bsnm{Pradisi}, \binits{G.}},
\bauthor{\bsnm{Sagnotti}, \binits{A.}},
\bauthor{\bsnm{Stanev}, \binits{Y.S.}}:
\batitle{{Completeness conditions for boundary operators in 2-D conformal field theory}}.
\bjtitle{Phys. Lett. B}
\bvolume{381},
\bfpage{97}--\blpage{104}
(\byear{1996})
\doiurl{10.1016/0370-2693(96)00578-3}
{\href{https://arxiv.org/abs/hep-th/9603097}{{arXiv:hep-th/9603097}}}
\end{barticle}
\endbibitem

\bibitem[\protect\citeauthoryear{Stanev}{2001}]{Stanev:2001na}
\begin{bchapter}
\bauthor{\bsnm{Stanev}, \binits{Y.S.}}:
\bctitle{{Two-dimensional conformal field theory on open and unoriented surfaces}}.
In: \bbtitle{4th SIGRAV Graduate School on Contemporary Relativity and Gravitational Physics and 2001 School on Algebraic Geometry and Physics: Geometry and Physics of Branes (SAGP 2001)},
pp. \bfpage{39}--\blpage{85}
(\byear{2001})
\end{bchapter}
\endbibitem

\bibitem[\protect\citeauthoryear{Gava et~al.}{1986}]{Gava:1986ei}
\begin{barticle}
\bauthor{\bsnm{Gava}, \binits{E.}},
\bauthor{\bsnm{Jengo}, \binits{R.}},
\bauthor{\bsnm{Jayaraman}, \binits{T.}},
\bauthor{\bsnm{Ramachandran}, \binits{R.}}:
\batitle{{Multiloop Divergences in the Closed Bosonic String Theory}}.
\bjtitle{Phys. Lett. B}
\bvolume{168},
\bfpage{207}--\blpage{211}
(\byear{1986})
\doiurl{10.1016/0370-2693(86)90964-0}
\end{barticle}
\endbibitem

\bibitem[\protect\citeauthoryear{Bianchi and Morales}{2000}]{Bianchi:2000de}
\begin{barticle}
\bauthor{\bsnm{Bianchi}, \binits{M.}},
\bauthor{\bsnm{Morales}, \binits{J.F.}}:
\batitle{{Anomalies {\textbackslash}{\&} tadpoles}}.
\bjtitle{JHEP}
\bvolume{03},
\bfpage{030}
(\byear{2000})
\doiurl{10.1088/1126-6708/2000/03/030}
{\href{https://arxiv.org/abs/hep-th/0002149}{{arXiv:hep-th/0002149}}}
\end{barticle}
\endbibitem

\bibitem[\protect\citeauthoryear{Aldazabal et~al.}{1999}]{Aldazabal:1999nu}
\begin{barticle}
\bauthor{\bsnm{Aldazabal}, \binits{G.}},
\bauthor{\bsnm{Badagnani}, \binits{D.}},
\bauthor{\bsnm{Ibanez}, \binits{L.E.}},
\bauthor{\bsnm{Uranga}, \binits{A.M.}}:
\batitle{{Tadpole versus anomaly cancellation in D = 4, D = 6 compact IIB orientifolds}}.
\bjtitle{JHEP}
\bvolume{06},
\bfpage{031}
(\byear{1999})
\doiurl{10.1088/1126-6708/1999/06/031}
{\href{https://arxiv.org/abs/hep-th/9904071}{{arXiv:hep-th/9904071}}}
\end{barticle}
\endbibitem

\bibitem[\protect\citeauthoryear{Witten}{2015}]{Witten:2013pra}
\begin{barticle}
\bauthor{\bsnm{Witten}, \binits{E.}}:
\batitle{{The Feynman $i \epsilon$ in String Theory}}.
\bjtitle{JHEP}
\bvolume{04},
\bfpage{055}
(\byear{2015})
\doiurl{10.1007/JHEP04(2015)055}
{\href{https://arxiv.org/abs/1307.5124}{{arXiv:1307.5124}}}
{[hep-th]}
\end{barticle}
\endbibitem

\bibitem[\protect\citeauthoryear{Eberhardt and Mizera}{2023}]{Eberhardt:2022zay}
\begin{barticle}
\bauthor{\bsnm{Eberhardt}, \binits{L.}},
\bauthor{\bsnm{Mizera}, \binits{S.}}:
\batitle{{Unitarity cuts of the worldsheet}}.
\bjtitle{SciPost Phys.}
\bvolume{14}(\bissue{2}),
\bfpage{015}
(\byear{2023})
\doiurl{10.21468/SciPostPhys.14.2.015}
{\href{https://arxiv.org/abs/2208.12233}{{arXiv:2208.12233}}}
{[hep-th]}
\end{barticle}
\endbibitem

\bibitem[\protect\citeauthoryear{Marcus}{1989}]{Marcus:1988vs}
\begin{barticle}
\bauthor{\bsnm{Marcus}, \binits{N.}}:
\batitle{{Unitarity and regularized divergences in string amplitudes}}.
\bjtitle{Phys. Lett. B}
\bvolume{219},
\bfpage{265}--\blpage{272}
(\byear{1989})
\doiurl{10.1016/0370-2693(89)90389-4}
\end{barticle}
\endbibitem

\bibitem[\protect\citeauthoryear{Baccianti et~al.}{2025}]{Baccianti:2025gll}
\begin{barticle}
\bauthor{\bsnm{Baccianti}, \binits{M.M.}},
\bauthor{\bsnm{Chandra}, \binits{J.}},
\bauthor{\bsnm{Eberhardt}, \binits{L.}},
\bauthor{\bsnm{Hartman}, \binits{T.}},
\bauthor{\bsnm{Mizera}, \binits{S.}}:
\batitle{{Rademacher expansion of modular integrals}}.
\bjtitle{SciPost Phys.}
\bvolume{19}(\bissue{4}),
\bfpage{103}
(\byear{2025})
\doiurl{10.21468/SciPostPhys.19.4.103}
{\href{https://arxiv.org/abs/2501.13827}{{arXiv:2501.13827}}}
{[hep-th]}
\end{barticle}
\endbibitem

\bibitem[\protect\citeauthoryear{Fradkin and Tseytlin}{1985a}]{Fradkin:1984pq}
\begin{barticle}
\bauthor{\bsnm{Fradkin}, \binits{E.S.}},
\bauthor{\bsnm{Tseytlin}, \binits{A.A.}}:
\batitle{{Effective Field Theory from Quantized Strings}}.
\bjtitle{Phys. Lett. B}
\bvolume{158},
\bfpage{316}--\blpage{322}
(\byear{1985})
\doiurl{10.1016/0370-2693(85)91190-6}
\end{barticle}
\endbibitem

\bibitem[\protect\citeauthoryear{Fradkin and Tseytlin}{1985b}]{Fradkin:1985ys}
\begin{barticle}
\bauthor{\bsnm{Fradkin}, \binits{E.S.}},
\bauthor{\bsnm{Tseytlin}, \binits{A.A.}}:
\batitle{{Quantum String Theory Effective Action}}.
\bjtitle{Nucl. Phys. B}
\bvolume{261},
\bfpage{1}--\blpage{27}
(\byear{1985})
\doiurl{10.1016/0550-3213(85)90559-0} .
\bcomment{[Erratum: Nucl.Phys.B 269, 745--745 (1986)]}
\end{barticle}
\endbibitem

\bibitem[\protect\citeauthoryear{Tseytlin}{1986}]{Tseytlin:1986ti}
\begin{barticle}
\bauthor{\bsnm{Tseytlin}, \binits{A.A.}}:
\batitle{{Vector Field Effective Action in the Open Superstring Theory}}.
\bjtitle{Nucl. Phys. B}
\bvolume{276},
\bfpage{391}
(\byear{1986})
\doiurl{10.1016/0550-3213(86)90303-2} .
\bcomment{[Erratum: Nucl.Phys.B 291, 876 (1987)]}
\end{barticle}
\endbibitem

\bibitem[\protect\citeauthoryear{Metsaev and Tseytlin}{1988}]{Metsaev:1987ju}
\begin{barticle}
\bauthor{\bsnm{Metsaev}, \binits{R.R.}},
\bauthor{\bsnm{Tseytlin}, \binits{A.A.}}:
\batitle{{On Loop Corrections To String Theory Effective Actions}}.
\bjtitle{Nucl. Phys. B}
\bvolume{298},
\bfpage{109}--\blpage{132}
(\byear{1988})
\doiurl{10.1016/0550-3213(88)90306-9}
\end{barticle}
\endbibitem

\bibitem[\protect\citeauthoryear{Tseytlin}{1989}]{Tseytlin:1988rr}
\begin{barticle}
\bauthor{\bsnm{Tseytlin}, \binits{A.A.}}:
\batitle{{Sigma Model Approach To String Theory}}.
\bjtitle{Int. J. Mod. Phys. A}
\bvolume{4},
\bfpage{1257}
(\byear{1989})
\doiurl{10.1142/S0217751X8900056X}
\end{barticle}
\endbibitem

\bibitem[\protect\citeauthoryear{Tseytlin}{1988}]{Tseytlin:1988mw}
\begin{barticle}
\bauthor{\bsnm{Tseytlin}, \binits{A.A.}}:
\batitle{{STRING THEORY EFFECTIVE ACTION: STRING LOOP CORRECTIONS}}.
\bjtitle{Int. J. Mod. Phys. A}
\bvolume{3},
\bfpage{365}--\blpage{395}
(\byear{1988})
\doiurl{10.1142/S0217751X88000138}
\end{barticle}
\endbibitem

\bibitem[\protect\citeauthoryear{Ahmadain and Wall}{2024}]{Ahmadain:2022tew}
\begin{barticle}
\bauthor{\bsnm{Ahmadain}, \binits{A.}},
\bauthor{\bsnm{Wall}, \binits{A.C.}}:
\batitle{{Off-shell strings I: S-matrix and action}}.
\bjtitle{SciPost Phys.}
\bvolume{17}(\bissue{1}),
\bfpage{005}
(\byear{2024})
\doiurl{10.21468/SciPostPhys.17.1.005}
{\href{https://arxiv.org/abs/2211.08607}{{arXiv:2211.08607}}}
{[hep-th]}
\end{barticle}
\endbibitem

\bibitem[\protect\citeauthoryear{Ahmadain et~al.}{2024}]{Ahmadain:2024hdp}
\begin{botherref}
\oauthor{\bsnm{Ahmadain}, \binits{A.}},
\oauthor{\bsnm{Frenkel}, \binits{A.}},
\oauthor{\bsnm{Wall}, \binits{A.C.}}:
{A Background-Independent Closed String Action at Tree Level}
(2024)
{\href{https://arxiv.org/abs/2410.11938}{{arXiv:2410.11938}}}
{[hep-th]}
\end{botherref}
\endbibitem

\bibitem[\protect\citeauthoryear{Sen and Zwiebach}{2024}]{Sen:2024nfd}
\begin{botherref}
\oauthor{\bsnm{Sen}, \binits{A.}},
\oauthor{\bsnm{Zwiebach}, \binits{B.}}:
{String Field Theory: A Review}
(2024)
{\href{https://arxiv.org/abs/2405.19421}{{arXiv:2405.19421}}}
{[hep-th]}
\end{botherref}
\endbibitem

\bibitem[\protect\citeauthoryear{Tseytlin}{1988a}]{Tseytlin:1987ww}
\begin{barticle}
\bauthor{\bsnm{Tseytlin}, \binits{A.A.}}:
\batitle{{Renormalization of Mobius Infinities and Partition Function Representation for String Theory Effective Action}}.
\bjtitle{Phys. Lett. B}
\bvolume{202},
\bfpage{81}--\blpage{88}
(\byear{1988})
\doiurl{10.1016/0370-2693(88)90857-X}
\end{barticle}
\endbibitem

\bibitem[\protect\citeauthoryear{Tseytlin}{1988b}]{Tseytlin:1988tv}
\begin{barticle}
\bauthor{\bsnm{Tseytlin}, \binits{A.A.}}:
\batitle{{Mobius Infinity Subtraction and Effective Action in $\sigma$ Model Approach to Closed String Theory}}.
\bjtitle{Phys. Lett. B}
\bvolume{208},
\bfpage{221}--\blpage{227}
(\byear{1988})
\doiurl{10.1016/0370-2693(88)90421-2}
\end{barticle}
\endbibitem

\bibitem[\protect\citeauthoryear{Tseytlin}{1993}]{Tseytlin:1993df}
\begin{barticle}
\bauthor{\bsnm{Tseytlin}, \binits{A.A.}}:
\batitle{{On field redefinitions and exact solutions in string theory}}.
\bjtitle{Phys. Lett. B}
\bvolume{317},
\bfpage{559}--\blpage{564}
(\byear{1993})
\doiurl{10.1016/0370-2693(93)91372-T}
{\href{https://arxiv.org/abs/hep-th/9308042}{{arXiv:hep-th/9308042}}}
\end{barticle}
\endbibitem

\bibitem[\protect\citeauthoryear{Banks}{1991}]{Banks:1991sg}
\begin{barticle}
\bauthor{\bsnm{Banks}, \binits{T.}}:
\batitle{{The Tachyon potential in string theory}}.
\bjtitle{Nucl. Phys. B}
\bvolume{361},
\bfpage{166}--\blpage{172}
(\byear{1991})
\doiurl{10.1016/0550-3213(91)90620-D}
\end{barticle}
\endbibitem

\bibitem[\protect\citeauthoryear{Tseytlin}{1991}]{Tseytlin:1991bu}
\begin{barticle}
\bauthor{\bsnm{Tseytlin}, \binits{A.A.}}:
\batitle{{On the tachyonic terms in the string effective action}}.
\bjtitle{Phys. Lett. B}
\bvolume{264},
\bfpage{311}--\blpage{318}
(\byear{1991})
\doiurl{10.1016/0370-2693(91)90355-T}
\end{barticle}
\endbibitem

\bibitem[\protect\citeauthoryear{Tseytlin}{2001}]{Tseytlin:2000mt}
\begin{barticle}
\bauthor{\bsnm{Tseytlin}, \binits{A.A.}}:
\batitle{{Sigma model approach to string theory effective actions with tachyons}}.
\bjtitle{J. Math. Phys.}
\bvolume{42},
\bfpage{2854}--\blpage{2871}
(\byear{2001})
\doiurl{10.1063/1.1376129}
{\href{https://arxiv.org/abs/hep-th/0011033}{{arXiv:hep-th/0011033}}}
\end{barticle}
\endbibitem

\bibitem[\protect\citeauthoryear{Witten}{1992}]{Witten:1992qy}
\begin{barticle}
\bauthor{\bsnm{Witten}, \binits{E.}}:
\batitle{{On background independent open string field theory}}.
\bjtitle{Phys. Rev. D}
\bvolume{46},
\bfpage{5467}--\blpage{5473}
(\byear{1992})
\doiurl{10.1103/PhysRevD.46.5467}
{\href{https://arxiv.org/abs/hep-th/9208027}{{arXiv:hep-th/9208027}}}
\end{barticle}
\endbibitem

\bibitem[\protect\citeauthoryear{Witten}{1993}]{Witten:1992cr}
\begin{barticle}
\bauthor{\bsnm{Witten}, \binits{E.}}:
\batitle{{Some computations in background independent off-shell string theory}}.
\bjtitle{Phys. Rev. D}
\bvolume{47},
\bfpage{3405}--\blpage{3410}
(\byear{1993})
\doiurl{10.1103/PhysRevD.47.3405}
{\href{https://arxiv.org/abs/hep-th/9210065}{{arXiv:hep-th/9210065}}}
\end{barticle}
\endbibitem

\bibitem[\protect\citeauthoryear{Shatashvili}{1993}]{Shatashvili:1993kk}
\begin{barticle}
\bauthor{\bsnm{Shatashvili}, \binits{S.L.}}:
\batitle{{Comment on the background independent open string theory}}.
\bjtitle{Phys. Lett. B}
\bvolume{311},
\bfpage{83}--\blpage{86}
(\byear{1993})
\doiurl{10.1016/0370-2693(93)90537-R}
{\href{https://arxiv.org/abs/hep-th/9303143}{{arXiv:hep-th/9303143}}}
\end{barticle}
\endbibitem

\bibitem[\protect\citeauthoryear{Shatashvili}{1994}]{Shatashvili:1993ps}
\begin{barticle}
\bauthor{\bsnm{Shatashvili}, \binits{S.L.}}:
\batitle{{On the problems with background independence in string theory}}.
\bjtitle{Alg. Anal.}
\bvolume{6},
\bfpage{215}--\blpage{226}
(\byear{1994})
\doiurl{10.1007/3-540-58453-6_12}
{\href{https://arxiv.org/abs/hep-th/9311177}{{arXiv:hep-th/9311177}}}
\end{barticle}
\endbibitem

\bibitem[\protect\citeauthoryear{Kutasov et~al.}{2000}]{Kutasov:2000qp}
\begin{barticle}
\bauthor{\bsnm{Kutasov}, \binits{D.}},
\bauthor{\bsnm{Marino}, \binits{M.}},
\bauthor{\bsnm{Moore}, \binits{G.W.}}:
\batitle{{Some exact results on tachyon condensation in string field theory}}.
\bjtitle{JHEP}
\bvolume{10},
\bfpage{045}
(\byear{2000})
\doiurl{10.1088/1126-6708/2000/10/045}
{\href{https://arxiv.org/abs/hep-th/0009148}{{arXiv:hep-th/0009148}}}
\end{barticle}
\endbibitem

\bibitem[\protect\citeauthoryear{Harvey et~al.}{2000}]{Harvey:2000na}
\begin{botherref}
\oauthor{\bsnm{Harvey}, \binits{J.A.}},
\oauthor{\bsnm{Kutasov}, \binits{D.}},
\oauthor{\bsnm{Martinec}, \binits{E.J.}}:
{On the relevance of tachyons}
(2000)
{\href{https://arxiv.org/abs/hep-th/0003101}{{arXiv:hep-th/0003101}}}
\end{botherref}
\endbibitem

\bibitem[\protect\citeauthoryear{Sen}{1998a}]{Sen:1998ii}
\begin{barticle}
\bauthor{\bsnm{Sen}, \binits{A.}}:
\batitle{{Stable nonBPS bound states of BPS D-branes}}.
\bjtitle{JHEP}
\bvolume{08},
\bfpage{010}
(\byear{1998})
\doiurl{10.1088/1126-6708/1998/08/010}
{\href{https://arxiv.org/abs/hep-th/9805019}{{arXiv:hep-th/9805019}}}
\end{barticle}
\endbibitem

\bibitem[\protect\citeauthoryear{Sen}{1998b}]{Sen:1998sm}
\begin{barticle}
\bauthor{\bsnm{Sen}, \binits{A.}}:
\batitle{{Tachyon condensation on the brane anti-brane system}}.
\bjtitle{JHEP}
\bvolume{08},
\bfpage{012}
(\byear{1998})
\doiurl{10.1088/1126-6708/1998/08/012}
{\href{https://arxiv.org/abs/hep-th/9805170}{{arXiv:hep-th/9805170}}}
\end{barticle}
\endbibitem

\bibitem[\protect\citeauthoryear{Sen and Zwiebach}{2000}]{Sen:1999nx}
\begin{barticle}
\bauthor{\bsnm{Sen}, \binits{A.}},
\bauthor{\bsnm{Zwiebach}, \binits{B.}}:
\batitle{{Tachyon condensation in string field theory}}.
\bjtitle{JHEP}
\bvolume{03},
\bfpage{002}
(\byear{2000})
\doiurl{10.1088/1126-6708/2000/03/002}
{\href{https://arxiv.org/abs/hep-th/9912249}{{arXiv:hep-th/9912249}}}
\end{barticle}
\endbibitem

\bibitem[\protect\citeauthoryear{Schnabl}{2006}]{Schnabl:2005gv}
\begin{barticle}
\bauthor{\bsnm{Schnabl}, \binits{M.}}:
\batitle{{Analytic solution for tachyon condensation in open string field theory}}.
\bjtitle{Adv. Theor. Math. Phys.}
\bvolume{10}(\bissue{4}),
\bfpage{433}--\blpage{501}
(\byear{2006})
\doiurl{10.4310/ATMP.2006.v10.n4.a1}
{\href{https://arxiv.org/abs/hep-th/0511286}{{arXiv:hep-th/0511286}}}
\end{barticle}
\endbibitem

\bibitem[\protect\citeauthoryear{Hellerman and Swanson}{2007}]{Hellerman:2007zz}
\begin{botherref}
\oauthor{\bsnm{Hellerman}, \binits{S.}},
\oauthor{\bsnm{Swanson}, \binits{I.}}:
{A Stable vacuum of the tachyonic $E_8$ string}
(2007)
{\href{https://arxiv.org/abs/0710.1628}{{arXiv:0710.1628}}}
{[hep-th]}
\end{botherref}
\endbibitem

\bibitem[\protect\citeauthoryear{Kaidi}{2021}]{Kaidi:2020jla}
\begin{barticle}
\bauthor{\bsnm{Kaidi}, \binits{J.}}:
\batitle{{Stable Vacua for Tachyonic Strings}}.
\bjtitle{Phys. Rev. D}
\bvolume{103}(\bissue{10}),
\bfpage{106026}
(\byear{2021})
\doiurl{10.1103/PhysRevD.103.106026}
{\href{https://arxiv.org/abs/2010.10521}{{arXiv:2010.10521}}}
{[hep-th]}
\end{barticle}
\endbibitem

\bibitem[\protect\citeauthoryear{Kaidi et~al.}{2023}]{Kaidi:2023tqo}
\begin{barticle}
\bauthor{\bsnm{Kaidi}, \binits{J.}},
\bauthor{\bsnm{Ohmori}, \binits{K.}},
\bauthor{\bsnm{Tachikawa}, \binits{Y.}},
\bauthor{\bsnm{Yonekura}, \binits{K.}}:
\batitle{{Nonsupersymmetric Heterotic Branes}}.
\bjtitle{Phys. Rev. Lett.}
\bvolume{131}(\bissue{12}),
\bfpage{121601}
(\byear{2023})
\doiurl{10.1103/PhysRevLett.131.121601}
{\href{https://arxiv.org/abs/2303.17623}{{arXiv:2303.17623}}}
{[hep-th]}
\end{barticle}
\endbibitem

\bibitem[\protect\citeauthoryear{Kaidi et~al.}{2025}]{Kaidi:2024cbx}
\begin{barticle}
\bauthor{\bsnm{Kaidi}, \binits{J.}},
\bauthor{\bsnm{Tachikawa}, \binits{Y.}},
\bauthor{\bsnm{Yonekura}, \binits{K.}}:
\batitle{{On non-supersymmetric heterotic branes}}.
\bjtitle{JHEP}
\bvolume{03},
\bfpage{211}
(\byear{2025})
\doiurl{10.1007/JHEP03(2025)211}
{\href{https://arxiv.org/abs/2411.04344}{{arXiv:2411.04344}}}
{[hep-th]}
\end{barticle}
\endbibitem

\bibitem[\protect\citeauthoryear{Dienes}{1994}]{Dienes:1994np}
\begin{barticle}
\bauthor{\bsnm{Dienes}, \binits{K.R.}}:
\batitle{{Modular invariance, finiteness, and misaligned supersymmetry: New constraints on the numbers of physical string states}}.
\bjtitle{Nucl. Phys. B}
\bvolume{429},
\bfpage{533}--\blpage{588}
(\byear{1994})
\doiurl{10.1016/0550-3213(94)90153-8}
{\href{https://arxiv.org/abs/hep-th/9402006}{{arXiv:hep-th/9402006}}}
\end{barticle}
\endbibitem

\bibitem[\protect\citeauthoryear{Rademacher}{1937a}]{Rademacher:1937a}
\begin{barticle}
\bauthor{\bsnm{Rademacher}, \binits{H.}}:
\batitle{{A Convergent Series for the Partition Function $p(n)$}}.
\bjtitle{Proc. Natl. Acad. Sci.}
\bvolume{23},
\bfpage{78}
(\byear{1937})
\end{barticle}
\endbibitem

\bibitem[\protect\citeauthoryear{Rademacher}{1937b}]{Rademacher:1937b}
\begin{barticle}
\bauthor{\bsnm{Rademacher}, \binits{H.}}:
\batitle{{On the Partition Function $p(n)$}}.
\bjtitle{Proc. Lond. Math. Soc.}
\bvolume{43},
\bfpage{241}
(\byear{1937})
\end{barticle}
\endbibitem

\bibitem[\protect\citeauthoryear{Rademacher}{1938}]{Rademacher:1938b}
\begin{barticle}
\bauthor{\bsnm{Rademacher}, \binits{H.}}:
\batitle{{ The Fourier Coefficients of the Modular Invariant $J (\tau )$}}.
\bjtitle{Am. J. Math.}
\bvolume{60},
\bfpage{501}
(\byear{1938})
\end{barticle}
\endbibitem

\bibitem[\protect\citeauthoryear{Dijkgraaf et~al.}{2000}]{Dijkgraaf:2000fq}
\begin{botherref}
\oauthor{\bsnm{Dijkgraaf}, \binits{R.}},
\oauthor{\bsnm{Maldacena}, \binits{J.M.}},
\oauthor{\bsnm{Moore}, \binits{G.W.}},
\oauthor{\bsnm{Verlinde}, \binits{E.P.}}:
{A Black hole Farey tail}
(2000)
{\href{https://arxiv.org/abs/hep-th/0005003}{{arXiv:hep-th/0005003}}}
\end{botherref}
\endbibitem

\bibitem[\protect\citeauthoryear{Manschot and Moore}{2010}]{Manschot:2007ha}
\begin{barticle}
\bauthor{\bsnm{Manschot}, \binits{J.}},
\bauthor{\bsnm{Moore}, \binits{G.W.}}:
\batitle{{A Modern Farey Tail}}.
\bjtitle{Commun. Num. Theor. Phys.}
\bvolume{4},
\bfpage{103}--\blpage{159}
(\byear{2010})
\doiurl{10.4310/CNTP.2010.v4.n1.a3}
{\href{https://arxiv.org/abs/0712.0573}{{arXiv:0712.0573}}}
{[hep-th]}
\end{barticle}
\endbibitem

\bibitem[\protect\citeauthoryear{Hardy and Ramanujan}{1918}]{Hardy}
\begin{barticle}
\bauthor{\bsnm{Hardy}, \binits{G.H.}},
\bauthor{\bsnm{Ramanujan}, \binits{S.}}:
\batitle{{Asymptotic Formula\ae\ in Combinatory Analysis}}.
\bjtitle{Proc. Lond. Math. Soc.}
\bvolume{17},
\bfpage{75}
(\byear{1918})
\end{barticle}
\endbibitem

\bibitem[\protect\citeauthoryear{Kani and Vafa}{1990}]{Kani:1989im}
\begin{barticle}
\bauthor{\bsnm{Kani}, \binits{I.}},
\bauthor{\bsnm{Vafa}, \binits{C.}}:
\batitle{{Asymptotic Mass Degeneracies in Conformal Field Theories}}.
\bjtitle{Commun. Math. Phys.}
\bvolume{130},
\bfpage{529}--\blpage{580}
(\byear{1990})
\doiurl{10.1007/BF02096934}
\end{barticle}
\endbibitem

\bibitem[\protect\citeauthoryear{Cribiori et~al.}{2021}]{Cribiori:2020sct}
\begin{barticle}
\bauthor{\bsnm{Cribiori}, \binits{N.}},
\bauthor{\bsnm{Parameswaran}, \binits{S.}},
\bauthor{\bsnm{Tonioni}, \binits{F.}},
\bauthor{\bsnm{Wrase}, \binits{T.}}:
\batitle{{Misaligned Supersymmetry and Open Strings}}.
\bjtitle{JHEP}
\bvolume{04},
\bfpage{099}
(\byear{2021})
\doiurl{10.1007/JHEP04(2021)099}
{\href{https://arxiv.org/abs/2012.04677}{{arXiv:2012.04677}}}
{[hep-th]}
\end{barticle}
\endbibitem

\bibitem[\protect\citeauthoryear{Angelantonj et~al.}{2023}]{Angelantonj:2023egh}
\begin{barticle}
\bauthor{\bsnm{Angelantonj}, \binits{C.}},
\bauthor{\bsnm{Florakis}, \binits{I.}},
\bauthor{\bsnm{Leone}, \binits{G.}}:
\batitle{{Tachyons and misaligned supersymmetry in closed string vacua}}.
\bjtitle{JHEP}
\bvolume{06},
\bfpage{174}
(\byear{2023})
\doiurl{10.1007/JHEP06(2023)174}
{\href{https://arxiv.org/abs/2301.13702}{{arXiv:2301.13702}}}
{[hep-th]}
\end{barticle}
\endbibitem

\bibitem[\protect\citeauthoryear{Cardy}{1982}]{Cardy:1981fd}
\begin{barticle}
\bauthor{\bsnm{Cardy}, \binits{J.L.}}:
\batitle{{Duality and the Theta Parameter in Abelian Lattice Models}}.
\bjtitle{Nucl. Phys. B}
\bvolume{205},
\bfpage{17}--\blpage{26}
(\byear{1982})
\doiurl{10.1016/0550-3213(82)90464-3}
\end{barticle}
\endbibitem

\bibitem[\protect\citeauthoryear{Leone}{2023}]{Leone:2023qfd}
\begin{barticle}
\bauthor{\bsnm{Leone}, \binits{G.}}:
\batitle{{Tachyons and Misaligned Supersymmetry in orientifold vacua}}.
\bjtitle{JHEP}
\bvolume{11},
\bfpage{066}
(\byear{2023})
\doiurl{10.1007/JHEP11(2023)066}
{\href{https://arxiv.org/abs/2308.09757}{{arXiv:2308.09757}}}
{[hep-th]}
\end{barticle}
\endbibitem

\bibitem[\protect\citeauthoryear{Sagnotti}{1997}]{Sagnotti:1996qj}
\begin{barticle}
\bauthor{\bsnm{Sagnotti}, \binits{A.}}:
\batitle{{Surprises in open string perturbation theory}}.
\bjtitle{Nucl. Phys. B Proc. Suppl.}
\bvolume{56},
\bfpage{332}--\blpage{343}
(\byear{1997})
\doiurl{10.1016/S0920-5632(97)00344-7}
{\href{https://arxiv.org/abs/hep-th/9702093}{{arXiv:hep-th/9702093}}}
\end{barticle}
\endbibitem

\bibitem[\protect\citeauthoryear{Lerche et~al.}{1986}]{Lerche:1986ae}
\begin{barticle}
\bauthor{\bsnm{Lerche}, \binits{W.}},
\bauthor{\bsnm{Lust}, \binits{D.}},
\bauthor{\bsnm{Schellekens}, \binits{A.N.}}:
\batitle{{Ten-dimensional Heterotic Strings From Niemeier Lattices}}.
\bjtitle{Phys. Lett. B}
\bvolume{181},
\bfpage{71}
(\byear{1986})
\doiurl{10.1016/0370-2693(86)91257-8}
\end{barticle}
\endbibitem

\bibitem[\protect\citeauthoryear{Niemeier}{1973}]{Niemeier:1973}
\begin{barticle}
\bauthor{\bsnm{Niemeier}, \binits{H.V.}}:
\batitle{{Definite quadratische Formen der Dimension 24 und Diskriminante 1,}}.
\bjtitle{J. Number Theory}
\bvolume{5},
\bfpage{142}--\blpage{178}
(\byear{1973})
\end{barticle}
\endbibitem

\bibitem[\protect\citeauthoryear{Blumenhagen and Plauschinn}{2009}]{Blumenhagen:2009zz}
\begin{bbook}
\bauthor{\bsnm{Blumenhagen}, \binits{R.}},
\bauthor{\bsnm{Plauschinn}, \binits{E.}}:
\bbtitle{Introduction to Conformal Field Theory : with Applications to String Theory}
vol. \bseriesno{779},
(\byear{2009}).
\doiurl{10.1007/978-3-642-00450-6}
\end{bbook}
\endbibitem

\bibitem[\protect\citeauthoryear{Markou and Skvortsov}{2023}]{Markou:2023ffh}
\begin{barticle}
\bauthor{\bsnm{Markou}, \binits{C.}},
\bauthor{\bsnm{Skvortsov}, \binits{E.}}:
\batitle{{An excursion into the string spectrum}}.
\bjtitle{JHEP}
\bvolume{12},
\bfpage{055}
(\byear{2023})
\doiurl{10.1007/JHEP12(2023)055}
{\href{https://arxiv.org/abs/2309.15988}{{arXiv:2309.15988}}}
{[hep-th]}
\end{barticle}
\endbibitem

\bibitem[\protect\citeauthoryear{Basile and Markou}{2024}]{Basile:2024uxn}
\begin{barticle}
\bauthor{\bsnm{Basile}, \binits{T.}},
\bauthor{\bsnm{Markou}, \binits{C.}}:
\batitle{{On the deep superstring spectrum}}.
\bjtitle{JHEP}
\bvolume{07},
\bfpage{184}
(\byear{2024})
\doiurl{10.1007/JHEP07(2024)184}
{\href{https://arxiv.org/abs/2405.18467}{{arXiv:2405.18467}}}
{[hep-th]}
\end{barticle}
\endbibitem

\bibitem[\protect\citeauthoryear{Alvarez-Gaume et~al.}{1986}]{Alvarez-Gaume:1986ghj}
\begin{barticle}
\bauthor{\bsnm{Alvarez-Gaume}, \binits{L.}},
\bauthor{\bsnm{Ginsparg}, \binits{P.H.}},
\bauthor{\bsnm{Moore}, \binits{G.W.}},
\bauthor{\bsnm{Vafa}, \binits{C.}}:
\batitle{{An O(16) x O(16) Heterotic String}}.
\bjtitle{Phys. Lett. B}
\bvolume{171},
\bfpage{155}--\blpage{162}
(\byear{1986})
\doiurl{10.1016/0370-2693(86)91524-8}
\end{barticle}
\endbibitem

\bibitem[\protect\citeauthoryear{Kawai et~al.}{1986}]{Kawai:1986vd}
\begin{barticle}
\bauthor{\bsnm{Kawai}, \binits{H.}},
\bauthor{\bsnm{Lewellen}, \binits{D.C.}},
\bauthor{\bsnm{Tye}, \binits{S.H.H.}}:
\batitle{{Classification of Closed Fermionic String Models}}.
\bjtitle{Phys. Rev. D}
\bvolume{34},
\bfpage{3794}
(\byear{1986})
\doiurl{10.1103/PhysRevD.34.3794}
\end{barticle}
\endbibitem

\bibitem[\protect\citeauthoryear{Dixon and Harvey}{1986}]{Dixon:1986iz}
\begin{barticle}
\bauthor{\bsnm{Dixon}, \binits{L.J.}},
\bauthor{\bsnm{Harvey}, \binits{J.A.}}:
\batitle{{String Theories in Ten-Dimensions Without Space-Time Supersymmetry}}.
\bjtitle{Nucl. Phys. B}
\bvolume{274},
\bfpage{93}--\blpage{105}
(\byear{1986})
\doiurl{10.1016/0550-3213(86)90619-X}
\end{barticle}
\endbibitem

\bibitem[\protect\citeauthoryear{Schellekens}{1993}]{Schellekens:1992db}
\begin{barticle}
\bauthor{\bsnm{Schellekens}, \binits{A.N.}}:
\batitle{{Meromorphic C = 24 conformal field theories}}.
\bjtitle{Commun. Math. Phys.}
\bvolume{153},
\bfpage{159}--\blpage{186}
(\byear{1993})
\doiurl{10.1007/BF02099044}
{\href{https://arxiv.org/abs/hep-th/9205072}{{arXiv:hep-th/9205072}}}
\end{barticle}
\endbibitem

\bibitem[\protect\citeauthoryear{Boyle~Smith et~al.}{2024}]{BoyleSmith:2023xkd}
\begin{barticle}
\bauthor{\bsnm{Boyle~Smith}, \binits{P.}},
\bauthor{\bsnm{Lin}, \binits{Y.-H.}},
\bauthor{\bsnm{Tachikawa}, \binits{Y.}},
\bauthor{\bsnm{Zheng}, \binits{Y.}}:
\batitle{{Classification of chiral fermionic CFTs of central charge $\le$ 16}}.
\bjtitle{SciPost Phys.}
\bvolume{16}(\bissue{2}),
\bfpage{058}
(\byear{2024})
\doiurl{10.21468/SciPostPhys.16.2.058}
{\href{https://arxiv.org/abs/2303.16917}{{arXiv:2303.16917}}}
{[hep-th]}
\end{barticle}
\endbibitem

\bibitem[\protect\citeauthoryear{Ginsparg and Vafa}{1987}]{Ginsparg:1986wr}
\begin{barticle}
\bauthor{\bsnm{Ginsparg}, \binits{P.H.}},
\bauthor{\bsnm{Vafa}, \binits{C.}}:
\batitle{{Toroidal Compactification of Nonsupersymmetric Heterotic Strings}}.
\bjtitle{Nucl. Phys. B}
\bvolume{289},
\bfpage{414}
(\byear{1987})
\doiurl{10.1016/0550-3213(87)90387-7}
\end{barticle}
\endbibitem

\bibitem[\protect\citeauthoryear{Narain et~al.}{1987}]{Narain:1986qm}
\begin{barticle}
\bauthor{\bsnm{Narain}, \binits{K.S.}},
\bauthor{\bsnm{Sarmadi}, \binits{M.H.}},
\bauthor{\bsnm{Vafa}, \binits{C.}}:
\batitle{{Asymmetric Orbifolds}}.
\bjtitle{Nucl. Phys. B}
\bvolume{288},
\bfpage{551}
(\byear{1987})
\doiurl{10.1016/0550-3213(87)90228-8}
\end{barticle}
\endbibitem

\bibitem[\protect\citeauthoryear{Narain et~al.}{1991}]{Narain:1990mw}
\begin{barticle}
\bauthor{\bsnm{Narain}, \binits{K.S.}},
\bauthor{\bsnm{Sarmadi}, \binits{M.H.}},
\bauthor{\bsnm{Vafa}, \binits{C.}}:
\batitle{{Asymmetric orbifolds: Path integral and operator formulations}}.
\bjtitle{Nucl. Phys. B}
\bvolume{356},
\bfpage{163}--\blpage{207}
(\byear{1991})
\doiurl{10.1016/0550-3213(91)90145-N}
\end{barticle}
\endbibitem

\bibitem[\protect\citeauthoryear{Aldazabal et~al.}{2025}]{Aldazabal:2025zht}
\begin{barticle}
\bauthor{\bsnm{Aldazabal}, \binits{G.}},
\bauthor{\bsnm{Andr{\'e}s}, \binits{E.}},
\bauthor{\bsnm{Font}, \binits{A.}},
\bauthor{\bsnm{Narain}, \binits{K.}},
\bauthor{\bsnm{Zadeh}, \binits{I.G.}}:
\batitle{{Asymmetric orbifolds, rank reduction and heterotic islands}}.
\bjtitle{JHEP}
\bvolume{08},
\bfpage{083}
(\byear{2025})
\doiurl{10.1007/JHEP08(2025)083}
{\href{https://arxiv.org/abs/2501.17228}{{arXiv:2501.17228}}}
{[hep-th]}
\end{barticle}
\endbibitem

\bibitem[\protect\citeauthoryear{Harvey et~al.}{1988}]{Harvey:1987da}
\begin{barticle}
\bauthor{\bsnm{Harvey}, \binits{J.A.}},
\bauthor{\bsnm{Moore}, \binits{G.W.}},
\bauthor{\bsnm{Vafa}, \binits{C.}}:
\batitle{{Quasirystalline Compactification}}.
\bjtitle{Nucl. Phys. B}
\bvolume{304},
\bfpage{269}--\blpage{290}
(\byear{1988})
\doiurl{10.1016/0550-3213(88)90627-X}
\end{barticle}
\endbibitem

\bibitem[\protect\citeauthoryear{Baykara et~al.}{2026}]{Baykara:2024tjr}
\begin{barticle}
\bauthor{\bsnm{Baykara}, \binits{Z.K.}},
\bauthor{\bsnm{Tarazi}, \binits{H.-C.}},
\bauthor{\bsnm{Vafa}, \binits{C.}}:
\batitle{{New Nonsupersymmetric Tachyon-Free Strings}}.
\bjtitle{Phys. Rev. Lett.}
\bvolume{136}(\bissue{6}),
\bfpage{061602}
(\byear{2026})
\doiurl{10.1103/3c5v-yv16}
{\href{https://arxiv.org/abs/2406.00185}{{arXiv:2406.00185}}}
{[hep-th]}
\end{barticle}
\endbibitem

\bibitem[\protect\citeauthoryear{Baykara et~al.}{2025}]{Baykara:2024vss}
\begin{barticle}
\bauthor{\bsnm{Baykara}, \binits{Z.K.}},
\bauthor{\bsnm{Tarazi}, \binits{H.-C.}},
\bauthor{\bsnm{Vafa}, \binits{C.}}:
\batitle{{Quasicrystalline string landscape}}.
\bjtitle{Phys. Rev. D}
\bvolume{111}(\bissue{8}),
\bfpage{086025}
(\byear{2025})
\doiurl{10.1103/PhysRevD.111.086025}
{\href{https://arxiv.org/abs/2406.00129}{{arXiv:2406.00129}}}
{[hep-th]}
\end{barticle}
\endbibitem

\bibitem[\protect\citeauthoryear{Angelantonj et~al.}{2024}]{Angelantonj:2024jtu}
\begin{barticle}
\bauthor{\bsnm{Angelantonj}, \binits{C.}},
\bauthor{\bsnm{Florakis}, \binits{I.}},
\bauthor{\bsnm{Leone}, \binits{G.}},
\bauthor{\bsnm{Perugini}, \binits{D.}}:
\batitle{{Non-supersymmetric non-tachyonic heterotic vacua with reduced rank in various dimensions}}.
\bjtitle{JHEP}
\bvolume{10},
\bfpage{216}
(\byear{2024})
\doiurl{10.1007/JHEP10(2024)216}
{\href{https://arxiv.org/abs/2407.09597}{{arXiv:2407.09597}}}
{[hep-th]}
\end{barticle}
\endbibitem

\bibitem[\protect\citeauthoryear{Basile et~al.}{2019}]{Basile:2018irz}
\begin{barticle}
\bauthor{\bsnm{Basile}, \binits{I.}},
\bauthor{\bsnm{Mourad}, \binits{J.}},
\bauthor{\bsnm{Sagnotti}, \binits{A.}}:
\batitle{{On Classical Stability with Broken Supersymmetry}}.
\bjtitle{JHEP}
\bvolume{01},
\bfpage{174}
(\byear{2019})
\doiurl{10.1007/JHEP01(2019)174}
{\href{https://arxiv.org/abs/1811.11448}{{arXiv:1811.11448}}}
{[hep-th]}
\end{barticle}
\endbibitem

\bibitem[\protect\citeauthoryear{Baykara et~al.}{2023}]{Baykara:2022cwj}
\begin{barticle}
\bauthor{\bsnm{Baykara}, \binits{Z.K.}},
\bauthor{\bsnm{Robbins}, \binits{D.}},
\bauthor{\bsnm{Sethi}, \binits{S.}}:
\batitle{{Non-supersymmetric AdS from string theory}}.
\bjtitle{SciPost Phys.}
\bvolume{15}(\bissue{6}),
\bfpage{224}
(\byear{2023})
\doiurl{10.21468/SciPostPhys.15.6.224}
{\href{https://arxiv.org/abs/2212.02557}{{arXiv:2212.02557}}}
{[hep-th]}
\end{barticle}
\endbibitem

\bibitem[\protect\citeauthoryear{Sugimoto}{1999}]{Sugimoto:1999tx}
\begin{barticle}
\bauthor{\bsnm{Sugimoto}, \binits{S.}}:
\batitle{{Anomaly cancellations in type I D-9 - anti-D-9 system and the USp(32) string theory}}.
\bjtitle{Prog. Theor. Phys.}
\bvolume{102},
\bfpage{685}--\blpage{699}
(\byear{1999})
\doiurl{10.1143/PTP.102.685}
{\href{https://arxiv.org/abs/hep-th/9905159}{{arXiv:hep-th/9905159}}}
\end{barticle}
\endbibitem

\bibitem[\protect\citeauthoryear{Dudas and Mourad}{2001}]{Dudas:2000nv}
\begin{barticle}
\bauthor{\bsnm{Dudas}, \binits{E.}},
\bauthor{\bsnm{Mourad}, \binits{J.}}:
\batitle{{Consistent gravitino couplings in nonsupersymmetric strings}}.
\bjtitle{Phys. Lett. B}
\bvolume{514},
\bfpage{173}--\blpage{182}
(\byear{2001})
\doiurl{10.1016/S0370-2693(01)00777-8}
{\href{https://arxiv.org/abs/hep-th/0012071}{{arXiv:hep-th/0012071}}}
\end{barticle}
\endbibitem

\bibitem[\protect\citeauthoryear{Volkov and Akulov}{1972}]{Volkov:1972jx}
\begin{barticle}
\bauthor{\bsnm{Volkov}, \binits{D.V.}},
\bauthor{\bsnm{Akulov}, \binits{V.P.}}:
\batitle{{Possible universal neutrino interaction}}.
\bjtitle{JETP Lett.}
\bvolume{16},
\bfpage{438}--\blpage{440}
(\byear{1972})
\end{barticle}
\endbibitem

\bibitem[\protect\citeauthoryear{Antoniadis et~al.}{1999}]{Antoniadis:1999xk}
\begin{barticle}
\bauthor{\bsnm{Antoniadis}, \binits{I.}},
\bauthor{\bsnm{Dudas}, \binits{E.}},
\bauthor{\bsnm{Sagnotti}, \binits{A.}}:
\batitle{{Brane supersymmetry breaking}}.
\bjtitle{Phys. Lett. B}
\bvolume{464},
\bfpage{38}--\blpage{45}
(\byear{1999})
\doiurl{10.1016/S0370-2693(99)01023-0}
{\href{https://arxiv.org/abs/hep-th/9908023}{{arXiv:hep-th/9908023}}}
\end{barticle}
\endbibitem

\bibitem[\protect\citeauthoryear{Aldazabal and Uranga}{1999}]{Aldazabal:1999jr}
\begin{barticle}
\bauthor{\bsnm{Aldazabal}, \binits{G.}},
\bauthor{\bsnm{Uranga}, \binits{A.M.}}:
\batitle{{Tachyon free nonsupersymmetric type IIB orientifolds via Brane - anti-brane systems}}.
\bjtitle{JHEP}
\bvolume{10},
\bfpage{024}
(\byear{1999})
\doiurl{10.1088/1126-6708/1999/10/024}
{\href{https://arxiv.org/abs/hep-th/9908072}{{arXiv:hep-th/9908072}}}
\end{barticle}
\endbibitem

\bibitem[\protect\citeauthoryear{Angelantonj et~al.}{2024}]{Angelantonj:2024iwi}
\begin{barticle}
\bauthor{\bsnm{Angelantonj}, \binits{C.}},
\bauthor{\bsnm{Condeescu}, \binits{C.}},
\bauthor{\bsnm{Dudas}, \binits{E.}},
\bauthor{\bsnm{Leone}, \binits{G.}}:
\batitle{{Rigid vacua with Brane Supersymmetry Breaking}}.
\bjtitle{JHEP}
\bvolume{04},
\bfpage{103}
(\byear{2024})
\doiurl{10.1007/JHEP04(2024)103}
{\href{https://arxiv.org/abs/2403.02392}{{arXiv:2403.02392}}}
{[hep-th]}
\end{barticle}
\endbibitem

\bibitem[\protect\citeauthoryear{Leone}{2025}]{Leone:2024hnr}
\begin{barticle}
\bauthor{\bsnm{Leone}, \binits{G.}}:
\batitle{{New comments on six-dimensional orientifold vacua with reduced rank and unitarity constraints}}.
\bjtitle{JHEP}
\bvolume{06},
\bfpage{062}
(\byear{2025})
\doiurl{10.1007/JHEP06(2025)062}
{\href{https://arxiv.org/abs/2412.19185}{{arXiv:2412.19185}}}
{[hep-th]}
\end{barticle}
\endbibitem

\bibitem[\protect\citeauthoryear{Klebanov and Tseytlin}{1999}]{Klebanov:1998yya}
\begin{barticle}
\bauthor{\bsnm{Klebanov}, \binits{I.R.}},
\bauthor{\bsnm{Tseytlin}, \binits{A.A.}}:
\batitle{{D-branes and dual gauge theories in type 0 strings}}.
\bjtitle{Nucl. Phys. B}
\bvolume{546},
\bfpage{155}--\blpage{181}
(\byear{1999})
\doiurl{10.1016/S0550-3213(99)00041-3}
{\href{https://arxiv.org/abs/hep-th/9811035}{{arXiv:hep-th/9811035}}}
\end{barticle}
\endbibitem

\bibitem[\protect\citeauthoryear{Dudas et~al.}{2002}]{Dudas:2001wd}
\begin{barticle}
\bauthor{\bsnm{Dudas}, \binits{E.}},
\bauthor{\bsnm{Mourad}, \binits{J.}},
\bauthor{\bsnm{Sagnotti}, \binits{A.}}:
\batitle{{Charged and uncharged D-branes in various string theories}}.
\bjtitle{Nucl. Phys. B}
\bvolume{620},
\bfpage{109}--\blpage{151}
(\byear{2002})
\doiurl{10.1016/S0550-3213(01)00552-1}
{\href{https://arxiv.org/abs/hep-th/0107081}{{arXiv:hep-th/0107081}}}
\end{barticle}
\endbibitem

\bibitem[\protect\citeauthoryear{Sagnotti}{1995}]{Sagnotti:1995ga}
\begin{bchapter}
\bauthor{\bsnm{Sagnotti}, \binits{A.}}:
\bctitle{{Some properties of open string theories}}.
In: \bbtitle{International Workshop on Supersymmetry and Unification of Fundamental Interactions (SUSY 95)},
pp. \bfpage{473}--\blpage{484}
(\byear{1995})
\end{bchapter}
\endbibitem

\bibitem[\protect\citeauthoryear{Fraiman et~al.}{2024}]{Fraiman:2023cpa}
\begin{barticle}
\bauthor{\bsnm{Fraiman}, \binits{B.}},
\bauthor{\bsnm{Gra\~na}, \binits{M.}},
\bauthor{\bsnm{Parra De~Freitas}, \binits{H.}},
\bauthor{\bsnm{Sethi}, \binits{S.}}:
\batitle{{Non-supersymmetric heterotic strings on a circle}}.
\bjtitle{JHEP}
\bvolume{12},
\bfpage{082}
(\byear{2024})
\doiurl{10.1007/JHEP12(2024)082}
{\href{https://arxiv.org/abs/2307.13745}{{arXiv:2307.13745}}}
{[hep-th]}
\end{barticle}
\endbibitem

\bibitem[\protect\citeauthoryear{De~Freitas}{2024}]{DeFreitas:2024ztt}
\begin{barticle}
\bauthor{\bsnm{De~Freitas}, \binits{H.P.}}:
\batitle{{Non-supersymmetric heterotic strings and chiral CFTs}}.
\bjtitle{JHEP}
\bvolume{11},
\bfpage{002}
(\byear{2024})
\doiurl{10.1007/JHEP11(2024)002}
{\href{https://arxiv.org/abs/2402.15562}{{arXiv:2402.15562}}}
{[hep-th]}
\end{barticle}
\endbibitem

\bibitem[\protect\citeauthoryear{Antoniadis et~al.}{2025}]{Antoniadis:2025twv}
\begin{barticle}
\bauthor{\bsnm{Antoniadis}, \binits{I.}},
\bauthor{\bsnm{Avalos}, \binits{A.R.D.}},
\bauthor{\bsnm{Faraggi}, \binits{A.E.}}:
\batitle{{A perturbatively stable non-supersymmetric string model with AdS vacuum}}.
\bjtitle{JHEP}
\bvolume{11},
\bfpage{068}
(\byear{2025})
\doiurl{10.1007/JHEP11(2025)068}
{\href{https://arxiv.org/abs/2504.19364}{{arXiv:2504.19364}}}
{[hep-th]}
\end{barticle}
\endbibitem

\bibitem[\protect\citeauthoryear{Angelantonj}{2000}]{Angelantonj:1999jh}
\begin{barticle}
\bauthor{\bsnm{Angelantonj}, \binits{C.}}:
\batitle{{Comments on open string orbifolds with a nonvanishing B(ab)}}.
\bjtitle{Nucl. Phys. B}
\bvolume{566},
\bfpage{126}--\blpage{150}
(\byear{2000})
\doiurl{10.1016/S0550-3213(99)00662-8}
{\href{https://arxiv.org/abs/hep-th/9908064}{{arXiv:hep-th/9908064}}}
\end{barticle}
\endbibitem

\bibitem[\protect\citeauthoryear{Angelantonj et~al.}{2000}]{Angelantonj:1999ms}
\begin{barticle}
\bauthor{\bsnm{Angelantonj}, \binits{C.}},
\bauthor{\bsnm{Antoniadis}, \binits{I.}},
\bauthor{\bsnm{D'Appollonio}, \binits{G.}},
\bauthor{\bsnm{Dudas}, \binits{E.}},
\bauthor{\bsnm{Sagnotti}, \binits{A.}}:
\batitle{{Type I vacua with brane supersymmetry breaking}}.
\bjtitle{Nucl. Phys. B}
\bvolume{572},
\bfpage{36}--\blpage{70}
(\byear{2000})
\doiurl{10.1016/S0550-3213(00)00052-3}
{\href{https://arxiv.org/abs/hep-th/9911081}{{arXiv:hep-th/9911081}}}
\end{barticle}
\endbibitem

\bibitem[\protect\citeauthoryear{Pradisi and Riccioni}{2001}]{Pradisi:2001yv}
\begin{barticle}
\bauthor{\bsnm{Pradisi}, \binits{G.}},
\bauthor{\bsnm{Riccioni}, \binits{F.}}:
\batitle{{Geometric couplings and brane supersymmetry breaking}}.
\bjtitle{Nucl. Phys. B}
\bvolume{615},
\bfpage{33}--\blpage{60}
(\byear{2001})
\doiurl{10.1016/S0550-3213(01)00441-2}
{\href{https://arxiv.org/abs/hep-th/0107090}{{arXiv:hep-th/0107090}}}
\end{barticle}
\endbibitem

\bibitem[\protect\citeauthoryear{Kitazawa}{2018}]{Kitazawa:2018zys}
\begin{barticle}
\bauthor{\bsnm{Kitazawa}, \binits{N.}}:
\batitle{{Brane SUSY Breaking and the Gravitino Mass}}.
\bjtitle{JHEP}
\bvolume{04},
\bfpage{081}
(\byear{2018})
\doiurl{10.1007/JHEP04(2018)081}
{\href{https://arxiv.org/abs/1802.03088}{{arXiv:1802.03088}}}
{[hep-th]}
\end{barticle}
\endbibitem

\bibitem[\protect\citeauthoryear{Polchinski}{1988}]{Polchinski:1988jq}
\begin{barticle}
\bauthor{\bsnm{Polchinski}, \binits{J.}}:
\batitle{{Factorization of Bosonic String Amplitudes}}.
\bjtitle{Nucl. Phys. B}
\bvolume{307},
\bfpage{61}--\blpage{92}
(\byear{1988})
\doiurl{10.1016/0550-3213(88)90522-6}
\end{barticle}
\endbibitem

\bibitem[\protect\citeauthoryear{Belavin and Knizhnik}{1986}]{Belavin:1986cy}
\begin{barticle}
\bauthor{\bsnm{Belavin}, \binits{A.A.}},
\bauthor{\bsnm{Knizhnik}, \binits{V.G.}}:
\batitle{{Algebraic Geometry and the Geometry of Quantum Strings}}.
\bjtitle{Phys. Lett. B}
\bvolume{168},
\bfpage{201}--\blpage{206}
(\byear{1986})
\doiurl{10.1016/0370-2693(86)90963-9}
\end{barticle}
\endbibitem

\bibitem[\protect\citeauthoryear{Moore}{1986}]{Moore:1986rh}
\begin{barticle}
\bauthor{\bsnm{Moore}, \binits{G.W.}}:
\batitle{{Modular Forms and Two Loop String Physics}}.
\bjtitle{Phys. Lett. B}
\bvolume{176},
\bfpage{369}--\blpage{379}
(\byear{1986})
\doiurl{10.1016/0370-2693(86)90180-2}
\end{barticle}
\endbibitem

\bibitem[\protect\citeauthoryear{Kato et~al.}{1986}]{Kato:1986wj}
\begin{barticle}
\bauthor{\bsnm{Kato}, \binits{A.}},
\bauthor{\bsnm{Matsuo}, \binits{Y.}},
\bauthor{\bsnm{Odake}, \binits{S.}}:
\batitle{{Modular Invariance and Two Loop Bosonic String Vacuum Amplitude}}.
\bjtitle{Phys. Lett. B}
\bvolume{179},
\bfpage{241}--\blpage{246}
(\byear{1986})
\doiurl{10.1016/0370-2693(86)90573-3}
\end{barticle}
\endbibitem

\bibitem[\protect\citeauthoryear{Alvarez-Gaume et~al.}{1986}]{Alvarez-Gaume:1986rcs}
\begin{barticle}
\bauthor{\bsnm{Alvarez-Gaume}, \binits{L.}},
\bauthor{\bsnm{Moore}, \binits{G.W.}},
\bauthor{\bsnm{Vafa}, \binits{C.}}:
\batitle{{Theta Functions, Modular Invariance and Strings}}.
\bjtitle{Commun. Math. Phys.}
\bvolume{106},
\bfpage{1}--\blpage{40}
(\byear{1986})
\doiurl{10.1007/BF01210925}
\end{barticle}
\endbibitem

\bibitem[\protect\citeauthoryear{Martinec}{1986}]{Martinec:1986wa}
\begin{barticle}
\bauthor{\bsnm{Martinec}, \binits{E.J.}}:
\batitle{{Nonrenormalization Theorems and Fermionic String Finiteness}}.
\bjtitle{Phys. Lett. B}
\bvolume{171},
\bfpage{189}
(\byear{1986})
\doiurl{10.1016/0370-2693(86)91529-7}
\end{barticle}
\endbibitem

\bibitem[\protect\citeauthoryear{Friedan et~al.}{1986}]{Friedan:1985ge}
\begin{barticle}
\bauthor{\bsnm{Friedan}, \binits{D.}},
\bauthor{\bsnm{Martinec}, \binits{E.J.}},
\bauthor{\bsnm{Shenker}, \binits{S.H.}}:
\batitle{{Conformal invariance, supersymmetry and string theory}}.
\bjtitle{Nucl. Phys. B}
\bvolume{271},
\bfpage{93}--\blpage{165}
(\byear{1986})
\doiurl{10.1016/S0550-3213(86)80006-2}
\end{barticle}
\endbibitem

\bibitem[\protect\citeauthoryear{Fischler and Susskind}{1986a}]{Fischler:1986ci}
\begin{barticle}
\bauthor{\bsnm{Fischler}, \binits{W.}},
\bauthor{\bsnm{Susskind}, \binits{L.}}:
\batitle{{Dilaton Tadpoles, String Condensates and Scale Invariance}}.
\bjtitle{Phys. Lett. B}
\bvolume{171},
\bfpage{383}--\blpage{389}
(\byear{1986})
\doiurl{10.1016/0370-2693(86)91425-5}
\end{barticle}
\endbibitem

\bibitem[\protect\citeauthoryear{Fischler and Susskind}{1986b}]{Fischler:1986tb}
\begin{barticle}
\bauthor{\bsnm{Fischler}, \binits{W.}},
\bauthor{\bsnm{Susskind}, \binits{L.}}:
\batitle{{Dilaton Tadpoles, String Condensates and Scale Invariance. 2.}}
\bjtitle{Phys. Lett. B}
\bvolume{173},
\bfpage{262}--\blpage{264}
(\byear{1986})
\doiurl{10.1016/0370-2693(86)90514-9}
\end{barticle}
\endbibitem

\bibitem[\protect\citeauthoryear{Lovelace}{1986}]{Lovelace:1986kr}
\begin{barticle}
\bauthor{\bsnm{Lovelace}, \binits{C.}}:
\batitle{{Stability of String Vacua. 1. A New Picture of the Renormalization Group}}.
\bjtitle{Nucl. Phys. B}
\bvolume{273},
\bfpage{413}--\blpage{467}
(\byear{1986})
\doiurl{10.1016/0550-3213(86)90253-1}
\end{barticle}
\endbibitem

\bibitem[\protect\citeauthoryear{Callan et~al.}{1987}]{Callan:1986bc}
\begin{barticle}
\bauthor{\bsnm{Callan}, \binits{C.G.} \bsuffix{Jr.}},
\bauthor{\bsnm{Lovelace}, \binits{C.}},
\bauthor{\bsnm{Nappi}, \binits{C.R.}},
\bauthor{\bsnm{Yost}, \binits{S.A.}}:
\batitle{{String Loop Corrections to beta Functions}}.
\bjtitle{Nucl. Phys. B}
\bvolume{288},
\bfpage{525}--\blpage{550}
(\byear{1987})
\doiurl{10.1016/0550-3213(87)90227-6}
\end{barticle}
\endbibitem

\bibitem[\protect\citeauthoryear{Das and Rey}{1987}]{Das:1986dy}
\begin{barticle}
\bauthor{\bsnm{Das}, \binits{S.R.}},
\bauthor{\bsnm{Rey}, \binits{S.-J.}}:
\batitle{{Dilaton Condensates and Loop Effects in Open and Closed Bosonic Strings}}.
\bjtitle{Phys. Lett. B}
\bvolume{186},
\bfpage{328}--\blpage{338}
(\byear{1987})
\doiurl{10.1016/0370-2693(87)90303-0}
\end{barticle}
\endbibitem

\bibitem[\protect\citeauthoryear{Callan et~al.}{1988a}]{Callan:1988st}
\begin{barticle}
\bauthor{\bsnm{Callan}, \binits{C.G.} \bsuffix{Jr.}},
\bauthor{\bsnm{Lovelace}, \binits{C.}},
\bauthor{\bsnm{Nappi}, \binits{C.R.}},
\bauthor{\bsnm{Yost}, \binits{S.A.}}:
\batitle{{Loop Corrections to Conformal Invariance for Type 1 Superstrings}}.
\bjtitle{Phys. Lett. B}
\bvolume{206},
\bfpage{41}--\blpage{46}
(\byear{1988})
\doiurl{10.1016/0370-2693(88)91259-2}
\end{barticle}
\endbibitem

\bibitem[\protect\citeauthoryear{Callan et~al.}{1988b}]{Callan:1988wz}
\begin{barticle}
\bauthor{\bsnm{Callan}, \binits{C.G.} \bsuffix{Jr.}},
\bauthor{\bsnm{Lovelace}, \binits{C.}},
\bauthor{\bsnm{Nappi}, \binits{C.R.}},
\bauthor{\bsnm{Yost}, \binits{S.A.}}:
\batitle{{Loop Corrections to Superstring Equations of Motion}}.
\bjtitle{Nucl. Phys. B}
\bvolume{308},
\bfpage{221}--\blpage{284}
(\byear{1988})
\doiurl{10.1016/0550-3213(88)90565-2}
\end{barticle}
\endbibitem

\bibitem[\protect\citeauthoryear{Russo and Tseytlin}{1990}]{Russo:1989kq}
\begin{barticle}
\bauthor{\bsnm{Russo}, \binits{J.}},
\bauthor{\bsnm{Tseytlin}, \binits{A.A.}}:
\batitle{{RENORMALIZATION OF MULTIPLE INFINITIES AND RENORMALIZATION GROUP IN STRING LOOPS}}.
\bjtitle{Nucl. Phys. B}
\bvolume{340},
\bfpage{113}--\blpage{147}
(\byear{1990})
\doiurl{10.1016/0550-3213(90)90159-B}
\end{barticle}
\endbibitem

\bibitem[\protect\citeauthoryear{Tseytlin}{1990}]{Tseytlin:1990mv}
\begin{barticle}
\bauthor{\bsnm{Tseytlin}, \binits{A.A.}}:
\batitle{{Renormalization group and string loops}}.
\bjtitle{Int. J. Mod. Phys. A}
\bvolume{5},
\bfpage{589}--\blpage{658}
(\byear{1990})
\doiurl{10.1142/S0217751X90000301}
\end{barticle}
\endbibitem

\bibitem[\protect\citeauthoryear{Dudas et~al.}{2005}]{Dudas:2004nd}
\begin{barticle}
\bauthor{\bsnm{Dudas}, \binits{E.}},
\bauthor{\bsnm{Pradisi}, \binits{G.}},
\bauthor{\bsnm{Nicolosi}, \binits{M.}},
\bauthor{\bsnm{Sagnotti}, \binits{A.}}:
\batitle{{On tadpoles and vacuum redefinitions in string theory}}.
\bjtitle{Nucl. Phys. B}
\bvolume{708},
\bfpage{3}--\blpage{44}
(\byear{2005})
\doiurl{10.1016/j.nuclphysb.2004.11.028}
{\href{https://arxiv.org/abs/hep-th/0410101}{{arXiv:hep-th/0410101}}}
\end{barticle}
\endbibitem

\bibitem[\protect\citeauthoryear{Pius et~al.}{2014}]{Pius:2014gza}
\begin{barticle}
\bauthor{\bsnm{Pius}, \binits{R.}},
\bauthor{\bsnm{Rudra}, \binits{A.}},
\bauthor{\bsnm{Sen}, \binits{A.}}:
\batitle{{String Perturbation Theory Around Dynamically Shifted Vacuum}}.
\bjtitle{JHEP}
\bvolume{10},
\bfpage{070}
(\byear{2014})
\doiurl{10.1007/JHEP10(2014)070}
{\href{https://arxiv.org/abs/1404.6254}{{arXiv:1404.6254}}}
{[hep-th]}
\end{barticle}
\endbibitem

\bibitem[\protect\citeauthoryear{Sen}{2015}]{Sen:2015uoa}
\begin{barticle}
\bauthor{\bsnm{Sen}, \binits{A.}}:
\batitle{{Supersymmetry Restoration in Superstring Perturbation Theory}}.
\bjtitle{JHEP}
\bvolume{12},
\bfpage{075}
(\byear{2015})
\doiurl{10.1007/JHEP12(2015)075}
{\href{https://arxiv.org/abs/1508.02481}{{arXiv:1508.02481}}}
{[hep-th]}
\end{barticle}
\endbibitem

\bibitem[\protect\citeauthoryear{Kitazawa}{2008}]{Kitazawa:2008hv}
\begin{barticle}
\bauthor{\bsnm{Kitazawa}, \binits{N.}}:
\batitle{{Tadpole Resummations in String Theory}}.
\bjtitle{Phys. Lett. B}
\bvolume{660},
\bfpage{415}--\blpage{421}
(\byear{2008})
\doiurl{10.1016/j.physletb.2008.01.028}
{\href{https://arxiv.org/abs/0801.1702}{{arXiv:0801.1702}}}
{[hep-th]}
\end{barticle}
\endbibitem

\bibitem[\protect\citeauthoryear{Dudas and Mourad}{2000}]{Dudas:2000ff}
\begin{barticle}
\bauthor{\bsnm{Dudas}, \binits{E.}},
\bauthor{\bsnm{Mourad}, \binits{J.}}:
\batitle{{Brane solutions in strings with broken supersymmetry and dilaton tadpoles}}.
\bjtitle{Phys. Lett. B}
\bvolume{486},
\bfpage{172}--\blpage{178}
(\byear{2000})
\doiurl{10.1016/S0370-2693(00)00734-6}
{\href{https://arxiv.org/abs/hep-th/0004165}{{arXiv:hep-th/0004165}}}
\end{barticle}
\endbibitem

\bibitem[\protect\citeauthoryear{Basile et~al.}{2022}]{Basile:2022ypo}
\begin{barticle}
\bauthor{\bsnm{Basile}, \binits{I.}},
\bauthor{\bsnm{Raucci}, \binits{S.}},
\bauthor{\bsnm{Thom{\'e}e}, \binits{S.}}:
\batitle{{Revisiting Dudas-Mourad Compactifications}}.
\bjtitle{Universe}
\bvolume{8}(\bissue{10}),
\bfpage{544}
(\byear{2022})
\doiurl{10.3390/universe8100544}
{\href{https://arxiv.org/abs/2209.10553}{{arXiv:2209.10553}}}
{[hep-th]}
\end{barticle}
\endbibitem

\bibitem[\protect\citeauthoryear{Blumenhagen and Font}{2001}]{Blumenhagen:2000dc}
\begin{barticle}
\bauthor{\bsnm{Blumenhagen}, \binits{R.}},
\bauthor{\bsnm{Font}, \binits{A.}}:
\batitle{{Dilaton tadpoles, warped geometries and large extra dimensions for nonsupersymmetric strings}}.
\bjtitle{Nucl. Phys. B}
\bvolume{599},
\bfpage{241}--\blpage{254}
(\byear{2001})
\doiurl{10.1016/S0550-3213(01)00028-1}
{\href{https://arxiv.org/abs/hep-th/0011269}{{arXiv:hep-th/0011269}}}
\end{barticle}
\endbibitem

\bibitem[\protect\citeauthoryear{Dudas et~al.}{2003}]{Dudas:2002dg}
\begin{barticle}
\bauthor{\bsnm{Dudas}, \binits{E.}},
\bauthor{\bsnm{Mourad}, \binits{J.}},
\bauthor{\bsnm{Timirgaziu}, \binits{C.}}:
\batitle{{Time and space dependent backgrounds from nonsupersymmetric strings}}.
\bjtitle{Nucl. Phys. B}
\bvolume{660},
\bfpage{3}--\blpage{24}
(\byear{2003})
\doiurl{10.1016/S0550-3213(03)00248-7}
{\href{https://arxiv.org/abs/hep-th/0209176}{{arXiv:hep-th/0209176}}}
\end{barticle}
\endbibitem

\bibitem[\protect\citeauthoryear{Pelliconi and Sagnotti}{2021}]{Pelliconi:2021eak}
\begin{barticle}
\bauthor{\bsnm{Pelliconi}, \binits{P.}},
\bauthor{\bsnm{Sagnotti}, \binits{A.}}:
\batitle{{Integrable Models and Supersymmetry Breaking}}.
\bjtitle{Nucl. Phys. B}
\bvolume{965},
\bfpage{115363}
(\byear{2021})
\doiurl{10.1016/j.nuclphysb.2021.115363}
{\href{https://arxiv.org/abs/2102.06184}{{arXiv:2102.06184}}}
{[hep-th]}
\end{barticle}
\endbibitem

\bibitem[\protect\citeauthoryear{Mourad and Sagnotti}{2021a}]{Mourad:2021qwf}
\begin{barticle}
\bauthor{\bsnm{Mourad}, \binits{J.}},
\bauthor{\bsnm{Sagnotti}, \binits{A.}}:
\batitle{{On warped string vacuum profiles and cosmologies. Part I. Supersymmetric strings}}.
\bjtitle{JHEP}
\bvolume{12},
\bfpage{137}
(\byear{2021})
\doiurl{10.1007/JHEP12(2021)137}
{\href{https://arxiv.org/abs/2109.06852}{{arXiv:2109.06852}}}
{[hep-th]}
\end{barticle}
\endbibitem

\bibitem[\protect\citeauthoryear{Mourad and Sagnotti}{2021b}]{Mourad:2021roa}
\begin{barticle}
\bauthor{\bsnm{Mourad}, \binits{J.}},
\bauthor{\bsnm{Sagnotti}, \binits{A.}}:
\batitle{{On warped string vacuum profiles and cosmologies. Part II. Non-supersymmetric strings}}.
\bjtitle{JHEP}
\bvolume{12},
\bfpage{138}
(\byear{2021})
\doiurl{10.1007/JHEP12(2021)138}
{\href{https://arxiv.org/abs/2109.12328}{{arXiv:2109.12328}}}
{[hep-th]}
\end{barticle}
\endbibitem

\bibitem[\protect\citeauthoryear{Mourad and Sagnotti}{2022}]{Mourad:2022loy}
\begin{barticle}
\bauthor{\bsnm{Mourad}, \binits{J.}},
\bauthor{\bsnm{Sagnotti}, \binits{A.}}:
\batitle{{A 4D IIB flux vacuum and supersymmetry breaking. Part I. Fermionic spectrum}}.
\bjtitle{JHEP}
\bvolume{08},
\bfpage{301}
(\byear{2022})
\doiurl{10.1007/JHEP08(2022)301}
{\href{https://arxiv.org/abs/2206.03340}{{arXiv:2206.03340}}}
{[hep-th]}
\end{barticle}
\endbibitem

\bibitem[\protect\citeauthoryear{Mourad and Sagnotti}{2023}]{Mourad:2023ppi}
\begin{barticle}
\bauthor{\bsnm{Mourad}, \binits{J.}},
\bauthor{\bsnm{Sagnotti}, \binits{A.}}:
\batitle{{A 4D IIB flux vacuum and supersymmetry breaking. Part II. Bosonic spectrum and stability}}.
\bjtitle{JHEP}
\bvolume{11},
\bfpage{061}
(\byear{2023})
\doiurl{10.1007/JHEP11(2023)061}
{\href{https://arxiv.org/abs/2309.04026}{{arXiv:2309.04026}}}
{[hep-th]}
\end{barticle}
\endbibitem

\bibitem[\protect\citeauthoryear{Mourad and Sagnotti}{2024}]{Mourad:2023loc}
\begin{barticle}
\bauthor{\bsnm{Mourad}, \binits{J.}},
\bauthor{\bsnm{Sagnotti}, \binits{A.}}:
\batitle{{Effective orientifolds from broken supersymmetry}}.
\bjtitle{J. Phys. A}
\bvolume{57}(\bissue{3}),
\bfpage{035401}
(\byear{2024})
\doiurl{10.1088/1751-8121/ad16f8}
{\href{https://arxiv.org/abs/2309.05268}{{arXiv:2309.05268}}}
{[hep-th]}
\end{barticle}
\endbibitem

\bibitem[\protect\citeauthoryear{Mourad et~al.}{2024}]{Mourad:2024dur}
\begin{barticle}
\bauthor{\bsnm{Mourad}, \binits{J.}},
\bauthor{\bsnm{Raucci}, \binits{S.}},
\bauthor{\bsnm{Sagnotti}, \binits{A.}}:
\batitle{{Brane-like solutions and other non-supersymmetric vacua}}.
\bjtitle{JHEP}
\bvolume{10},
\bfpage{054}
(\byear{2024})
\doiurl{10.1007/JHEP10(2024)054}
{\href{https://arxiv.org/abs/2406.14926}{{arXiv:2406.14926}}}
{[hep-th]}
\end{barticle}
\endbibitem

\bibitem[\protect\citeauthoryear{Mourad and Sagnotti}{2020}]{Mourad:2020cjq}
\begin{barticle}
\bauthor{\bsnm{Mourad}, \binits{J.}},
\bauthor{\bsnm{Sagnotti}, \binits{A.}}:
\batitle{{On boundaries, charges and Fermi fields}}.
\bjtitle{Phys. Lett. B}
\bvolume{804},
\bfpage{135368}
(\byear{2020})
\doiurl{10.1016/j.physletb.2020.135368}
{\href{https://arxiv.org/abs/2002.05372}{{arXiv:2002.05372}}}
{[hep-th]}
\end{barticle}
\endbibitem

\bibitem[\protect\citeauthoryear{Raucci}{2022}]{Raucci:2022jgw}
\begin{barticle}
\bauthor{\bsnm{Raucci}, \binits{S.}}:
\batitle{{On codimension-one vacua and string theory}}.
\bjtitle{Nucl. Phys. B}
\bvolume{985},
\bfpage{116002}
(\byear{2022})
\doiurl{10.1016/j.nuclphysb.2022.116002}
{\href{https://arxiv.org/abs/2206.06399}{{arXiv:2206.06399}}}
{[hep-th]}
\end{barticle}
\endbibitem

\bibitem[\protect\citeauthoryear{Mourad and Sagnotti}{2023}]{Mourad:2023wjg}
\begin{barticle}
\bauthor{\bsnm{Mourad}, \binits{J.}},
\bauthor{\bsnm{Sagnotti}, \binits{A.}}:
\batitle{{Non-supersymmetric vacua and self-adjoint extensions}}.
\bjtitle{JHEP}
\bvolume{08},
\bfpage{041}
(\byear{2023})
\doiurl{10.1007/JHEP08(2023)041}
{\href{https://arxiv.org/abs/2305.09587}{{arXiv:2305.09587}}}
{[hep-th]}
\end{barticle}
\endbibitem

\bibitem[\protect\citeauthoryear{McNamara and Vafa}{2019}]{McNamara:2019rup}
\begin{botherref}
\oauthor{\bsnm{McNamara}, \binits{J.}},
\oauthor{\bsnm{Vafa}, \binits{C.}}:
{Cobordism Classes and the Swampland}
(2019)
{\href{https://arxiv.org/abs/1909.10355}{{arXiv:1909.10355}}}
{[hep-th]}
\end{botherref}
\endbibitem

\bibitem[\protect\citeauthoryear{Antonelli and Basile}{2019}]{Antonelli:2019nar}
\begin{barticle}
\bauthor{\bsnm{Antonelli}, \binits{R.}},
\bauthor{\bsnm{Basile}, \binits{I.}}:
\batitle{{Brane annihilation in non-supersymmetric strings}}.
\bjtitle{JHEP}
\bvolume{11},
\bfpage{021}
(\byear{2019})
\doiurl{10.1007/JHEP11(2019)021}
{\href{https://arxiv.org/abs/1908.04352}{{arXiv:1908.04352}}}
{[hep-th]}
\end{barticle}
\endbibitem

\bibitem[\protect\citeauthoryear{Buratti et~al.}{2021a}]{Buratti:2021yia}
\begin{barticle}
\bauthor{\bsnm{Buratti}, \binits{G.}},
\bauthor{\bsnm{Delgado}, \binits{M.}},
\bauthor{\bsnm{Uranga}, \binits{A.M.}}:
\batitle{{Dynamical tadpoles, stringy cobordism, and the SM from spontaneous compactification}}.
\bjtitle{JHEP}
\bvolume{06},
\bfpage{170}
(\byear{2021})
\doiurl{10.1007/JHEP06(2021)170}
{\href{https://arxiv.org/abs/2104.02091}{{arXiv:2104.02091}}}
{[hep-th]}
\end{barticle}
\endbibitem

\bibitem[\protect\citeauthoryear{Buratti et~al.}{2021b}]{Buratti:2021fiv}
\begin{barticle}
\bauthor{\bsnm{Buratti}, \binits{G.}},
\bauthor{\bsnm{Calder{\'o}n-Infante}, \binits{J.}},
\bauthor{\bsnm{Delgado}, \binits{M.}},
\bauthor{\bsnm{Uranga}, \binits{A.M.}}:
\batitle{{Dynamical Cobordism and Swampland Distance Conjectures}}.
\bjtitle{JHEP}
\bvolume{10},
\bfpage{037}
(\byear{2021})
\doiurl{10.1007/JHEP10(2021)037}
{\href{https://arxiv.org/abs/2107.09098}{{arXiv:2107.09098}}}
{[hep-th]}
\end{barticle}
\endbibitem

\bibitem[\protect\citeauthoryear{Angius et~al.}{2022}]{Angius:2022aeq}
\begin{barticle}
\bauthor{\bsnm{Angius}, \binits{R.}},
\bauthor{\bsnm{Calder{\'o}n-Infante}, \binits{J.}},
\bauthor{\bsnm{Delgado}, \binits{M.}},
\bauthor{\bsnm{Huertas}, \binits{J.}},
\bauthor{\bsnm{Uranga}, \binits{A.M.}}:
\batitle{{At the end of the world: Local Dynamical Cobordism}}.
\bjtitle{JHEP}
\bvolume{06},
\bfpage{142}
(\byear{2022})
\doiurl{10.1007/JHEP06(2022)142}
{\href{https://arxiv.org/abs/2203.11240}{{arXiv:2203.11240}}}
{[hep-th]}
\end{barticle}
\endbibitem

\bibitem[\protect\citeauthoryear{Blumenhagen et~al.}{2022}]{Blumenhagen:2022mqw}
\begin{barticle}
\bauthor{\bsnm{Blumenhagen}, \binits{R.}},
\bauthor{\bsnm{Cribiori}, \binits{N.}},
\bauthor{\bsnm{Kneissl}, \binits{C.}},
\bauthor{\bsnm{Makridou}, \binits{A.}}:
\batitle{{Dynamical cobordism of a domain wall and its companion defect 7-brane}}.
\bjtitle{JHEP}
\bvolume{08},
\bfpage{204}
(\byear{2022})
\doiurl{10.1007/JHEP08(2022)204}
{\href{https://arxiv.org/abs/2205.09782}{{arXiv:2205.09782}}}
{[hep-th]}
\end{barticle}
\endbibitem

\bibitem[\protect\citeauthoryear{Angius et~al.}{2022}]{Angius:2022mgh}
\begin{barticle}
\bauthor{\bsnm{Angius}, \binits{R.}},
\bauthor{\bsnm{Delgado}, \binits{M.}},
\bauthor{\bsnm{Uranga}, \binits{A.M.}}:
\batitle{{Dynamical Cobordism and the beginning of time: supercritical strings and tachyon condensation}}.
\bjtitle{JHEP}
\bvolume{08},
\bfpage{285}
(\byear{2022})
\doiurl{10.1007/JHEP08(2022)285}
{\href{https://arxiv.org/abs/2207.13108}{{arXiv:2207.13108}}}
{[hep-th]}
\end{barticle}
\endbibitem

\bibitem[\protect\citeauthoryear{Blumenhagen et~al.}{2023}]{Blumenhagen:2023abk}
\begin{barticle}
\bauthor{\bsnm{Blumenhagen}, \binits{R.}},
\bauthor{\bsnm{Kneissl}, \binits{C.}},
\bauthor{\bsnm{Wang}, \binits{C.}}:
\batitle{{Dynamical Cobordism Conjecture: solutions for end-of-the-world branes}}.
\bjtitle{JHEP}
\bvolume{05},
\bfpage{123}
(\byear{2023})
\doiurl{10.1007/JHEP05(2023)123}
{\href{https://arxiv.org/abs/2303.03423}{{arXiv:2303.03423}}}
{[hep-th]}
\end{barticle}
\endbibitem

\bibitem[\protect\citeauthoryear{Angius et~al.}{2023}]{Angius:2023xtu}
\begin{barticle}
\bauthor{\bsnm{Angius}, \binits{R.}},
\bauthor{\bsnm{Huertas}, \binits{J.}},
\bauthor{\bsnm{Uranga}, \binits{A.M.}}:
\batitle{{Small black hole explosions}}.
\bjtitle{JHEP}
\bvolume{06},
\bfpage{070}
(\byear{2023})
\doiurl{10.1007/JHEP06(2023)070}
{\href{https://arxiv.org/abs/2303.15903}{{arXiv:2303.15903}}}
{[hep-th]}
\end{barticle}
\endbibitem

\bibitem[\protect\citeauthoryear{Huertas and Uranga}{2023}]{Huertas:2023syg}
\begin{barticle}
\bauthor{\bsnm{Huertas}, \binits{J.}},
\bauthor{\bsnm{Uranga}, \binits{A.M.}}:
\batitle{{Aspects of dynamical cobordism in AdS/CFT}}.
\bjtitle{JHEP}
\bvolume{08},
\bfpage{140}
(\byear{2023})
\doiurl{10.1007/JHEP08(2023)140}
{\href{https://arxiv.org/abs/2306.07335}{{arXiv:2306.07335}}}
{[hep-th]}
\end{barticle}
\endbibitem

\bibitem[\protect\citeauthoryear{Angius et~al.}{2024}]{Angius:2023uqk}
\begin{barticle}
\bauthor{\bsnm{Angius}, \binits{R.}},
\bauthor{\bsnm{Makridou}, \binits{A.}},
\bauthor{\bsnm{Uranga}, \binits{A.M.}}:
\batitle{{Intersecting end of the world branes}}.
\bjtitle{JHEP}
\bvolume{03},
\bfpage{110}
(\byear{2024})
\doiurl{10.1007/JHEP03(2024)110}
{\href{https://arxiv.org/abs/2312.16286}{{arXiv:2312.16286}}}
{[hep-th]}
\end{barticle}
\endbibitem

\bibitem[\protect\citeauthoryear{Angius}{2024}]{Angius:2024zjv}
\begin{barticle}
\bauthor{\bsnm{Angius}, \binits{R.}}:
\batitle{{End of the world brane networks for infinite distance limits in CY moduli space}}.
\bjtitle{JHEP}
\bvolume{09},
\bfpage{178}
(\byear{2024})
\doiurl{10.1007/JHEP09(2024)178}
{\href{https://arxiv.org/abs/2404.14486}{{arXiv:2404.14486}}}
{[hep-th]}
\end{barticle}
\endbibitem

\bibitem[\protect\citeauthoryear{Ruiz}{2020}]{Ruiz:2020jjz}
\begin{barticle}
\bauthor{\bsnm{Ruiz}, \binits{I.}}:
\batitle{{Morse-Bott inequalities, topology change and cobordisms to nothing}}.
\bjtitle{JHEP}
\bvolume{25}(\bissue{6}),
\bfpage{30}
(\byear{2020})
\doiurl{10.1007/JHEP06(2025)030}
{\href{https://arxiv.org/abs/2410.21372}{{arXiv:2410.21372}}}
{[hep-th]}
\end{barticle}
\endbibitem

\bibitem[\protect\citeauthoryear{Huertas and Uranga}{2025}]{Huertas:2024mvy}
\begin{barticle}
\bauthor{\bsnm{Huertas}, \binits{J.}},
\bauthor{\bsnm{Uranga}, \binits{A.M.}}:
\batitle{{End of the world brane dynamics in holographic 4d $ \mathcal{N} $ = 4 SU(N) with 3d $ \mathcal{N} $ = 2 boundary conditions}}.
\bjtitle{JHEP}
\bvolume{01},
\bfpage{002}
(\byear{2025})
\doiurl{10.1007/JHEP01(2025)002}
{\href{https://arxiv.org/abs/2410.05368}{{arXiv:2410.05368}}}
{[hep-th]}
\end{barticle}
\endbibitem

\bibitem[\protect\citeauthoryear{Angius et~al.}{2025}]{Angius:2024pqk}
\begin{barticle}
\bauthor{\bsnm{Angius}, \binits{R.}},
\bauthor{\bsnm{Uranga}, \binits{A.M.}},
\bauthor{\bsnm{Wang}, \binits{C.}}:
\batitle{{End of the world boundaries for chiral quantum gravity theories}}.
\bjtitle{JHEP}
\bvolume{03},
\bfpage{064}
(\byear{2025})
\doiurl{10.1007/JHEP03(2025)064}
{\href{https://arxiv.org/abs/2410.07322}{{arXiv:2410.07322}}}
{[hep-th]}
\end{barticle}
\endbibitem

\bibitem[\protect\citeauthoryear{Wald}{1980}]{Wald:1980jn}
\begin{barticle}
\bauthor{\bsnm{Wald}, \binits{R.M.}}:
\batitle{{DYNAMICS IN NONGLOBALLY HYPERBOLIC, STATIC SPACE-TIMES}}.
\bjtitle{J. Math. Phys.}
\bvolume{21},
\bfpage{2802}--\blpage{2805}
(\byear{1980})
\doiurl{10.1063/1.524403}
\end{barticle}
\endbibitem

\bibitem[\protect\citeauthoryear{Horowitz and Marolf}{1995}]{Horowitz:1995gi}
\begin{barticle}
\bauthor{\bsnm{Horowitz}, \binits{G.T.}},
\bauthor{\bsnm{Marolf}, \binits{D.}}:
\batitle{{Quantum probes of space-time singularities}}.
\bjtitle{Phys. Rev. D}
\bvolume{52},
\bfpage{5670}--\blpage{5675}
(\byear{1995})
\doiurl{10.1103/PhysRevD.52.5670}
{\href{https://arxiv.org/abs/gr-qc/9504028}{{arXiv:gr-qc/9504028}}}
\end{barticle}
\endbibitem

\bibitem[\protect\citeauthoryear{Gubser}{2000}]{Gubser:2000nd}
\begin{barticle}
\bauthor{\bsnm{Gubser}, \binits{S.S.}}:
\batitle{{Curvature singularities: The Good, the bad, and the naked}}.
\bjtitle{Adv. Theor. Math. Phys.}
\bvolume{4},
\bfpage{679}--\blpage{745}
(\byear{2000})
\doiurl{10.4310/ATMP.2000.v4.n3.a6}
{\href{https://arxiv.org/abs/hep-th/0002160}{{arXiv:hep-th/0002160}}}
\end{barticle}
\endbibitem

\bibitem[\protect\citeauthoryear{Mourad et~al.}{2024}]{Mourad:2024mpg}
\begin{barticle}
\bauthor{\bsnm{Mourad}, \binits{J.}},
\bauthor{\bsnm{Raucci}, \binits{S.}},
\bauthor{\bsnm{Sagnotti}, \binits{A.}}:
\batitle{{Brane profiles of non-supersymmetric strings}}.
\bjtitle{JHEP}
\bvolume{09},
\bfpage{019}
(\byear{2024})
\doiurl{10.1007/JHEP09(2024)019}
{\href{https://arxiv.org/abs/2406.16327}{{arXiv:2406.16327}}}
{[hep-th]}
\end{barticle}
\endbibitem

\bibitem[\protect\citeauthoryear{Dudas and Mourad}{2001}]{Dudas:2000sn}
\begin{barticle}
\bauthor{\bsnm{Dudas}, \binits{E.}},
\bauthor{\bsnm{Mourad}, \binits{J.}}:
\batitle{{D-branes in nontachyonic 0B orientifolds}}.
\bjtitle{Nucl. Phys. B}
\bvolume{598},
\bfpage{189}--\blpage{224}
(\byear{2001})
\doiurl{10.1016/S0550-3213(00)00781-1}
{\href{https://arxiv.org/abs/hep-th/0010179}{{arXiv:hep-th/0010179}}}
\end{barticle}
\endbibitem

\bibitem[\protect\citeauthoryear{Dudas et~al.}{2011}]{Dudas:2010gi}
\begin{barticle}
\bauthor{\bsnm{Dudas}, \binits{E.}},
\bauthor{\bsnm{Kitazawa}, \binits{N.}},
\bauthor{\bsnm{Sagnotti}, \binits{A.}}:
\batitle{{On Climbing Scalars in String Theory}}.
\bjtitle{Phys. Lett. B}
\bvolume{694},
\bfpage{80}--\blpage{88}
(\byear{2011})
\doiurl{10.1016/j.physletb.2010.09.040}
{\href{https://arxiv.org/abs/1009.0874}{{arXiv:1009.0874}}}
{[hep-th]}
\end{barticle}
\endbibitem

\bibitem[\protect\citeauthoryear{Condeescu and Dudas}{2013}]{Condeescu:2013gaa}
\begin{barticle}
\bauthor{\bsnm{Condeescu}, \binits{C.}},
\bauthor{\bsnm{Dudas}, \binits{E.}}:
\batitle{{Kasner solutions, climbing scalars and big-bang singularity}}.
\bjtitle{JCAP}
\bvolume{08},
\bfpage{013}
(\byear{2013})
\doiurl{10.1088/1475-7516/2013/08/013}
{\href{https://arxiv.org/abs/1306.0911}{{arXiv:1306.0911}}}
{[hep-th]}
\end{barticle}
\endbibitem

\bibitem[\protect\citeauthoryear{Dudas et~al.}{2012}]{Dudas:2012vv}
\begin{barticle}
\bauthor{\bsnm{Dudas}, \binits{E.}},
\bauthor{\bsnm{Kitazawa}, \binits{N.}},
\bauthor{\bsnm{Patil}, \binits{S.P.}},
\bauthor{\bsnm{Sagnotti}, \binits{A.}}:
\batitle{{CMB Imprints of a Pre-Inflationary Climbing Phase}}.
\bjtitle{JCAP}
\bvolume{05},
\bfpage{012}
(\byear{2012})
\doiurl{10.1088/1475-7516/2012/05/012}
{\href{https://arxiv.org/abs/1202.6630}{{arXiv:1202.6630}}}
{[hep-th]}
\end{barticle}
\endbibitem

\bibitem[\protect\citeauthoryear{Sagnotti}{2013}]{Sagnotti:2013ica}
\begin{botherref}
\oauthor{\bsnm{Sagnotti}, \binits{A.}}:
{Brane SUSY breaking and inflation: implications for scalar fields and CMB distortion},
465--473
(2013)
\doiurl{10.1134/S1547477114070395}
{\href{https://arxiv.org/abs/1303.6685}{{arXiv:1303.6685}}}
{[hep-th]}
\end{botherref}
\endbibitem

\bibitem[\protect\citeauthoryear{Kitazawa and Sagnotti}{2014}]{Kitazawa:2014dya}
\begin{barticle}
\bauthor{\bsnm{Kitazawa}, \binits{N.}},
\bauthor{\bsnm{Sagnotti}, \binits{A.}}:
\batitle{{Pre-inflationary clues from String Theory?}}
\bjtitle{JCAP}
\bvolume{04},
\bfpage{017}
(\byear{2014})
\doiurl{10.1088/1475-7516/2014/04/017}
{\href{https://arxiv.org/abs/1402.1418}{{arXiv:1402.1418}}}
{[hep-th]}
\end{barticle}
\endbibitem

\bibitem[\protect\citeauthoryear{Kitazawa and Sagnotti}{2015a}]{Kitazawa:2014mca}
\begin{barticle}
\bauthor{\bsnm{Kitazawa}, \binits{N.}},
\bauthor{\bsnm{Sagnotti}, \binits{A.}}:
\batitle{{String theory clues for the low{\textendash}$\ell$ CMB ?}}
\bjtitle{EPJ Web Conf.}
\bvolume{95},
\bfpage{03031}
(\byear{2015})
\doiurl{10.1051/epjconf/20159503031}
{\href{https://arxiv.org/abs/1411.6396}{{arXiv:1411.6396}}}
{[hep-th]}
\end{barticle}
\endbibitem

\bibitem[\protect\citeauthoryear{Kitazawa and Sagnotti}{2015b}]{Kitazawa:2015uda}
\begin{barticle}
\bauthor{\bsnm{Kitazawa}, \binits{N.}},
\bauthor{\bsnm{Sagnotti}, \binits{A.}}:
\batitle{{A string-inspired model for the low-$\ell$ CMB}}.
\bjtitle{Mod. Phys. Lett. A}
\bvolume{30}(\bissue{28}),
\bfpage{1550137}
(\byear{2015})
\doiurl{10.1142/S0217732315501370}
{\href{https://arxiv.org/abs/1503.04483}{{arXiv:1503.04483}}}
{[hep-th]}
\end{barticle}
\endbibitem

\bibitem[\protect\citeauthoryear{Angelantonj and Armoni}{2000a}]{Angelantonj:1999qg}
\begin{barticle}
\bauthor{\bsnm{Angelantonj}, \binits{C.}},
\bauthor{\bsnm{Armoni}, \binits{A.}}:
\batitle{{Nontachyonic type 0B orientifolds, nonsupersymmetric gauge theories and cosmological RG flow}}.
\bjtitle{Nucl. Phys. B}
\bvolume{578},
\bfpage{239}--\blpage{258}
(\byear{2000})
\doiurl{10.1016/S0550-3213(00)00136-X}
{\href{https://arxiv.org/abs/hep-th/9912257}{{arXiv:hep-th/9912257}}}
\end{barticle}
\endbibitem

\bibitem[\protect\citeauthoryear{Angelantonj and Armoni}{2000b}]{Angelantonj:2000kh}
\begin{barticle}
\bauthor{\bsnm{Angelantonj}, \binits{C.}},
\bauthor{\bsnm{Armoni}, \binits{A.}}:
\batitle{{RG flow, Wilson loops and the dilaton tadpole}}.
\bjtitle{Phys. Lett. B}
\bvolume{482},
\bfpage{329}--\blpage{336}
(\byear{2000})
\doiurl{10.1016/S0370-2693(00)00475-5}
{\href{https://arxiv.org/abs/hep-th/0003050}{{arXiv:hep-th/0003050}}}
\end{barticle}
\endbibitem

\bibitem[\protect\citeauthoryear{Basile}{2021}]{Basile:2021mkd}
\begin{barticle}
\bauthor{\bsnm{Basile}, \binits{I.}}:
\batitle{{Supersymmetry breaking, brane dynamics and Swampland conjectures}}.
\bjtitle{JHEP}
\bvolume{10},
\bfpage{080}
(\byear{2021})
\doiurl{10.1007/JHEP10(2021)080}
{\href{https://arxiv.org/abs/2106.04574}{{arXiv:2106.04574}}}
{[hep-th]}
\end{barticle}
\endbibitem

\bibitem[\protect\citeauthoryear{Basile et~al.}{2025}]{Basile:2025lek}
\begin{barticle}
\bauthor{\bsnm{Basile}, \binits{I.}},
\bauthor{\bsnm{Borys}, \binits{A.}},
\bauthor{\bsnm{Masias}, \binits{J.}}:
\batitle{{Dynamical dark energy in 0{\textquoteright}B braneworlds}}.
\bjtitle{Eur. Phys. J. C}
\bvolume{85}(\bissue{9}),
\bfpage{986}
(\byear{2025})
\doiurl{10.1140/epjc/s10052-025-14706-9}
{\href{https://arxiv.org/abs/2502.20438}{{arXiv:2502.20438}}}
{[hep-th]}
\end{barticle}
\endbibitem

\bibitem[\protect\citeauthoryear{Gubser and Mitra}{2002}]{Gubser:2001zr}
\begin{barticle}
\bauthor{\bsnm{Gubser}, \binits{S.S.}},
\bauthor{\bsnm{Mitra}, \binits{I.}}:
\batitle{{Some interesting violations of the Breitenlohner-Freedman bound}}.
\bjtitle{JHEP}
\bvolume{07},
\bfpage{044}
(\byear{2002})
\doiurl{10.1088/1126-6708/2002/07/044}
{\href{https://arxiv.org/abs/hep-th/0108239}{{arXiv:hep-th/0108239}}}
\end{barticle}
\endbibitem

\bibitem[\protect\citeauthoryear{Mourad and Sagnotti}{2017}]{Mourad:2016xbk}
\begin{barticle}
\bauthor{\bsnm{Mourad}, \binits{J.}},
\bauthor{\bsnm{Sagnotti}, \binits{A.}}:
\batitle{{$AdS$ Vacua from Dilaton Tadpoles and Form Fluxes}}.
\bjtitle{Phys. Lett. B}
\bvolume{768},
\bfpage{92}--\blpage{96}
(\byear{2017})
\doiurl{10.1016/j.physletb.2017.02.053}
{\href{https://arxiv.org/abs/1612.08566}{{arXiv:1612.08566}}}
{[hep-th]}
\end{barticle}
\endbibitem

\bibitem[\protect\citeauthoryear{Raucci}{2023}]{Raucci:2022bjw}
\begin{barticle}
\bauthor{\bsnm{Raucci}, \binits{S.}}:
\batitle{{On new vacua of non-supersymmetric strings}}.
\bjtitle{Phys. Lett. B}
\bvolume{837},
\bfpage{137663}
(\byear{2023})
\doiurl{10.1016/j.physletb.2022.137663}
{\href{https://arxiv.org/abs/2209.06537}{{arXiv:2209.06537}}}
{[hep-th]}
\end{barticle}
\endbibitem

\bibitem[\protect\citeauthoryear{Basile}{2023}]{Basile:2022zee}
\begin{barticle}
\bauthor{\bsnm{Basile}, \binits{I.}}:
\batitle{{Emergent Strings at an Infinite Distance with Broken Supersymmetry}}.
\bjtitle{Astronomy}
\bvolume{2}(\bissue{3}),
\bfpage{206}--\blpage{225}
(\byear{2023})
\doiurl{10.3390/astronomy2030015}
{\href{https://arxiv.org/abs/2201.08851}{{arXiv:2201.08851}}}
{[hep-th]}
\end{barticle}
\endbibitem

\bibitem[\protect\citeauthoryear{De~Luca et~al.}{2022}]{DeLuca:2021pej}
\begin{barticle}
\bauthor{\bsnm{De~Luca}, \binits{G.B.}},
\bauthor{\bsnm{Silverstein}, \binits{E.}},
\bauthor{\bsnm{Torroba}, \binits{G.}}:
\batitle{{Hyperbolic compactification of M-theory and de Sitter quantum gravity}}.
\bjtitle{SciPost Phys.}
\bvolume{12}(\bissue{3}),
\bfpage{083}
(\byear{2022})
\doiurl{10.21468/SciPostPhys.12.3.083}
{\href{https://arxiv.org/abs/2104.13380}{{arXiv:2104.13380}}}
{[hep-th]}
\end{barticle}
\endbibitem

\bibitem[\protect\citeauthoryear{Luca et~al.}{2023}]{Luca:2022inb}
\begin{barticle}
\bauthor{\bsnm{Luca}, \binits{G.B.D.}},
\bauthor{\bsnm{De~Ponti}, \binits{N.}},
\bauthor{\bsnm{Mondino}, \binits{A.}},
\bauthor{\bsnm{Tomasiello}, \binits{A.}}:
\batitle{{Gravity from thermodynamics: Optimal transport and negative effective dimensions}}.
\bjtitle{SciPost Phys.}
\bvolume{15}(\bissue{2}),
\bfpage{039}
(\byear{2023})
\doiurl{10.21468/SciPostPhys.15.2.039}
{\href{https://arxiv.org/abs/2212.02511}{{arXiv:2212.02511}}}
{[hep-th]}
\end{barticle}
\endbibitem

\bibitem[\protect\citeauthoryear{Valeixo~Bento and Montero}{2026}]{ValeixoBento:2025yhz}
\begin{barticle}
\bauthor{\bsnm{Valeixo~Bento}, \binits{B.}},
\bauthor{\bsnm{Montero}, \binits{M.}}:
\batitle{{An M-theory dS maximum from Casimir energies on Riemann-flat manifolds}}.
\bjtitle{JHEP}
\bvolume{01},
\bfpage{099}
(\byear{2026})
\doiurl{10.1007/JHEP01(2026)099}
{\href{https://arxiv.org/abs/2507.02037}{{arXiv:2507.02037}}}
{[hep-th]}
\end{barticle}
\endbibitem

\bibitem[\protect\citeauthoryear{Dall'Agata and Zwirner}{2025}]{DallAgata:2025jii}
\begin{barticle}
\bauthor{\bsnm{Dall'Agata}, \binits{G.}},
\bauthor{\bsnm{Zwirner}, \binits{F.}}:
\batitle{{Supersymmetry-breaking compactifications on Riemann-flat manifolds}}.
\bjtitle{JHEP}
\bvolume{12},
\bfpage{131}
(\byear{2025})
\doiurl{10.1007/JHEP12(2025)131}
{\href{https://arxiv.org/abs/2507.02339}{{arXiv:2507.02339}}}
{[hep-th]}
\end{barticle}
\endbibitem

\bibitem[\protect\citeauthoryear{Aparici et~al.}{2025}]{Aparici:2025kjj}
\begin{botherref}
\oauthor{\bsnm{Aparici}, \binits{M.}},
\oauthor{\bsnm{Basile}, \binits{I.}},
\oauthor{\bsnm{Risso}, \binits{N.}}:
{Instabilities in scale-separated Casimir vacua}
(2025)
{\href{https://arxiv.org/abs/2507.17802}{{arXiv:2507.17802}}}
{[hep-th]}
\end{botherref}
\endbibitem

\bibitem[\protect\citeauthoryear{Basile and Lanza}{2020}]{Basile:2020mpt}
\begin{barticle}
\bauthor{\bsnm{Basile}, \binits{I.}},
\bauthor{\bsnm{Lanza}, \binits{S.}}:
\batitle{{de Sitter in non-supersymmetric string theories: no-go theorems and brane-worlds}}.
\bjtitle{JHEP}
\bvolume{10},
\bfpage{108}
(\byear{2020})
\doiurl{10.1007/JHEP10(2020)108}
{\href{https://arxiv.org/abs/2007.13757}{{arXiv:2007.13757}}}
{[hep-th]}
\end{barticle}
\endbibitem

\bibitem[\protect\citeauthoryear{Maldacena and Nunez}{2001}]{Maldacena:2000mw}
\begin{barticle}
\bauthor{\bsnm{Maldacena}, \binits{J.M.}},
\bauthor{\bsnm{Nunez}, \binits{C.}}:
\batitle{{Supergravity description of field theories on curved manifolds and a no go theorem}}.
\bjtitle{Int. J. Mod. Phys. A}
\bvolume{16},
\bfpage{822}--\blpage{855}
(\byear{2001})
\doiurl{10.1142/S0217751X01003937}
{\href{https://arxiv.org/abs/hep-th/0007018}{{arXiv:hep-th/0007018}}}
\end{barticle}
\endbibitem

\bibitem[\protect\citeauthoryear{Fabinger and Horava}{2000}]{Fabinger:2000jd}
\begin{barticle}
\bauthor{\bsnm{Fabinger}, \binits{M.}},
\bauthor{\bsnm{Horava}, \binits{P.}}:
\batitle{{Casimir effect between world branes in heterotic M theory}}.
\bjtitle{Nucl. Phys. B}
\bvolume{580},
\bfpage{243}--\blpage{263}
(\byear{2000})
\doiurl{10.1016/S0550-3213(00)00255-8}
{\href{https://arxiv.org/abs/hep-th/0002073}{{arXiv:hep-th/0002073}}}
\end{barticle}
\endbibitem

\bibitem[\protect\citeauthoryear{McGreevy and Silverstein}{2005}]{McGreevy:2005ci}
\begin{barticle}
\bauthor{\bsnm{McGreevy}, \binits{J.}},
\bauthor{\bsnm{Silverstein}, \binits{E.}}:
\batitle{{The Tachyon at the end of the universe}}.
\bjtitle{JHEP}
\bvolume{08},
\bfpage{090}
(\byear{2005})
\doiurl{10.1088/1126-6708/2005/08/090}
{\href{https://arxiv.org/abs/hep-th/0506130}{{arXiv:hep-th/0506130}}}
\end{barticle}
\endbibitem

\bibitem[\protect\citeauthoryear{Adams et~al.}{2005}]{Adams:2005rb}
\begin{barticle}
\bauthor{\bsnm{Adams}, \binits{A.}},
\bauthor{\bsnm{Liu}, \binits{X.}},
\bauthor{\bsnm{McGreevy}, \binits{J.}},
\bauthor{\bsnm{Saltman}, \binits{A.}},
\bauthor{\bsnm{Silverstein}, \binits{E.}}:
\batitle{{Things fall apart: Topology change from winding tachyons}}.
\bjtitle{JHEP}
\bvolume{10},
\bfpage{033}
(\byear{2005})
\doiurl{10.1088/1126-6708/2005/10/033}
{\href{https://arxiv.org/abs/hep-th/0502021}{{arXiv:hep-th/0502021}}}
\end{barticle}
\endbibitem

\bibitem[\protect\citeauthoryear{Delgado}{2024}]{Delgado:2023uqk}
\begin{barticle}
\bauthor{\bsnm{Delgado}, \binits{M.}}:
\batitle{{The bubble of nothing under T-duality}}.
\bjtitle{JHEP}
\bvolume{05},
\bfpage{333}
(\byear{2024})
\doiurl{10.1007/JHEP05(2024)333}
{\href{https://arxiv.org/abs/2312.09291}{{arXiv:2312.09291}}}
{[hep-th]}
\end{barticle}
\endbibitem

\bibitem[\protect\citeauthoryear{Witten}{1982}]{Witten:1981gj}
\begin{barticle}
\bauthor{\bsnm{Witten}, \binits{E.}}:
\batitle{{Instability of the Kaluza-Klein Vacuum}}.
\bjtitle{Nucl. Phys. B}
\bvolume{195},
\bfpage{481}--\blpage{492}
(\byear{1982})
\doiurl{10.1016/0550-3213(82)90007-4}
\end{barticle}
\endbibitem

\bibitem[\protect\citeauthoryear{Garc{\'\i}a~Etxebarria et~al.}{2020}]{GarciaEtxebarria:2020xsr}
\begin{barticle}
\bauthor{\bsnm{Garc{\'\i}a~Etxebarria}, \binits{I.}},
\bauthor{\bsnm{Montero}, \binits{M.}},
\bauthor{\bsnm{Sousa}, \binits{K.}},
\bauthor{\bsnm{Valenzuela}, \binits{I.}}:
\batitle{{Nothing is certain in string compactifications}}.
\bjtitle{JHEP}
\bvolume{12},
\bfpage{032}
(\byear{2020})
\doiurl{10.1007/JHEP12(2020)032}
{\href{https://arxiv.org/abs/2005.06494}{{arXiv:2005.06494}}}
{[hep-th]}
\end{barticle}
\endbibitem

\bibitem[\protect\citeauthoryear{Witten}{1981}]{Witten:1981mf}
\begin{barticle}
\bauthor{\bsnm{Witten}, \binits{E.}}:
\batitle{{A Simple Proof of the Positive Energy Theorem}}.
\bjtitle{Commun. Math. Phys.}
\bvolume{80},
\bfpage{381}
(\byear{1981})
\doiurl{10.1007/BF01208277}
\end{barticle}
\endbibitem

\bibitem[\protect\citeauthoryear{Nester}{1981}]{Nester:1981bjx}
\begin{barticle}
\bauthor{\bsnm{Nester}, \binits{J.A.}}:
\batitle{{A New gravitational energy expression with a simple positivity proof}}.
\bjtitle{Phys. Lett. A}
\bvolume{83},
\bfpage{241}
(\byear{1981})
\doiurl{10.1016/0375-9601(81)90972-5}
\end{barticle}
\endbibitem

\bibitem[\protect\citeauthoryear{Boucher}{1984}]{Boucher:1984yx}
\begin{barticle}
\bauthor{\bsnm{Boucher}, \binits{W.}}:
\batitle{{POSITIVE ENERGY WITHOUT SUPERSYMMETRY}}.
\bjtitle{Nucl. Phys. B}
\bvolume{242},
\bfpage{282}--\blpage{296}
(\byear{1984})
\doiurl{10.1016/0550-3213(84)90394-8}
\end{barticle}
\endbibitem

\bibitem[\protect\citeauthoryear{Townsend}{1984}]{Townsend:1984iu}
\begin{barticle}
\bauthor{\bsnm{Townsend}, \binits{P.K.}}:
\batitle{{Positive Energy and the Scalar Potential in Higher Dimensional (Super)gravity Theories}}.
\bjtitle{Phys. Lett. B}
\bvolume{148},
\bfpage{55}--\blpage{59}
(\byear{1984})
\doiurl{10.1016/0370-2693(84)91610-1}
\end{barticle}
\endbibitem

\bibitem[\protect\citeauthoryear{Skenderis and Townsend}{1999}]{Skenderis:1999mm}
\begin{barticle}
\bauthor{\bsnm{Skenderis}, \binits{K.}},
\bauthor{\bsnm{Townsend}, \binits{P.K.}}:
\batitle{{Gravitational stability and renormalization group flow}}.
\bjtitle{Phys. Lett. B}
\bvolume{468},
\bfpage{46}--\blpage{51}
(\byear{1999})
\doiurl{10.1016/S0370-2693(99)01212-5}
{\href{https://arxiv.org/abs/hep-th/9909070}{{arXiv:hep-th/9909070}}}
\end{barticle}
\endbibitem

\bibitem[\protect\citeauthoryear{Freedman et~al.}{2004}]{Freedman:2003ax}
\begin{barticle}
\bauthor{\bsnm{Freedman}, \binits{D.Z.}},
\bauthor{\bsnm{Nunez}, \binits{C.}},
\bauthor{\bsnm{Schnabl}, \binits{M.}},
\bauthor{\bsnm{Skenderis}, \binits{K.}}:
\batitle{{Fake supergravity and domain wall stability}}.
\bjtitle{Phys. Rev. D}
\bvolume{69},
\bfpage{104027}
(\byear{2004})
\doiurl{10.1103/PhysRevD.69.104027}
{\href{https://arxiv.org/abs/hep-th/0312055}{{arXiv:hep-th/0312055}}}
\end{barticle}
\endbibitem

\bibitem[\protect\citeauthoryear{Townsend}{2008}]{Townsend:2007aw}
\begin{barticle}
\bauthor{\bsnm{Townsend}, \binits{P.K.}}:
\batitle{{Hamilton-Jacobi mechanics from pseudo-supersymmetry}}.
\bjtitle{Class. Quant. Grav.}
\bvolume{25},
\bfpage{045017}
(\byear{2008})
\doiurl{10.1088/0264-9381/25/4/045017}
{\href{https://arxiv.org/abs/0710.5178}{{arXiv:0710.5178}}}
{[hep-th]}
\end{barticle}
\endbibitem

\bibitem[\protect\citeauthoryear{Trigiante et~al.}{2012}]{Trigiante:2012eb}
\begin{barticle}
\bauthor{\bsnm{Trigiante}, \binits{M.}},
\bauthor{\bsnm{Van~Riet}, \binits{T.}},
\bauthor{\bsnm{Vercnocke}, \binits{B.}}:
\batitle{{Fake supersymmetry versus Hamilton-Jacobi}}.
\bjtitle{JHEP}
\bvolume{05},
\bfpage{078}
(\byear{2012})
\doiurl{10.1007/JHEP05(2012)078}
{\href{https://arxiv.org/abs/1203.3194}{{arXiv:1203.3194}}}
{[hep-th]}
\end{barticle}
\endbibitem

\bibitem[\protect\citeauthoryear{Danielsson et~al.}{2016}]{Danielsson:2016rmq}
\begin{barticle}
\bauthor{\bsnm{Danielsson}, \binits{U.H.}},
\bauthor{\bsnm{Dibitetto}, \binits{G.}},
\bauthor{\bsnm{Vargas}, \binits{S.C.}}:
\batitle{{Universal isolation in the AdS landscape}}.
\bjtitle{Phys. Rev. D}
\bvolume{94}(\bissue{12}),
\bfpage{126002}
(\byear{2016})
\doiurl{10.1103/PhysRevD.94.126002}
{\href{https://arxiv.org/abs/1605.09289}{{arXiv:1605.09289}}}
{[hep-th]}
\end{barticle}
\endbibitem

\bibitem[\protect\citeauthoryear{Raucci}{2023}]{Raucci:2023xgx}
\begin{barticle}
\bauthor{\bsnm{Raucci}, \binits{S.}}:
\batitle{{Fake supersymmetry with tadpole potentials}}.
\bjtitle{JHEP}
\bvolume{07},
\bfpage{078}
(\byear{2023})
\doiurl{10.1007/JHEP07(2023)078}
{\href{https://arxiv.org/abs/2304.12717}{{arXiv:2304.12717}}}
{[hep-th]}
\end{barticle}
\endbibitem

\bibitem[\protect\citeauthoryear{Giri et~al.}{2022}]{Giri:2021eob}
\begin{barticle}
\bauthor{\bsnm{Giri}, \binits{S.}},
\bauthor{\bsnm{Martucci}, \binits{L.}},
\bauthor{\bsnm{Tomasiello}, \binits{A.}}:
\batitle{{On the stability of string theory vacua}}.
\bjtitle{JHEP}
\bvolume{04},
\bfpage{054}
(\byear{2022})
\doiurl{10.1007/JHEP04(2022)054}
{\href{https://arxiv.org/abs/2112.10795}{{arXiv:2112.10795}}}
{[hep-th]}
\end{barticle}
\endbibitem

\bibitem[\protect\citeauthoryear{Menet}{2024a}]{Menet:2023rnt}
\begin{barticle}
\bauthor{\bsnm{Menet}, \binits{V.}}:
\batitle{{New non-supersymmetric flux vacua from generalised calibrations}}.
\bjtitle{JHEP}
\bvolume{05},
\bfpage{100}
(\byear{2024})
\doiurl{10.1007/JHEP05(2024)100}
{\href{https://arxiv.org/abs/2311.12115}{{arXiv:2311.12115}}}
{[hep-th]}
\end{barticle}
\endbibitem

\bibitem[\protect\citeauthoryear{Menet}{2024b}]{Menet:2023rml}
\begin{barticle}
\bauthor{\bsnm{Menet}, \binits{V.}}:
\batitle{{D-terms in generalised complex geometry}}.
\bjtitle{JHEP}
\bvolume{07},
\bfpage{071}
(\byear{2024})
\doiurl{10.1007/JHEP07(2024)071}
{\href{https://arxiv.org/abs/2312.04517}{{arXiv:2312.04517}}}
{[hep-th]}
\end{barticle}
\endbibitem

\bibitem[\protect\citeauthoryear{Menet and Tomasiello}{2025}]{Menet:2025nbf}
\begin{barticle}
\bauthor{\bsnm{Menet}, \binits{V.}},
\bauthor{\bsnm{Tomasiello}, \binits{A.}}:
\batitle{{Stability of non-supersymmetric vacua from calibrations}}.
\bjtitle{JHEP}
\bvolume{11},
\bfpage{070}
(\byear{2025})
\doiurl{10.1007/JHEP11(2025)070}
{\href{https://arxiv.org/abs/2507.02787}{{arXiv:2507.02787}}}
{[hep-th]}
\end{barticle}
\endbibitem

\end{thebibliography}

\end{document}